# VLA OBSERVATIONS OF 9 EXTENDED GREEN OBJECTS IN THE MILKY WAY: UBIQUITOUS WEAK, COMPACT CONTINUUM EMISSION, AND MULTI-EPOCH EMISSION FROM $CH_3OH$, $H_2O$, AND $NH_3$ MASERS

A. P. M. TOWNER[1,2], C. L. BROGAN[2], T. R. HUNTER[2], C. J. CYGANOWSKI[3]

[1]Department of Astronomy, University of Florida, 211 Bryant Space Science Center, P.O. Box 112055, Gainesville, FL 32611, USA
[2]National Radio Astronomy Observatory, 520 Edgemont Rd, Charlottesville, VA 22903, USA
[3]Scottish Universities Physics Alliance (SUPA), School of Physics and Astronomy, University of St. Andrews, North Haugh, St Andrews, Fife KY16 9SS, UK

## ABSTRACT

We have observed a sample of 9 Extended Green Objects (EGOs) at 1.3 and 5 cm with the VLA with subarcsecond resolution and ~7-14 μJy beam$^{-1}$-sensitivities in order to characterize centimeter continuum emission as it first appears in these massive protoclusters. We find EGO-associated continuum emission – within 1″ of the extended 4.5 μm emission – in every field, which is typically faint (order $10^1$-$10^2$ μJy) and compact (unresolved at 0.″3-0.″5). The derived spectral indices of our 36 total detections are consistent with a wide array of physical processes, including both non-thermal (19% of detections) and thermal free-free processes (e.g. ionized jets and compact H II regions, 78% of sample), and warm dust (1 source). We also find EGO-associated 6.7 GHz $CH_3OH$ and 22 GHz $H_2O$ maser emission in 100% of the sample, and $NH_3$ (3,3) masers in ~45%; we do not detect any $NH_3$ (6,6) masers at ~ 5.6 mJy beam$^{-1}$ sensitivity. We find statistically-significant correlations between $L_{radio}$ and $L_{bol}$ at two physical scales and three frequencies, consistent with thermal emission from ionized jets, but no correlation between $L_{H_2O}$ and $L_{radio}$ for our sample. From these data, we conclude that EGOs likely host multiple different centimeter continuum-producing processes simultaneously. Additionally, at our ~1000 au resolution, we find that all EGOs except G18.89–0.47 contain 1~2 massive sources based on the presence of $CH_3OH$ maser groups, which is consistent with our previous work suggesting that these are typical massive protoclusters in which only one to a few of the YSOs are massive.

## 1. INTRODUCTION

Most stars form in groups called protoclusters (Lada & Lada 2003). Massive protoclusters - clusters of protostars in which at least one member has $M \gtrsim 8M_\odot$ - differ from low-mass star-forming regions in their total mass and bolometric luminosity, and in the internal distribution of protostars within the clump. Unlike low-mass protostars, which can be found forming in relative isolation within a star-forming group, massive protostars are consistently found with close ($\lesssim$0.05 pc, 10,000 au) protostellar companions of varying masses within massive protoclusters (Cyganowski et al. 2017; Beuther et al. 2018; Hunter et al. 2017). Theoretical explanations for these differences include Hierarchical Collapse models of protocluster formation, in which nearby protostars compete for the same overall gas reservoir and form sub-clusters within a protocluster (Bonnell & Bate 2006; Vázquez-Semadeni et al. 2017), and Monolithic Collapse models, in which individual stars or binary systems form from single cores which largely do not interact with each other as they evolve but whose initial distribution may still be environment-dependent (McKee & Tan 2002; Banerjee & Kroupa 2017).

Distinguishing between these theoretical possibilities requires observing massive star-forming regions in a state of evolution late enough that the presence of massive protostars can be confirmed but early enough that the natal clump remains intact. Crucially, these observations must have high enough sensitivity and angular resolution that the small spatial scales in question ($\lesssim$5,000 au) can be directly observed. Interferometric centimeter continuum observations have become a common tool for observing massive protoclusters, both because of the high sensitivities and angular resolutions available and because multiple processes important to massive star formation emit continuum radiation in the centimeter regime.

The primary sources of centimeter continuum emission are expected to be: 1) emission from nascent, radiatively-ionized H II regions, 2) thermal free-free emission from ionized jets or stellar winds, 3) the Rayleigh-Jeans tail of warm dust emission, and 4) non-thermal synchrotron emission. Nascent H II regions have been predicted to exhibit a range of morphologies from bipolar to spherical, or even sizes and morphologies that are variable with time (Keto 2003, 2007; Klessen et al. 2011); additionally, some emission could also come not from an expanding radiatively-ionized region, but from radiatively-ionized accretion flows onto the central protostar (Tanaka et al. 2016). Despite this broad range of potential origins, centimeter continuum emission from very young

H II regions is generally expected to be weak ($\lesssim$1 mJy) and compact ($\lesssim$2,000 au), as the ionized regions themselves are still quite small (Purser et al. 2016; Rosero et al. 2016, and references therein). It may exhibit a spectral index $\alpha$ (where $S_\nu \sim \nu^\alpha$) anywhere between $\alpha$ = -0.1 (optically thin free-free emission) and $\alpha$ = 2 (optically thick thermal emission), depending on free-free optical depth ($\tau_{ff}$, which is a product of size, electron density ($n_e$), and the density gradient)..

Thermal free-free emission from protostellar jets is generally assumed to come from gas which was originally ionized in shocks (Curiel et al. 1987), and it can exhibit a similar range of spectral indices as nascent H II regions. Reynolds (1986) show in detail how an unresolved, partially-ionized bipolar jet can produce a spectral index between $-0.1 \leq \alpha \leq 2$, depending on ionization fraction, acceleration or recombination within the flow, and density gradients. Measurements of jet properties such as mass-loss rate and total momentum can help to constrain important properties of both individual protostars (e.g. accretion rate) and energy feedback into the protocluster system as a whole. Additionally, some theories of H II region formation require the formation of jets before the formation of H II regions (Tanaka et al. 2016), while some observations have shown a contemporaneous jet and H II region in the same object (Guzmán et al. 2016). Given the causal link between accretion and ejection (Frank et al. 2014), jets are also one potential avenue for probing massive accretion disks around Massive Young Stellar Objects (MYSOs), of which there are currently very few direct observations (see, e.g., Johnston et al. 2020; Ilee et al. 2018). Jets have largely been treated as distinguishable from HC H II regions in the radio regime through their comparatively elongated morphologies in spatially-resolved observations, but their predicted aggregate properties (weak centimeter continuum emission, compact except at very high resolutions, and centimeter-regime spectral indices) are quite similar to those predicted for nascent H II regions.

While the Rayleigh-Jeans tail of thermal emission from warm dust may be optically thick in the millimeter and submillimeter regimes, this is unlikely to be true at centimeter wavelengths. Instead, one should expect $\alpha \sim 2 + \beta$, where $\beta$ is the grain opacity spectral index (generally taken to be between 1.5 and 2; see, e.g, Towner et al. 2019; Tigé et al. 2017; Brogan et al. 2016). Non-thermal emission will have $\alpha <$ -0.1, and could potentially come from the magnetospheres of pre-main sequence (PMS) stars, or from synchrotron emission from relativistically-accelerated electrons in shocks driven by strong jets (Carrasco-González et al. 2010; Reid et al. 1995).

Recent interferometric observations of massive star-forming regions have shown ubiquitous weak, compact centimeter continuum emission that is consistent with thermal free-free emission. Purser et al. (2016) targeted 49 massive star-forming regions at 5.5, 9.0, 17.0, and 22.8 GHz using the Australia Telescope Compact Array (ATCA), with typical noise levels of 17, 20, 40, and $\sim$85 μJy beam$^{-1}$ in the four bands, respectively, and sensitivity to 0.″5-16.″5 angular scales. They report a high incidence of jets in their data: based on source spectral index and morphology, line emission, and radio and bolometric luminosities, they identify 26 of their 45 detections as jets or jet candidates, 14 as H II regions, and the others as either disk winds or of ambiguous origin. Rosero et al. (2016) use the VLA to examine 58 high-mass star forming regions at 6 and 1.3 cm with $\sim$0.″3 resolution at both bands and 5 and 9 μJy beam$^{-1}$ sensitivities, respectively. The sample was selected to span a range of evolutionary states, based on weak (<1 mJy) centimeter continuum emission, overall clump temperature, and mid-infrared properties. Rosero et al. (2016) detect a total of 70 centimeter-continuum sources in 34 of their targets. Rosero et al. (2019b) report that 80% of these detections have spectral indices consistent with thermal free-free emission from ionized gas ($-0.1 \leq \alpha \leq 2$). They conclude that at least 30% of their detections are likely to be ionized jets associated with massive protostars but note that, for the most compact sources, they cannot rule out the possibility of pressure-confined or gravitationally-trapped HC H II regions with the current data.

Finally, the Protostellar Outflows and the EarliesT Stages (POETS) survey (Moscadelli et al. 2016; Sanna et al. 2018, and subsequent papers) specifically targets outflows and jets in massive star-forming regions. This sample of 36 sources was drawn from data from the Bar and Spiral Structure Legacy (BeSSeL) Survey as published in Reid et al. (2014). Sources were selected on the basis of having rich $H_2O$-maser emission (>10 sites of $H_2O$ masers stronger than 1 Jy) in the BeSSeL data, $L_{bol}$ corresponding to ZAMS stellar type B3 to O7, weak or no prior radio-continuum detections (i.e. below 50 mJy in flux density and 1″ in size), and a heliocentric distance within 9 kpc. Observations were performed with 0.″1 resolution and $\sim$10 μJy beam$^{-1}$ sensitivity at 6, 15, and 22 GHz using the VLA. Sanna et al. (2018) report a total of 33 centimeter continuum sites in 25 regions with a spectral index range of $-0.1 \leq \alpha \leq 1.3$, and conclude that the centimeter continuum emission in their sample is predominantly produced by ionized gas in stellar winds and jets.

### 1.1. *The EGO Sample*

In this paper, we present VLA 1.3 and 5 cm continuum and $H_2O$, $CH_3OH$, and $NH_3$ maser-line observations of 9 Extended Green Objects (EGOs; Cyganowski et al. 2008). EGOs are thought to be signposts of massive protoclusters in a specific stage of evolution just prior to the formation of HC H II regions. Their extended emission at 4.5 μm is generally accepted to be due to shocked $H_2$ (Marston et al. 2004) in protostellar outflows, and their strong association with Class I (collisionally-pumped) $CH_3OH$ masers at 44 and 25 GHz lends additional support to this conclusion (Cyganowski et al. 2009; Towner et al. 2017). EGOs are also strongly correlated with 6.7 GHz Class II $CH_3OH$ masers - which are pumped by infrared radiation and require high $CH_3OH$ column densities and dust temperatures (Sobolev et al. 1997), and are exclusively associated with massive protostars (Minier et al. 2003) - and with IRDCs (Cyganowski et al. 2009). EGOs, then, are objects in which the presence of massive protostars can be confirmed but for which the host clump is still largely intact. In other words, they are in a unique and extremely useful stage of evolution for distinguishing between different models of protocluster formation.

The specific EGOs we present in this paper are part of a subsample of the >300 EGOs originally presented in Cyganowski et al. (2008). We have been conducting high-angular resolution, multi-wavelength observations of this subsample in order to accurately constrain their properties over a broad range of wavelengths (5 cm to 3.6 μm). In particular, previous observations at 1.3 and 3.6 cm with the VLA revealed weak (<2 mJy beam$^{-1}$) or no centimeter continuum emission in these regions (Towner et al. 2017; Cyganowski et al. 2011). Dedicated observations at 19.7 and 37.1 μm with the Stratospheric Observatory for Infrared Astronomy (SOFIA), combined with archival data, have allowed us to model the Spectral Energy Distributions (SEDs) of these sources from 3.6 to 870 μm, and constrain overall properties of the protoclusters such as mass, temperature, and luminosity (Towner et al. 2019). The general properties of these EGOs are presented in Table 1.

**Table 1**. EGO Source Properties

| EGO Name | Coordinates (J2000) RA ($^{h\ m\ s}$) | Dec (° ′ ″) | $V^a_{LSR}$ (km s$^{-1}$) | Distance$^b$ (kpc) | $M^c_{clump}$ ($M_\odot$) | $T^d_{dust}$ (K) | $L_{bol}{}^e$ ($10^3\ L_\odot$) | $L/M^f$ ($L_\odot/M_\odot$) |
|---|---|---|---|---|---|---|---|---|
| G10.29−0.13 | 18:08:49.3 | -20:05:57 | 14 | 1.9 (0.3) | 84 | 24 (1) | 2.53 (2.06) | 30 |
| G10.34−0.14 | 18:09:00.0 | -20:03:35 | 12 | 1.6 (0.2) | 59 | 26 (1) | 1.85 (0.65) | 31 |
| G12.91−0.03 | 18:13:48.2 | -17:45:39 | 57 | 4.5 (0.7) | 638 | 23 (1) | 5.61 (0.62) | 9 |
| G14.33−0.64 | 18:18:54.4 | -16:47:46 | 23 | $1.13^{+0.14}_{-0.11}$ (2.3±0.3) | 146 | 29 (2) | 3.85 (0.67) | 26 |
| G14.63−0.58 | 18:19:15.4 | -16:30:07 | 19 | $1.83^{+0.08}_{-0.07}$ (1.9±0.3) | 244 | 22 (1) | 1.29 (0.34) | 5 |
| G18.89−0.47 | 18:27:07.9 | -12:41:36 | 66 | 4.2 (0.6) | 752 | 22 (1) | 2.53 (0.33) | 3 |
| G19.36−0.03 | 18:26:25.8 | -12:03:57 | 27 | 2.2 (0.3) | 151 | 26 (1) | 3.09 (0.44) | 20 |
| G22.04+0.22 | 18:30:34.7 | -09:34:47 | 51 | 3.4 (0.5) | 253 | 26 (1) | 4.97 (0.26) | 20 |
| G28.83−0.25 | 18:44:51.3 | -03:45:48 | 87 | 4.8 (0.7) | 806 | 26 (1) | 28.1 (4.1) | 35 |

[a] LSRK velocities are derived from single-component fits to $NH_3$ (1,1) emission and originally published in Cyganowski et al. (2013) (see their Table 3).

[b] Distances are estimated from the LSRK velocity and the Galactic rotation curve parameters from Reid et al. (2014) for all sources except G14.33–0.64 and G14.63–0.58. All kinematic distances are the near distance, and the uncertainty on each assumed to be 15%, as in Towner et al. (2019). Parallax distances (with their uncertainties) are given for G14.33–0.64 and G14.63–0.58 from Reid et al. (2014) and references therein, with the kinematic distances in parentheses for comparison.

[c] Masses are the average of the greybody-derived and $NH_3$ temperature-derived masses presented in Towner et al. (2019); see their Table 8 for values and their § 4.1 for procedure. We have compared these distances to those returned by the Monte-Carlo tool presented in Wenger et al. (2018) and found that this tool returns distances which agree with our current values within errors, with similar error magnitudes.

[d] Dust temperatures come from Table 8 of Towner et al. (2019). Uncertainties are listed in parentheses. In Towner et al. (2019), temperatures are derived from single-component graybody fits to far-IR (70, 160, & 870 μm) data.

[e] Towner et al. (2019) derive four luminosities for each EGO: one using single-component greybody fitting (see their Table 8) and three using radiative-transfer modeling packages (see their Table 9). In this work, we adopt as $L_{bol}$ for each EGO the median of these four values. Uncertainties, listed in parentheses, are the median absolute deviation from the median.

[f] $L_{bol}$ (column 6, this table) divided by $M_{clump}$ (column 4, this table).

By observing these 9 EGOs with sub-arcsecond resolutions and at μJy beam$^{-1}$ continuum sensitivities, we seek to address the question of the origins of centimeter-continuum emission when it first appears in massive protoclusters. As multiple processes can produce centimeter continuum emission in massive protoclusters, we examine both continuum-source spectral indices and source morphologies to distinguish between the possibilities (nascent H II regions, ionized jets, warm dust, and non-thermal emission). The maser-line data provide additional constraints on the possible emission mechanisms, such as by identifying definitively-massive continuum sources (6.7 GHz $CH_3OH$) and allowing us to explore the relationship between maser (primarily $H_2O$) and centimeter continuum luminosities, for which relationships at lower masses are relatively well understood (see Anglada et al. 2018, for a recent review of jet emission in star-forming regions). Additionally, the $NH_3$ maser data allow us to both add to the body of knowledge of this rare maser transition and, in future work, to disentangle non-thermal (maser) and thermal emission in $NH_3$ (3,3) and (6,6) observations of these regions with the VLA. We present our observations and data reduction methods in § 2, our results including high angular-resolution images in § 3, our analysis of spectral indices, maser emission, and luminosity correlations in § 4, and conclusions and future work in § 5.

## 2. OBSERVATIONS AND DATA REDUCTION

We used the Karl G. Jansky Very Large Array (VLA) to observe our sample of 9 EGOs at 1.3 cm (22 GHz, K-band) and 5 cm (6 GHz, C-band) under projects 17B-323 and 18A-249. These observations include continuum data at 1.3 and 5 cm, and line data for the 22 GHz $H_2O$, 23 and 25 GHz $NH_3$ (3,3) & (6,6), and 6.7 GHz Class II $CH_3OH$ maser species. Data for project 17B-323 were taken between 30 December 2017 and 13 February 2018, which we will refer to as observing epoch 2018.1. All observations for project 17B-323 were taken at 1.3 cm. 5 cm data for project 18A-249 were taken in May and June of 2018 (observing epoch 2018.4), and 1.3 cm data for 18A-249 were taken in July 2019 (observing epoch 2019.6). All 1.3 cm data were taken in either B- or BnA-configuration and all 5 cm data were taken in A-configuration. Median continuum beam sizes are $\sim$0.″27×0.″21 at 1.3 cm and $\sim$0.″50×0.″27 at 5 cm. Tables 2 and 3 summarize the parameters of the 1.3 cm and 5 cm observations, respectively. At 5 cm, all sources were observed in two observing blocks, approximately two days apart; the data from each block were combined to create final images for each source. Most sources were observed only once at 1.3 cm; those that were observed twice were observed in epochs 2018.1 and 2019.6. See § 2.2 and § 2.1 for discussions of the imaging procedures for the 5 cm and 1.3 cm data, respectively, including how data from different observation dates were combined.

**Table 2**. VLA K-band (22 GHz/1.3 cm) Observing Parameters

| EGO[a] | Observation[b] Date | Phase Cal | Cont. Synth. Beam ″ × ″ (°) | $\sigma$ Cont.[c] ($\mu$Jy/beam) | $\sigma$ $H_2O$[d] (mJy/beam) | $\sigma$ $NH_3$[d] (mJy/beam) | TOS[e] (min.) |
|---|---|---|---|---|---|---|---|
| G10.29–0.13 | 2018 Feb 13 & 2019 Jun 27 | J1832-2039 | 0.24×0.16 (83) | 13.9 | 7.9 & 15 | 7.5 & 10 | 40.6 |
| G10.34–0.14 | 2017 Dec 30 | J1832-2039 | 0.41×0.23 (8) | 11.9 | 4.5 | 4.2 | 33.0 |
| G12.91–0.03 | 2018 Jan 08 & 2019 Jul 09 | J1832-2039 | 0.23×0.17 (83) | 13.1 | 8.4 & 13 | 6.8 & 7.6 | 42.6 |
| G14.33–0.64 | 2017 Dec 30 | J1832-2039 | 0.52×0.34 (5) | 10.3 | 4.3 | 4.0 | 33.0 |
| G14.63–0.58 | 2019 Jun 27 | J1832-2039 | 0.23×0.15 (87) | 22.2 | 12 | 11 | 36.8 |
| G18.89–0.47 | 2018 Jan 27 | J1832-1035 | 0.33×0.21 (-4) | 13.6 | 4.2 | 4.4 | 26.4 |
| G19.36–0.03 | 2018 Jan 06 & 2019 Jul 11 | J1832-1035 | 0.33×0.30 (44) | 10.4 | 11 & 13 | 5.4 & 8.8 | 62.8 |
| G22.04+0.22 | 2018 Jan 06 & 2019 Jul 11 | J1832-1035 | 0.25×0.17 (78) | 14.2 | 6.2 & 13 | 5.6 & 8.8 | 62.7 |
| G28.83–0.25 | 2018 Jan 27 | J1832-1035 | 0.27×0.22 (-9) | 13.8 | 3.8 | 3.7 | 33.5 |

[a] The J2000 pointing center for each source is listed in Table 1 (columns 2 and 3).

[b] Four sources were (G10.29–0.13, G12.91–0.03, G19.36–0.03, G22.04+0.22) were observed twice at 1.3 cm due to either significant RFI or shortened observation times in epoch 2018.1, resulting in high noise levels in the first epoch of observations.

[c] These 1$\sigma$ values are 1.482×MAD, the scaled median absolute deviation from the median (MAD). We measure the MAD in a representative emission-free region in each image.

[d] These 1$\sigma$ values are measured in a representative line-free channel in each data cube for the resampled (channel width: 0.25 km $s^{-1}$) data. For sources observed during two different epochs, the 1$\sigma$ values are listed separately in order of epoch. For the $H_2O$ cubes, the min, max, and median 1$\sigma$ values in Kelvin are 118 K, 966 K, and 454 K, respectively. The values reported for $NH_3$ are for the $NH_3$ (3,3) data cubes only; no maser emission was detected in the $NH_3$ (6,6) data. The min, max, and median 1$\sigma$ values for $NH_3$ (3,3) in Kelvin are 96 K, 684 K, and 301 K, respectively. The $NH_3$ (6,6) 1$\sigma$ values over all data sets range from 3.5 to 8.7 mJy $beam^{-1}$ (79 to 491 K), with a median of 5.6 mJy $beam^{-1}$ (242 K).

[e] For sources observed in two epochs, this is the total time combined across both epochs.

The 1.3 cm spectral setup used 61 wide-band (128 MHz bandwidth, 1 MHz channel width) spectral windows spanning the range 18 to 26 GHz. Three narrow-band (16 MHz) windows covered the $H_2O$ $6_{1,6}$–$5_{2,3}$ transition at 22.235 GHz, and the $NH_3$ (3,3) and (6,6) metastable transitions at 23.870 and 25.056 GHz, respectively. The 22 GHz $H_2O$ band had a native channel width of 0.101 km $s^{-1}$ (7.81 kHz), and the $NH_3$ (3,3) and (6,6) bands had native channel widths of 0.203 and 0.187 km $s^{-1}$, respectively (15.6 kHz in both cases). All three data sets were resampled at 0.25 km $s^{-1}$; the line sensitivities reported in Table 2 are for the resampled data. The 5 cm spectral setup used 32 wide-band (128 MHz bandwidth, 1 MHz channel width) spectral windows spanning 4 to 8 GHz with one narrow-band (4.0 MHz) window covering the $CH_3OH$ $5_{1,5}$–$6_{0,6}$ $A^+$ transition at 6.669 GHz. The narrow-band 6.7 GHz $CH_3OH$ window had a native channel width of 0.087 km $s^{-1}$ (1.95kHz), which was then resampled at 0.25 km $s^{-1}$; the line sensitivities reported in Table 3 are for the resampled data.

All observations were performed in dual-polarization mode. In addition to the standard calibrations applied by the VLA pipeline, we self-calibrated both the line and continuum data using $H_2O$ maser emission at 1.3 cm and $CH_3OH$ maser emission at 5 cm. This calibration is described in more detail in the imaging subsections, below. The line and continuum $\sigma$ reported for each band are the scaled MAD values (1.482×MAD), where MAD is the median absolute deviation from the median. The 1.3 cm data have typical continuum noise levels of ∼14 $\mu$Jy beam$^{-1}$ and line noise of ∼4-15 mJy beam$^{-1}$. Typical 5 cm noise levels are ∼7 $\mu$Jy beam$^{-1}$ in the continuum and ∼3 mJy beam$^{-1}$ for the 6.7 GHz $CH_3OH$ line.

The full width at half max (FWHM) of the 25 m VLA dishes at 1.3 cm is ∼1.9′, and at 5 cm is ∼7′. While the nominal largest angular scales of our data are 7.″9 at 1.3 cm and 8.″9 at 5 cm[1], radio-frequency interference (RFI) flagging during the calibration and imaging process (see below) resulted in effective Maximum Recoverable Scales (MRS) for our sample that are significantly smaller. At 1.3 cm, our MRS – calculated from the $5^{th}$ percentile of projected baseline lengths for each source – ranges from 4.″4 to 7.″2, with a median value of 6.″1. At 5 cm, our MRS ranges from 5.″9 to 7.″6, with a median value of 7.″3. The phase calibrators are J1832-1035, J1832-2039, and J1851+0035 (see Tables 2 and 3). The flux and bandpass calibrators for all sources were 3C286 (J1331+3030) and J1924-2914, respectively.

**Table 3**. VLA C-Band (6 GHz/5 cm) Observing Parameters

| EGO | Observation[a] Date | Phase Cal | Cont. Synth. Beam ″ × ″ (°) | $\sigma$ Cont.[b] ($\mu$Jy/beam) | $\sigma$ $CH_3OH$[c] (mJy/beam) | TOS[d] (min.) |
|---|---|---|---|---|---|---|
| G10.29–0.13 | 2018 May 01 & May 02 | J1832-2039 | 0.53×0.27 (-12) | 7.3 | 3.2 | 34.6 |
| G10.34–0.14 | 2018 May 01 & May 02 | J1832-2039 | 0.53×0.27 (-11) | 7.0 | 3.1 | 34.5 |
| G12.91–0.03 | 2018 Jun 04 & Jun 07 | J1832-2039 | 0.40×0.23 (-3) | 7.9 | 3.0 | 33.7 |
| G14.33–0.64 | 2018 May 01 & May 02 | J1832-2039 | 0.64×0.37 (-14) | 5.4 | 3.1 | 34.6 |
| G14.63–0.58 | 2018 May 01 & May 02 | J1832-2039 | 0.50×0.29 (-14) | 5.9 | 2.9 | 34.6 |
| G18.89–0.47 | 2018 Jun 04 & Jun 07 | J1832-1035 | 0.35×0.24 (1) | 9.1 | 3.5 | 41.1 |
| G19.36–0.03 | 2018 May 01 & May 02 | J1832-2039 | 0.56×0.35 (-14) | 6.3 | 3.0 | 33.7 |
| G22.04+0.22 | 2018 Jun 04 & Jun 07 | J1832-1035 | 0.47×0.35 (1) | 6.4 | 3.0 | 38.4 |
| G28.83–0.25 | 2018 Jun 04 & Jun 07 | J1851+0035 | 0.28×0.27 (0) | 8.0 | 2.9 | 33.8 |

[a] All sources were observed in C-band during observing epoch 2018.4 in A-configuration.

[b] These 1$\sigma$ values are 1.482×MAD, measured in a representative emission-free region in each image.

[c] These 1$\sigma$ values are measured in a representative line-free channel in each data cube. The min, max, and median 1$\sigma$ values in Kelvin are 547 K, 1145 K, and 838 K, respectively. The sensitivities reported are for the resampled (channel width: 0.25 km $s^{-1}$) data.

[d] The total on-source time for each field combined across the two days of observations.

### 2.1. *1.3 cm Line and Continuum Imaging*

The 22 GHz $H_2O$ maser emission was strong enough to obtain good self-calibration solutions (phase and amplitude) in all sources except G19.36–0.03, which had neither any line nor any centimeter continuum emission strong enough for this purpose. Therefore, all 1.3 cm data for G19.36–0.03 have the standard pipeline-reduction calibration applied, but no additional, self-calibration corrections. Our self-calibration method was as follows: two rounds of phase-only self-calibration, followed by one round of amplitude+phase self-calibration. We required that all solution intervals have valid solutions for at least 6 baselines per antenna, and that these solutions have a minimum signal-to-noise ratio of 3.0. For all rounds of self-cal, we required that the solution intervals be integer multiples of the integration time, and optimized solution invervals for each source separately (i.e. different `solint` were used for different sources as necessary). Additionally, for the ‘ap’ self-calibration

[1] From the Observational Status Summaries for semesters 2017B and 2018A; see https://science.nrao.edu/facilities/vla/docs/manuals

round, we required that the variation in amplitude with time in the solutions be $\leq 10\%$.

In epoch 2018.1, persistent radio-frequency interference (RFI) in the G19.36–0.03 and G22.04+0.22 1.3 cm data required significant flagging of both the bandpass calibrators and the science targets, which led to noise levels $\sim 2\times$ those of the rest of the sample in this epoch. Sources G10.29–0.13 and G12.91–0.03 suffered from similarly high 1.3 cm noise values in this epoch due to shorter observation times ($\sim 0.6\times$) than the other sources. Therefore, these four targets were observed again at 1.3 cm in the 2019.6 observing epoch. G14.63–0.58 was observed entirely during the 2019.6 epoch; significant RFI and subsequent flagging resulted in a 1.3 cm continuum noise level for this source of $\sim 1.5\times$ the rest of the sample. The continuum data were imaged using line-free regions of the wide-band spectral windows, with the $H_2O$-derived self-calibration applied. For those sources which were observed twice at 1.3 cm, we first imaged the continuum data for each epoch separately (with the appropriate $H_2O$-derived self-calibration corrections applied) to check for flux variation between the two..In no case did we find a significant ($>4\sigma$) variation in continuum flux density between the two epochs. We then combined the two epochs to create a single continuum image for these EGOs with improved signal-to-noise compared to the single-epoch data. All continuum imaging at 1.3 cm was performed with nterms=2 and briggs weighting. We use robust=0.0 by default, but use robust=2.0 for G14.33–0.64 and G19.36–0.03 due to the presence of somewhat extended continuum emission whose structure was not adequately recovered with robust=0. The beam and noise statistics listed in Table 2 for these two sources are for the robust=2.0 images.

The $H_2O$ and $NH_3$ cubes were both self-calibrated using the $H_2O$ maser emission. For those sources with two epochs of data at 1.3 cm, we did not combine the data from each epoch to create the $H_2O$ and $NH_3$ cubes due to the possibility of a change in the kinematic or spatial properties of the emission during the 18-month period between the two observations. For these sources, we report each epoch's results separately. All line imaging was performed with robust=0.0. All imaging – both line and continuum – was performed with a cell size at least $5\times$ smaller than the beam minor axis, and all images were primary beam-corrected.

### 2.2. *5 cm Line and Continuum Imaging*

All 5 cm data were self-calibrated using the 6.7 GHz $CH_3OH$ maser emission for each source, following the procedure described in § 2.1. All 5 cm science targets were observed in two scheduling blocks, taken on two different days (see Table 3 for details). Unlike the $H_2O$ and $NH_3$ data, we do combine datasets for the 6.7 GHz $CH_3OH$ maser observations, due to the relatively short time difference between observations and in order to improve the signal-to-noise (S/N) of our data. The 6.7 GHz $CH_3OH$ maser emission was sufficient for self-calibration (phase and amplitude) in all sources. Each dataset was flagged and self-calibrated separately, and final images were created using the combined data for each source. The final 6.7 GHz cubes were sampled with a velocity channel width of 0.25 km $s^{-1}$. All line imaging was performed with robust=0.0.

All continuum imaging was performed with nterms=2 and briggs weighting, with robust=0.0 by default. However, significant imaging artifacts from partially resolved-out sources in the FOV necessitated changing the robust value from 0.0 to -1.0 for G10.29–0.13, G10.34–0.14, and G14.63–0.58. For G10.29–0.13, G10.34–0.14, and G28.83–0.25, we restricted the uv-range to $>35k\lambda$ for the same reason. Finally, G14.33–0.64, G19.36–0.03, and G22.04+0.22 were imaged with robust=2.0 in order to better recover the flux and morphology of several more extended (but <MRS) structures in each field.

All images were primary beam-corrected, and use a cell size at least $5\times$ smaller than the beam minor axis.

## 3. RESULTS

We report and discuss all emission at both wavelengths that is found within the FWHM of the 1.3 cm continuum primary beam for each source. For our sample, this FWHM ($1'.9$) corresponds to a physical diameter of 0.62 to 2.65 pc depending on source distance. This is large enough in all cases to encompass sources we might reasonably expect to be associated with each EGO protocluster, given typical ATLASGAL clump sizes of order $\sim$0.1-1 pc (see Urquhart et al. 2014; Towner et al. 2019, their Figures 1-4). Our source-identification and photometry procedures are described in § 3.1 and 3.2, below, and our continuum and maser detections are summarized in Tables 4 & 5 and Table 6, respectively.

Figure 1 shows three-color images for each EGO, with. *Spitzer* IRAC data with 8.0, 4.5, and 3.6 μm mapped to R, G, and B, respectively. The positions of the $CH_3OH$, $H_2O$, and $NH_3$ (3,3) masers (if present) in each source are indicated with color-coded symbols. For the sources observed twice at 1.3 cm, we show only the $H_2O$ and $NH_3$ emission from the first epoch of observations (2018.1) in this set of images; figures showing kinematic details of the maser emission for each source, including a comparison of multi-epoch observations, are presented in § 3.2. For each EGO, the left-hand panel shows a 120″ FOV of the target, with SOFIA 37 μm emission overlaid in black[2] and APEX LABOCA 870 μm contours overlaid in silver[3]. These panels show the full extent of the extended 4.5 μm emission by which each EGO was originally identified, the 870 μm dust clumps in which they reside, and the associated bright 37 μm emission which may be indicative of outflow cavities (see De Buizer et al. 2012) or internally-heated envelopes from massive protostars. The right-hand panels show a 14.″4 FOV for all sources, corresponding to a physical size of 0.04 to 0.17 pc. The 1.3 cm and 5 cm continuum emission contours are overlaid in dark red and orange, respectively, and 37 μm contours and maser symbols are overlaid as in the left-hand panels.

[2] The SOFIA 37 μm data were taken with the Faint Object infraRed CAmera for the SOFIA Telescope (FORCAST, Herter et al. 2012) under Plan ID 04_0159; see Towner et al. (2019) for observing details (§2.1) and photometry procedures (§3.1).

[3] The APEX data were part of the APEX Telescope Large Area Survey of the Galaxy (ATLASGAL, Schuller et al. 2009) and were retrieved from the ATLASGAL Database Server at http://atlasgal.mpifr-bonn.mpg.de/cgi-

Across our nine EGOs, we find a total of 41 centimeter continuum detections (>4$\sigma$) within the FWHM of the primary beam. In Table 4, we report the position, flux densities, angular size (where possible), and derived 5-1.3 cm spectral index $\alpha$ for the 36 detections which have well-constrained flux densities. We discuss these detections in greater detail in the subsections below. In Table 5, we report the remaining five detections whose flux densities cannot be fully recovered in our data because some of their emission is resolved out. All of these latter sources are coincident with H II regions previously identified in the literature. For these partially resolved-out sources, we report approximate center positions, measured peak intensities, and approximate sizes at each band. Sizes are estimated from the morphology in our data, and therefore may be underestimates due to spatial filtering. As we cannot completely recover their flux densities, we do not discuss these sources further in this paper; Table 5 lists literature references for each partially resolved-out source.

## 3.1. *Centimeter Continuum Emission: Photometry Procedure and Properties*

### 3.1.1. *Photometry Procedure*

We determine the total flux density and size (if possible) of the 36 well-constrained sources using either aperture photometry or the `imfit` task in the Common Astronomy Software Applications (CASA) version 5.4.1-32 (McMullin et al. 2007). For sources with irregular or non-Gaussian morphology, we use aperture photometry: we measure the total flux density inside the 2–3$\sigma$ contour[4], and then correct for any background flux by subtracting the product of the aperture size (in beams) and the background noise level (in mJy beam$^{-1}$). We performed this background subtraction on all aperture photometry-measured sources including those whose local background appeared to be a gaussian distribution about zero; in cases where the background is truly zero, the procedure will have a minimal effect on the final flux densities. Compact sources were fit with 2-dimensional gaussians using `imfit`. Fits were considered ‘good’ if the pixel values of the residual image were less than 3 times the rms[5] of the residual as returned by `imfit`, and if the fitted integrated flux density was greater than the fitted peak intensity. Fits which did not meet these standards were iteratively improved either until they did or until it was determined that no such fit was possible. Usually, fits were improved by holding either the source position or the major and minor axes and position angle of the convolved size fixed to the beam size during the fit.

For sources for which it was not clear whether aperture photometry or *imfit* would be more appropriate, we measured the integrated flux density using both aperture photometry and `imfit`, and compared results. If the `imfit` result did not meet the 3×rms criterion despite iterative improvements, we used the aperture-photometry results for that source as long as the peak intensity was larger than the integrated flux density. All flux density uncertainties reported in Table 4 are the quadrature sum of the measurement uncertainty for that source and the absolute flux density uncertainty of the VLA in the relevant bands[6].

If a source could be successfully deconvolved from the beam with `imfit`, we report that deconvolved size in milliarcseconds (mas) in Table 4. If a source’s angular size could not be deconvolved with `imfit`, its size is listed as ‘unres.’ If we could only obtain a good fit for a source by fixing its major and minor axes (and/or position angle) to those of the synthesized beam, its size is listed as ‘fixed.’ We consider ‘fixed’ sources to be unresolved. If a source’s flux density was measured using aperture photometry, we report an upper limit on its size; this upper limit is the size of the maximum extent of the polygonal aperture used to measure the flux density.

### 3.1.2. *Continuum Source Naming Convention*

For each EGO, the numbering order in Table 4 is determined by the 1.3 cm integrated flux density of each detection. The brightest 1.3 cm source in an EGO is CM1, the second-brightest is CM2, etc. Sources which are detected only at 5 cm and not at 1.3 cm are numbered after all 1.3 cm sources, in decreasing order of 5 cm flux density.

### 3.1.3. *Properties of the Centimeter Continuum Emission*

We detect at least one source of centimeter continuum emission in every EGO at each wavelength, for a detection rate of 100% at both 1.3 cm and 5 cm. The total number of continuum detections per EGO ranges from 1 to 12, with a median of 3. These statistics include sources detected at only

bin/ATLASGAL_DATABASE.cgi

[4] Exact contour levels were chosen based on source morphology and separability from other nearby sources; generally sources are well-separated.

[5] `imfit` returns the rms of a residual image automatically, but not the MAD. In order to automate this process as much as possible, we use rms for this goodness-of-fit test. However, for all images, rms and MAD are identical to within a few percent, so the use of rms in this procedure did not significantly the quality of these fits.

[6] See https://science.nrao.edu/facilities/vla/docs/manuals/oss/performance/fdscale

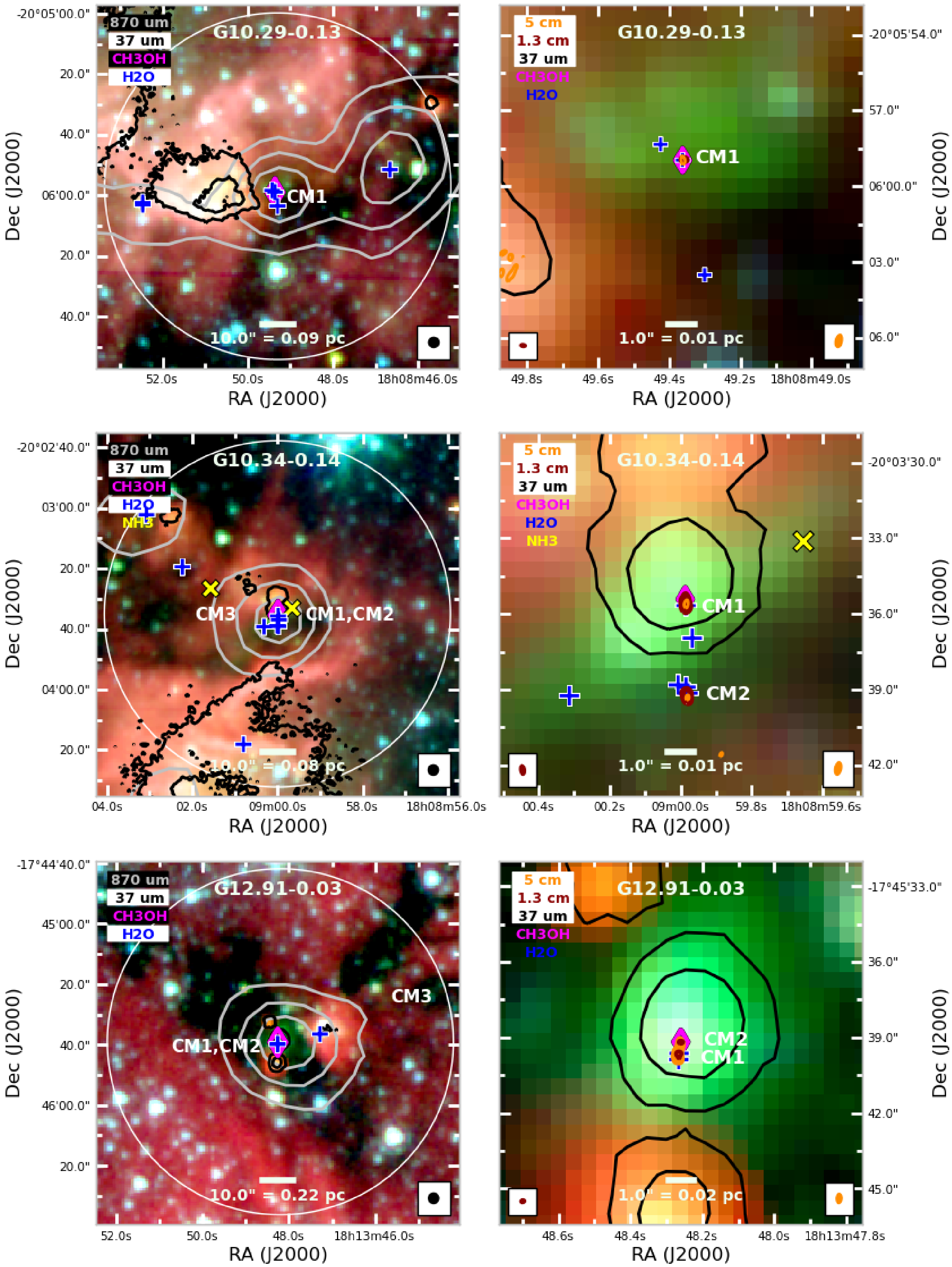

**Figure 1.** RGB mid-infrared images for the EGO sources G10.29–0.13, G10.34–0.14, and G12.91–0.03. $CH_3OH$ 6.7 GHz masers are shown as magenta diamonds (⋄), 22 GHz $H_2O$ masers as blue +, and $NH_3$ (3,3) masers, if detected, with yellow ×. *Left:* 120″ FOV of each EGO with ATLASGAL 870 μm contours overlaid in silver, SOFIA FORCAST 37.1 μm contours in black, and the 1.3 cm FWHM shown as a white circle. Background image shows *Spitzer* IRAC 8.0, 4.5, and 3.6 μm data mapped to R, G, and B, respectively. ATLASGAL contour levels are [0.25,0.50,0.75]×$I_{peak}$, where $I_{peak}$=[3.31,4.4,2.78] Jy beam$^{-1}$ for G10.29–0.13, G10.34–0.14, G12.91–0.03, respectively; SOFIA contour levels are [5,15,45,125,250]×$\sigma$, where $\sigma$=[0.26,0.30,0.18] Jy beam$^{-1}$, respectively. SOFIA 37.1 μm beam is shown in the lower right. *Right:* 14.″4 FOV of each EGO with 1.3 cm continuum emission overlaid in red contours and 5 cm continuum overlaid in orange contours (levels: [4,8,16]×$\sigma$ for both). SOFIA 37 μm contours are overlaid in black with the same contour levels as in the left-hand panels. 1.3 cm beam is shown in the lower left, 5 cm beam in the lower right.

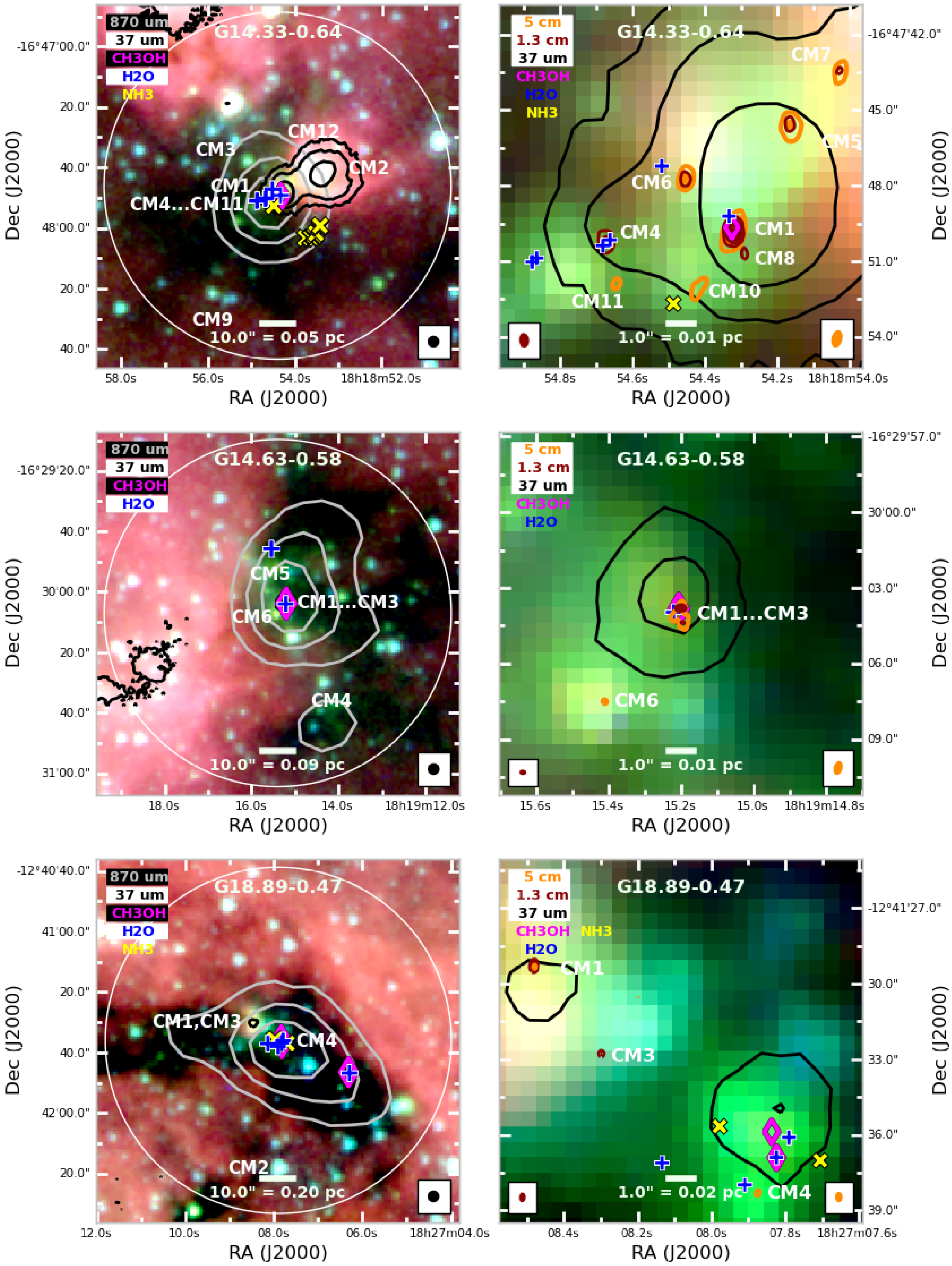

**Figure 1**. Same as previous but for EGO sources G14.33–0.64, G14.63–0.58, and G18.89–0.47. ATLASGAL $I_{peak}$=[12.98,4.35,3.30] Jy beam$^{-1}$ for G14.33–0.64, G14.63–0.58, and G18.89–0.47, respectively; SOFIA 37 µm $\sigma$=[0.24,0.22,0.25] Jy beam$^{-1}$, respectively. 1.3 and 5 cm contour levels are [4,8,20,50,100]$\times\sigma$ for G14.33–0.64, and [4,8,16]$\times\sigma$ for G14.63–0.58 and G18.89–0.47.

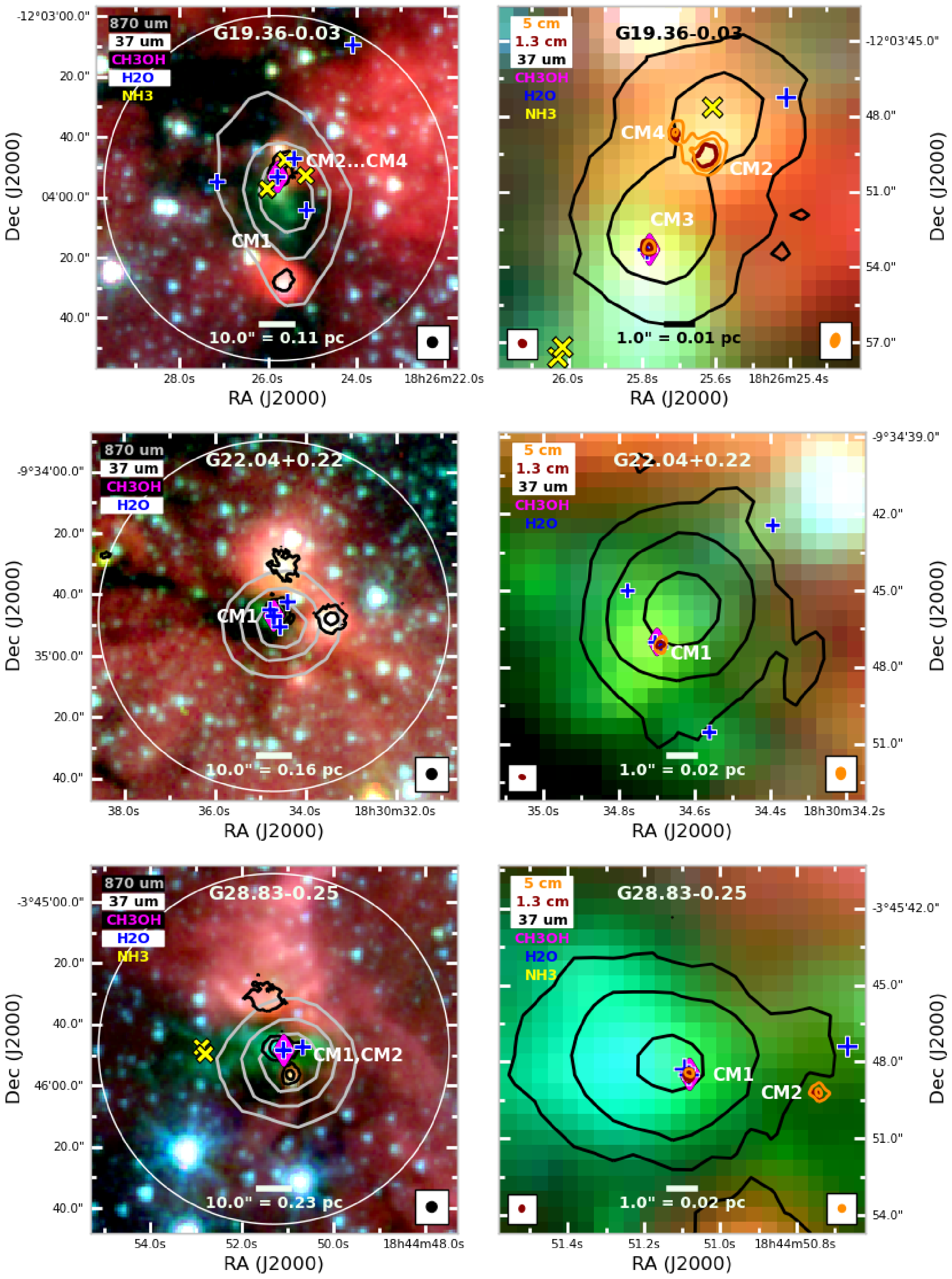

**Figure 1**. Same as previous but for EGO sources G19.36–0.03, G22.04+0.22, and G28.83–0.25. ATLASGAL $I_{peak}$=[2.90,3.33,4.08] Jy beam$^{-1}$ for G19.36–0.03, G22.04+0.22, and G28.83–0.25, respectively; SOFIA 37 μm $\sigma$=[0.46,0.21,0.26] Jy beam$^{-1}$, respectively. 1.3 and 5 cm contour levels are [4,8,20,50,100]$\times\sigma$ for G19.36–0.03 and [4,8,16]$\times\sigma$ for G22.04+0.22 and G28.83–0.25.

one wavelength. (Considering each wavelength separately, we detect a median of 2 sources per EGO at both 1.3 cm and 5 cm.) This is a factor of 3 increase from Towner et al. (2017), who detect a median of one 1.3 cm continuum source per EGO for their sample of 20 EGOs (including all 9 EGOs examined in this work), and a factor of ∼3.5 increase over Cyganowski et al. (2011), who detect 12 centimeter continuum sources at 3.6 and 1.3 cm in a sample of 14 EGOs (including 5 examined in this work) for an average of 0.85 sources per EGO. This is likely a result in the improvement in both angular resolution (∼10×) and continuum sensitivity (∼3-25×) in this work compared to both previous examinations.

Across our entire EGO sample, we detect 22 continuum sources at both 1.3 cm and 5 cm, 4 continuum sources at 1.3 cm only, and 10 sources at 5 cm only. There is a general trend that sources detected at both wavelengths tend to be strong detections at both wavelengths, and sources detected in only one wavelength tend to be weak detections in that single wavelength - i.e. it is uncommon for a source to be a very strong detection at 5 cm and a weak or non-detection at 1.3 cm (such as might occur with a steep, negative spectral index), and vice versa. For the total sample of 36 centimeter continuum sources, 1.3 cm and 5 cm median flux densities are 158 µJy and 58 µJy, respectively. For the sources detected at both wavelengths, the median 1.3 cm and 5 cm flux densities are 179 µJy and 77 µJy, respectively. The median flux density of sources detected only at 1.3 cm is 86 µJy, and the median flux density of sources detected only at 5 cm is 41 µJy.

The continuum sources are generally compact – either point sources or less than ∼1″ in diameter. In 7 EGOs, the majority of the centimeter continuum emission is coincident with or within a few arcseconds of the brightest source at 37 µm. The exceptions are G10.29–0.13, which was a non-detection at 37 µm, and G18.89–0.47, in which there are centimeter continuum detections within the 5$\sigma$ contour of the *second*-brightest 37 µm source (NE), but none coincident with the brightest 37 µm source (SW). Likewise, nearly all continuum detections are within the 5$\sigma$ contour of the 870 µm emission (30/36, 83%); the sources that are not within the 870 µm emission are preferentially detected only at 5 cm (4/6, 67%).

**Table 4**. Properties of Centimeter Continuum Sources

| EGO | Source[a] | Position[b] (J2000) | | 1.3 cm Properties[c] | | 5 cm Properties[c] | | Spectral |
|---|---|---|---|---|---|---|---|---|
| | Name | RA | Dec | Flux | Size | Flux | Size | Index |
| | | ($h$ $m$ $s$) | (° ′ ″) | ($\mu$Jy) | mas × mas (pa) | ($\mu$Jy) | mas × mas (pa) | ($\alpha$) |
| G10.29 | CM1 | 18:08:49.359 (0.001) | −20:05:58.938 (0.009) | 82 (14) | fixed | 41 (7) | fixed | 0.54 (0.19) |
| G10.34 | CM1 | 18:08:59.9859 (0.0006) | −20:03:35.61 (0.03) | 209 (31) | 357 × 146 (179) | 35 (6) | fixed | 1.38 (0.17) |
| | CM2 | 18:08:59.983 (0.001) | −20:03:39.24 (0.03) | 195 (36) | 379 × 225 (34) | 33 (6) | fixed | 1.38 (0.20) |
| | CM3 | 18:09:01.455 (0.001) | −20:03:38.71 (0.02) | 88 (12) | fixed | < 28 | ··· | > 0.89 |
| G12.91 | CM1 | 18:13:48.265 (0.002) | −17:45:39.66 (0.03) | 110 (19) | < 360 | 130 (16) | unres | −0.13 (0.16) |
| | CM2 | 18:13:48.260 (0.001) | −17:45:39.19 (0.01) | 77 (14) | fixed | 52 (8) | fixed | 0.30 (0.18) |
| | CM3 | 18:13:44.5899 (0.0008) | −17:45:21.17 (0.02) | < 52 | ··· | 82 (14) | unres | < −0.35 |
| G14.33 | CM1 | 18:18:54.319 (0.003) | −16:47:49.80 (0.04) | 1784 (92) | <1300 | 792 (27) | < 2100 | 0.63 (0.05) |
| | CM2 | 18:18:53.464 (0.003) | −16:47:43.72 (0.04) | 1751 (115) | <5500 | 2147 (77) | < 5600 | −0.16 (0.06) |
| | CM3 | 18:18:55.3714 (0.0002) | −16:47:31.326 (0.006) | 433 (32) | unres | 139 (10) | unres | 0.88 (0.08) |
| | CM4 | 18:18:54.672 (0.003) | −16:47:50.22 (0.01) | 393 (29) | 301×225 | < 22 | ··· | > 2.25 |
| | CM5 | 18:18:54.166 (0.001) | −16:47:45.58 (0.04) | 133 (25) | 395×175 | 203 (14) | < 1800 | −0.33 (0.16) |
| | CM6 | 18:18:54.4543 (0.0007) | −16:47:47.76 (0.02) | 127 (12) | unres | 141 (11) | unres | −0.08 (0.09) |
| | CM7 | 18:18:54.030 (0.004) | −16:47:43.5 (0.1) | 116 (39) | 669×440 | 108 (16) | fixed | 0.06 (0.28) |
| | CM8 | 18:18:54.290 (0.001) | −16:47:50.70 (0.03) | 79 (11) | unres | 31 (5) | fixed | 0.72 (0.17) |
| | CM9 | 18:18:55.912 (0.001) | −16:48:32.03 (0.03) | < 41 | ··· | 61 (5) | fixed | < −0.30 |
| | CM10 | 18:18:54.418 (0.001) | −16:47:52.08 (0.04) | < 41 | ··· | 44 (7) | fixed | < −0.05 |

*Table 4 continued*

Table 4 *(continued)*

| EGO | Source[a] Name | Position[b] (J2000) RA ($h$ $m$ $s$) | Dec ($^\circ$ $'$ $''$) | 1.3 cm Properties[c] Flux ($\mu$Jy) | Size mas × mas (pa) | 5 cm Properties[c] Flux ($\mu$Jy) | Size mas × mas (pa) | Spectral Index ($\alpha$) |
|---|---|---|---|---|---|---|---|---|
| | CM11 | 18:18:54.645 (0.003) | -16:47:51.96 (0.04) | < 41 | ··· | 38 (5) | fixed | < 0.06 |
| | CM12 | 18:18:53.580 (0.003) | −16:47:38.19 (0.03) | < 41 | ··· | 36 (10) | unres | < 0.10 |
| G14.63 | CM1 | 18:19:15.2004 (0.0005) | −16:30:03.802 (0.003) | 418 (43) | 154 × 36 (100) | 54 (5) | fixed | 1.58 (0.11) |
| | CM2 | 18:19:15.195 (0.002) | −16:30:04.38 (0.03) | 72 (19) | < 280 | 48 (8) | < 870 | 0.31 (0.24) |
| | CM3 | 18:19:15.216 (0.001) | −16:30:04.13 (0.02) | < 89 | ··· | 44 (5) | fixed | < 0.54 |
| | CM4 | 18:19:14.2033 (0.0008) | −16:30:38.00 (0.03) | < 89 | ··· | 35 (5) | fixed | < 0.72 |
| | CM5 | 18:19:15.411 (0.002) | −16:30:07.52 (0.03) | < 89 | ··· | 33 (10) | unres | < 0.77 |
| | CM6 | 18:19:15.5382 (0.0006) | −16:29:52.87 (0.02) | < 89 | ··· | 28 (5) | fixed | < 0.89 |
| G18.89 | CM1 | 18:27:08.4814 (0.0004) | −12:41:29.30 (0.02) | 137 (14) | fixed | 39 (9) | fixed | 0.97 (0.19) |
| | CM2 | 18:27:08.7530 (0.0007) | −12:42:20.34 (0.03) | 83 (13) | fixed | < 36 | ··· | > 0.64 |
| | CM3 | 18:27:08.2986 (0.0009) | −12:41:32.76 (0.03) | 63 (12) | fixed | < 36 | ··· | > 0.43 |
| | CM4 | 18:27:07.8752 (0.0009) | −12:41:38.29 (0.03) | < 54 | ··· | 54 (16) | fixed | < 0.01 |
| G19.36 | CM1 | 18:26:26.37745 (0.00004) | −12:04:19.9316 (0.0005) | 2633 (133) | 69 × 34 (85) | 987 (32) | 73 × 47 (155) | 0.76 (0.05) |
| | CM2 | 18:26:25.626 (0.002) | −12:03:49.56 (0.04) | 567 (56) | <1800 | 755 (34) | <1950 | −0.22 (0.09) |
| | CM3 | 18:26:25.7831 (0.0006) | −12:03:53.220 (0.009) | 250 (26) | 192 × 145 (175) | 66 (17) | unres | 1.03 (0.22) |
| | CM4 | 18:26:25.712 (0.002) | −12:03:48.66 (0.04) | 87 (17) | <680 | 74 (10) | <880 | 0.13 (0.18) |
| G22.04 | CM1 | 18:30:34.690 (0.002) | −09:34:47.12 (0.03) | 179 (24) | < 600 | 71 (18) | < 1100 | 0.72 (0.22) |
| G28.83 | CM1 | 18:44:51.084 (0.003) | −03:45:48.52 (0.04) | 264 (20) | < 760 | 79 (15) | unres | 0.93 (0.16) |
| | CM2 | 18:44:50.744 (0.003) | −03:45:49.32 (0.04) | 178 (20) | < 890 | 306 (22) | < 1150 | −0.42 (0.10) |

[a] See § 3.1.2 for source-naming convention.

[b] All positions are derived from the 1.3 cm data, unless the source is not detected at 1.3 cm, in which case the 5 cm position is used. For fitted sources, the position and position uncertainty are those returned by `imfit`; for aperture-photometry sources, the position is that of the peak intensity and position uncertainty is the cell size.

[c] Sources with a listed angular size in mas, 'unres', or 'fixed' were fit with `imfit`; sources with an upper limit were measured using aperture photometry.

In addition to flux densities and sizes, Table 4 also reports the calculated 1.3–5 cm (6–22 GHz) spectral index ($\alpha$, where $S_\nu \propto \nu^\alpha$) for each centimeter continuum detection. All spectral indices were calculated using Monte-Carlo analysis to determine $\alpha$ and its uncertainty given two continuum flux density values, their total uncertainties, and their central frequencies. For sources detected at both wavelengths, $\alpha \pm \Delta\alpha$ comes directly from the measured flux densities. For sources detected only at 1.3 cm, we calculate $\alpha$ using the 4$\sigma$ upper limit at 5 cm as the 5 cm 'flux density,' and report the resulting spectral index as a lower limit. For sources detected only at 5 cm, we use the 1.3 cm 4$\sigma$ upper limit as the 1.3 cm 'flux density,' and report the resulting spectral index as an upper limit. In total, for sources detected at both 1.3 cm and 5 cm, the spectral indices span -0.42 $< \alpha <$ 1.58, with a median value of $\alpha$ = 0.59.

**Table 5**. Extended Centimeter Continuum Sources: Flux Density Not Recoverable in One or Both Bands

| EGO | Source[a] Name | 1.3 cm Properties[b] RA (h m s) | Dec (° ′ ″) | Approx. Size (″ × ″) | 5 cm Properties[b] RA (h m s) | Dec (° ′ ″) | Approx. Size (″ × ″) | Reference[c] |
|---|---|---|---|---|---|---|---|---|
| G10.29–0.13 | NR-CM1 | 18:08:51 | -20:06:01 | 22×15 | 18:08:51 | -20:06:01 | 22×15 | Cyganowski et al. (2011) |
| G10.34–0.14 | NR-CM1 | 18:09:01 | -20:04:29 | 48×24 | 18:09:01 | -20:04:36 | 58×31 | Goedhart et al. (2002) |
| | NR-CM2 | 18:09:00 | -20:04:03 | 26×12 | ... | ... | ... | Goedhart et al. (2000) |
| | NR-CM3 | 18:09:02 | -20:03:03 | 7×4 | 18:09:02 | -20:03:04 | 7×3 | Deharveng et al. (2015) |
| G28.83–0.25 | NR-CM1 | ... | ... | ... | 18:44:51 | -03:45:31 | 15×7 | Deharveng et al. (2010) |

[a] Sources are listed here as "NR-CM1, NR-CM2, ..." to indicate that their flux density is not fully recoverable from our data. As we cannot fully recover their flux density or morphology, we do not analyze these sources further in this paper.

[b] Positions reported are for the peak intensity in each band. Sizes are based on the apparent morphology of the source in our data, and are measured from the polygonal aperture used to define that morphology. They may therefore be underestimates in one or more directions due to spatial filtering, and should be treated as approximate.

[c] Representative reference for previous detection within our estimated size for each source. This list is not comprehensive; interested readers should consult references in and to these publications for additional information.

In Figure 2 and the associated figure set, we show each centimeter continuum detection as a contour plot, with both 1.3 and 5 cm emission shown on the same panel. For continuum sources detected at only one wavelength, contours, beam, and labels are shown only for the detected wavelength. Most panels show only one centimeter-continuum source each, but sometimes continuum sources which are near neighbors are shown in the same plot. In these cases, each continuum source is labeled. Contour levels are [4,10]$\times\sigma$ by default, and all contour levels are also listed in the figure caption.

### 3.1.4. *Expected Contamination From Background Sources*

We estimate the expected contamination from background radio sources at both 5 and 1.3 cm following the population-flux density relation derived by Fomalont et al. (1991):

$$N(S) = (23.2 \pm 2.8) S^{-1.18\pm0.19} \quad (1)$$

where $N(S)$ is the number of sources (arcmin)$^{-2}$ above $S$ µJy at 6 cm (5 GHz). This approach assumes that the same radio populations are responsible for the background emission at 5, 6, and 22 GHz.

For our median noise level of 7 µJy beam$^{-1}$ at 5 cm, a 4$\sigma$ detection limit corresponds to 28 µJy. We scale our 5 cm upper limit to a 6 cm limit using the median source spectral index derived by Fomalont et al. (1991): $\alpha$ = -0.38, where $S \propto \nu^{\alpha}$. This gives us an expected background source rate of 0.42 sources (arcmin)$^{-2}$.For each field, we investigate only those sources within the FWHM of the 1.3 cm primary beam, which is 2.8 arcmin$^2$ in area ($1\farcm9$ in diameter). Therefore, we should expect 1.2 background sources per field at 5 cm, or, 10.6 background sources in the entire sample.

Our median noise at 1.3 cm is 14 µJy beam$^{-1}$, so 4$\sigma$ = 56 µJy. Scaling this to 5 GHz (again assuming $\alpha$ = -0.38) yields expected source counts of 0.10 (arcmin)$^{-2}$ or 0.28 sources per target. For the EGO 9 targets in our sample, we should expect a total of 2.6 background sources at 1.3 cm.

We note that our estimated source count per arcmin$^2$ at 1.3 cm is nearly identical to that of Rosero et al. (2016), who use the population-specific models of de Zotti et al. (2005) - scaled by the results of the 10C Survey at 15.7 GHz (AMI Consortium et al. 2011) - to derive an expected background source density of 0.12 (arcmin)$^{-2}$ with $S_{25.5GHz} > 45$ µJy.

## 3.2. *Maser Emission: Fitting Procedure and Properties*

### 3.2.1. *Fitting Procedure*

For each maser species, we perform 2D gaussian fitting to all maser emission, in each channel, which lies within the FWHM of the 1.3 cm continuum primary beam. We define a "maser spot" as a single location of maser emission in a single channel. All maser spots were fit as as point sources, i.e., they have their major/minor axis ratio and position angle fixed to match those of the synthesized beam; see Tables 2 and 3 for synthesized beam parameters. This choice reflects our assumption that the maser emission is unresolved at the current angular resolution. (See, e.g., Moscadelli et al. 2020, for an example of maser emission observed with VLBI in a sample of similar sources.) As with the centimeter continuum emission, the peak of the residual image was required to be $<3\times$rms of the residual for the fit of each maser spot to be considered good[7]. In most cases, iterative fitting was not necessary. However, some fields did have collections of maser spots which overlapped with each other, and these cases required iteratively fitting the maser spots while holding the fitted position (RA, Dec, or both) of one or more maser spots fixed during the fit. In these cases, all maser spots in the field are fit simultaneously, and the "iterative" nature of the fit reflects by-hand optimization of the fixed fit parameters. We report all detections whose fitted peaks are $>4\sigma$, where $\sigma$ is the rms of the specific channel in which each fit was performed.

[7] Again, the use of rms instead of scaled MAD is due to the fact that `imfit` automatically returns the rms of the residual image, and not MAD; as before, the rms and scaled MAD values are identical within a few percent in these images, so the use of one over the other for this goodness-of-fit criterion is unlikely to be significant.

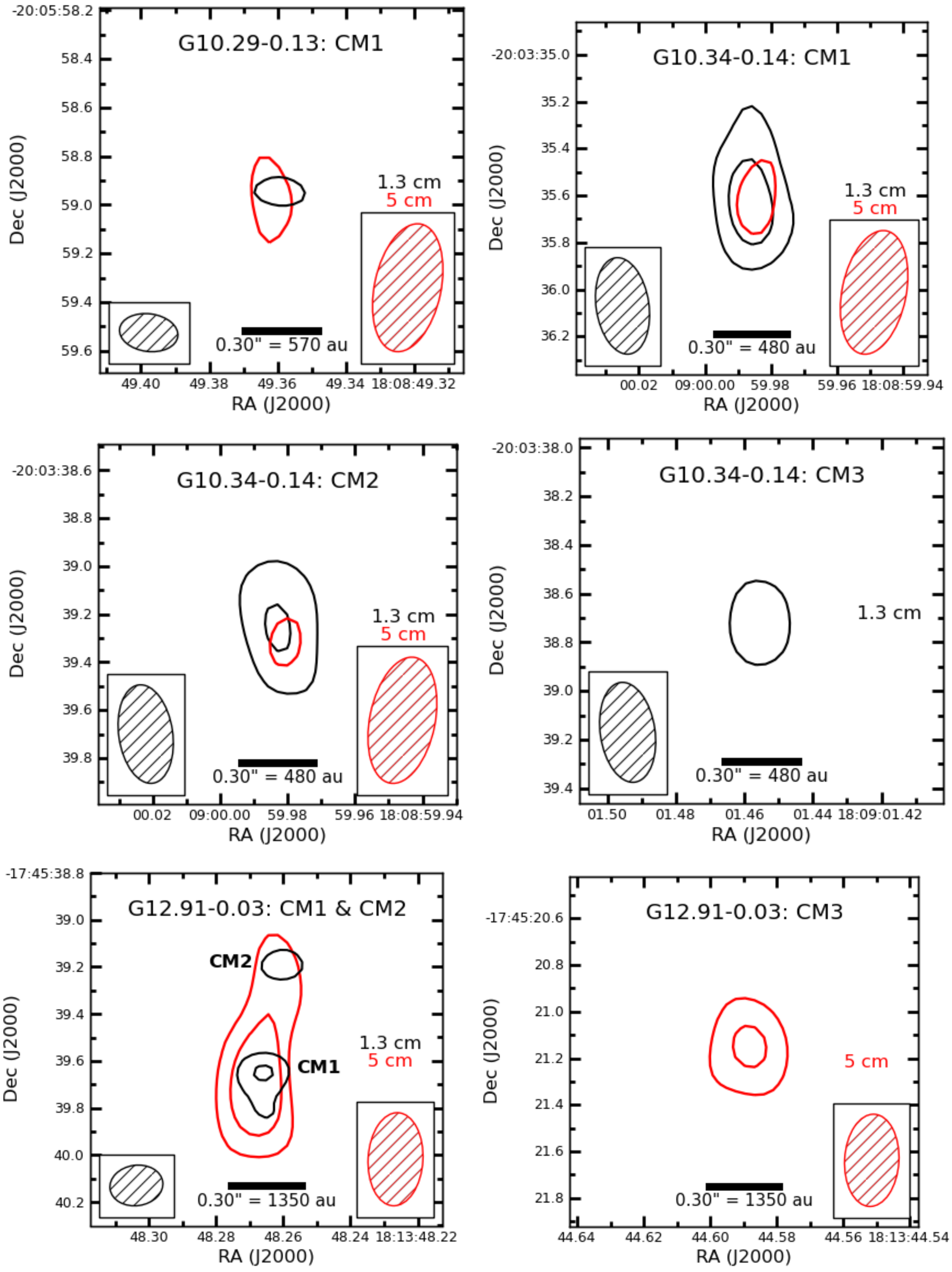

**Figure 2**. High-zoom contour plots of our continuum detections. 1.3 cm detections are shown in black, and 5 cm detections are shown in red. Beams are shown in the lower left and right corners, respectively; if a source is a non-detection at a particular wavelength, that beam is not shown. Contour levels are [4,10]$\times\sigma$. FOV is $1\farcs$ by default and scalebar is always 1/5 of the FOV; alternate FOV sizes (in special cases) are reflected in the scalebar. When multiple centimeter continuum sources are visible in the same field of view, each source is labeled individually and the title of the panel will reflect which sources are displayed (e.g. 'G12.91–0.03: CM1 & CM2'). The complete figure set (31 images) is available in the online journal.

#### 3.2.2. *Maser Naming Convention*

Table 6 summarizes the overall maser properties for each EGO. Within each EGO, every maser spot that we report is assigned to a "Maser Group." Typically, a maser group will be composed of maser spots at the same physical location emitting across multiple channels (e.g., a maser emitting from 5 km s$^{-1}$ to 10 km s$^{-1}$ at position (x,y) would be defined as a "maser group" of 20 individual maser spots, one in each channel). However, some maser groups do contain multiple maser spots within a single channel. In these cases, these maser spots are still in the same maser group if they are all associated with the same centimeter continuum source; we do not divide maser emission into subgroups.

Each maser group is named for the centimeter continuum source with which it is associated, and for its maser species (e.g. the group of $H_2O$ masers associated with the G14.63–0.58 centimeter continuum source CM1 is denoted "CM1-W1," while the $CH_3OH$ emission associated with that same continuum source is denoted "CM1-M1"). A maser group is always "associated" with the closest centimeter continuum source, up to a distance of 1″ (1130 to 4800 au for our sample).

If a maser group is not within 1″ of any continuum emission, it is denoted by "NC" (for "no centimeter continuum") and the maser species; maser spots of a single species are assigned to an "NC" maser group if they are all within 1″ of each other. This is a criterion with no corner cases - maser-spot separations are typically either well below or well above 1″. The "NC" numbering scheme is separate for each species and increases with increasing Right Ascension within a given EGO (i.e. an EGO might have both "NC-W1" and "NC-A1," which are not co-located). We do not associate any "NC" groups of different species with each other even if they are within 1 ″ (1130 to 4800 au), in order to preserve the RA- and species-based numbering scheme, although this does not mean that a shared physical origin between two "NC" maser groups of different species is impossible. For instance, the G18.89–0.47 maser groups NC-W3 and NC-M2 are very nearly co-located but have different "NC" numbers as they are different species.

For sources with two epochs of $H_2O$ and $NH_3$ maser observations, maser groups have the same name for each epoch as long as they are within 1″ of each other. In practice, however, this is another criterion with no corner cases: all second-epoch maser groups are either within 0.″3 of their original position or are far removed (>2″) from any other maser groups. If a given maser species and group is listed as " ··· " this means that that species+group combination was not detected during that observing epoch.

**Table 6**. Summary of Maser Emission

| EGO | Maser[a] Species | Obs[b] Epoch | Group[c] Name | Centroid Position[d] RA (h m s) (″) | Dec (° ′ ″) (″) | Angular Spread[e] (RA, Dec) (″) | $V_{Min}$, $V^{f}_{Max}$ (km/s) | $V^{g}_{Peak}$ (km/s) | $S^{h}_{Peak}$ (Jy) |
|---|---|---|---|---|---|---|---|---|---|
| G10.29–0.13 | $CH_3OH$ | 2018.4 | CM1-M1 | 18:08:49.363 (0.060) | -20:05:58.96 (0.03) | 0.061, 0.035 | 1.25, 20.75 | 8.00 | 3.89 (0.01) |
| | $H_2O$ | 2018.1 | CM1-W1 | 18:08:49.364 (0.013) | -20:05:58.98 (0.01) | 0.012, 0.010 | 5.50, 14.00 | 10.25 | 1.67 (0.02) |
| | | | NC-W1 | 18:08:46.669 (0.002) | -20:05:51.470 (0.001) | 0.008, 0.003 | 6.50, 21.25 | 12.50 | 23.1 (0.1) |
| | | | NC-W2 | 18:08:49.300 (0.011) | -20:06:03.53 (0.01) | 0.014, 0.027 | -12.00, 15.75 | 15.00 | 0.43 (0.01) |
| | | | NC-W3 | 18:08:49.424 (0.011) | -20:05:58.340 (0.001) | 0.047, 0.006 | 14.25, 17.75 | 15.00 | 4.18 (0.01) |
| | | | NC-W4 | 18:08:52.464 (0.012) | -20:06:02.62 (0.36) | 0.017, 0.437 | 6.50, 22.00 | 8.00 | 9.64 (0.02) |
| | $H_2O$ | 2019.6 | CM1-W1 | 18:08:49.366 (0.004) | -20:05:58.990 (0.002) | 0.008, 0.007 | 5.75, 12.00 | 6.50 | 2.88 (0.03) |
| | | | NC-W1 | 18:08:46.670 (0.003) | -20:05:51.470 (0.001) | 0.006, 0.003 | 2.75, 13.00 | 9.50 | 8.53 (0.04) |
| | | | NC-W2 | 18:08:49.301 (0.013) | -20:06:03.54 (0.01) | 0.014, 0.015 | -2.25, 9.50 | 0.75 | 0.38 (0.03) |
| | | | NC-W3 | 18:08:49.421 (0.009) | -20:05:58.34 (0.01) | 0.010, 0.007 | 17.00, 17.50 | 17.25 | 0.28 (0.03) |
| | | | NC-W4 | 18:08:52.465 (0.013) | -20:06:02.77 (0.38) | 0.023, 0.376 | 8.00, 21.75 | 21.00 | 3.75 (0.03) |
| G10.34–0.14 | $CH_3OH$ | 2018.4 | CM1-M1 | 18:08:59.986 (0.041) | -20:03:35.45 (0.10) | 0.039, 0.168 | 4.50, 20.50 | 11.50 | 10.2 (0.1) |
| | $H_2O$ | 2018.1 | CM1-W1 | 18:08:59.986 (0.012) | -20:03:35.67 (0.02) | 0.015, 0.038 | -1.00, 23.75 | 17.75 | 13.0 (0.1) |
| | | | CM2-W1 | 18:08:59.990 (0.186) | -20:03:39.00 (0.17) | 0.182, 0.201 | -34.25, 49.25 | 13.25 | 9.86 (0.01) |
| | | | NC-W1 | 18:08:59.967 (0.001) | -20:03:36.970 (0.004) | 0.008, 0.030 | 13.00, 15.50 | 13.75 | 6.25 (0.01) |
| | | | NC-W2 | 18:09:00.311 (0.009) | -20:03:39.24 (0.04) | 0.010, 0.045 | 14.50, 16.00 | 15.25 | 0.100 (0.01) |
| | | | NC-W3 | 18:09:00.792 (0.100) | -20:04:18.00 (0.13) | 0.107, 0.135 | 0.50, 23.25 | 10.25 | 0.497 (0.008) |

*Table 6 continued*

Table 6 *(continued)*

| EGO | Maser[a] Species | Obs[b] Epoch | Group[c] Name | Centroid Position[d] RA (h m s) (″) | Dec (° ′ ″) (″) | Angular Spread[e] (RA, Dec) (″) | $V_{Min}$, $V^{f}_{Max}$ (km/s) | $V^{g}_{Peak}$ (km/s) | $S^{h}_{Peak}$ (Jy) |
|---|---|---|---|---|---|---|---|---|---|
| | | | NC-W4 | 18:09:02.320 (0.004) | -20:03:19.41 (0.01) | 0.010, 0.012 | 5.50, 20.75 | 11.50 | 1.40 (0.01) |
| | | | NC-W5 | 18:09:03.074 (0.006) | -20:03:02.27 (0.01) | 0.006, 0.018 | 40.75, 41.75 | 41.25 | 0.093 (0.007) |
| | $NH_3$ | 2018.1 | NC-A1 | 18:08:59.653 (0.005) | -20:03:33.14 (0.01) | 0.007, 0.005 | 14.25, 15.00 | 14.50 | 0.168 (0.008) |
| | | | NC-A2 | 18:09:01.567 (0.015) | -20:03:26.60 (0.04) | 0.017, 0.051 | 12.00, 12.50 | 12.25 | 0.08 (0.01) |
| G12.91–0.03 | $CH_3OH$ | 2018.4 | CM2-M1 | 18:13:48.260 (0.009) | -17:45:39.18 (0.01) | 0.021, 0.032 | 50.5, 61.75 | 59.5 | 23.1 (0.1) |
| | $H_2O$ | 2018.1 | CM1-W1 | 18:13:48.265 (0.004) | -17:45:39.79 (0.10) | 0.006, 0.096 | 61.00, 71.75 | 64.75 | 3.77 (0.02) |
| | | | NC-W1 | ··· | ··· | ··· | ··· | ··· | |
| | | | NC-W2 | 18:13:47.282 (0.023) | -17:45:36.47 (0.01) | 0.048, 0.014 | -4.75, 96.50 | 30.75 | 3.59 (0.02) |
| | | | NC-W3 | ··· | ··· | ··· | ··· | ··· | |
| | | | NC-W4 | ··· | ··· | ··· | ··· | ··· | |
| | $H_2O$ | 2019.6 | CM1-W1 | 18:13:48.265 (0.005) | -17:45:39.67 (0.07) | 0.013, 0.120 | 62.50, 73.75 | 70.75 | 3.94 (0.02) |
| | | | NC-W1 | 18:13:46.665 (0.009) | -17:45:25.81 (0.01) | 0.017, 0.010 | 49.00, 60.00 | 59.25 | 0.998 (0.015) |
| | | | NC-W2 | 18:13:47.282 (0.012) | -17:45:36.46 (0.01) | 0.015, 0.009 | 20.25, 40.75 | 20.75 | 0.24 (0.02) |
| | | | NC-W3 | 18:13:47.499 (0.004) | -17:45:42.110 (0.001) | 0.007, 0.002 | 44.50, 46.75 | 46.00 | 1.47 (0.02) |
| | | | NC-W4 | 18:13:48.323 (0.008) | -17:45:37.45 (0.01) | 0.020, 0.011 | 11.75, 14.50 | 13.50 | 1.04 (0.02) |
| G14.33–0.64 | $CH_3OH$ | 2018.4 | CM1-M1 | 18:18:54.324 (0.032) | -16:47:49.66 (0.02) | 0.095, 0.037 | 20.25, 24.25 | 22.25 | 0.640 (0.005) |
| | $H_2O$ | 2018.1 | CM1-W1 | 18:18:54.334 (0.005) | -16:47:49.190 (0.004) | 0.011, 0.007 | 20.50, 22.25 | 21.25 | 0.298 (0.008) |
| | | | CM4-W1 | 18:18:54.667 (0.099) | -16:47:50.20 (0.07) | 0.115, 0.073 | 15.75, 33.75 | 32.75 | 2.84 (0.01) |
| | | | NC-W1 | 18:18:54.519 (0.008) | -16:47:48.23 (0.01) | 0.010, 0.016 | 6.25, 16.75 | 14.75 | 0.235 (0.008) |
| | | | NC-W2 | 18:18:54.871 (0.057) | -16:47:50.95 (0.07) | 0.057, 0.073 | 21.00, 23.50 | 21.50 | 0.073 (0.007) |
| | $NH_3$ | 2018.1 | NC-A1 | 18:18:53.524 (2.645) | -16:48:01.27 (2.34) | 3.405, 2.757 | 23.00, 24.50 | 23.25 | 0.169 (0.053) |
| | | | NC-A2 | 18:18:54.487 (0.020) | -16:47:52.66 (0.01) | 0.022, 0.010 | 20.25, 21.50 | 21.00 | 0.027 (0.006) |
| G14.63–0.58 | $CH_3OH$ | 2018.4 | CM1-M1 | 18:19:15.205 (0.003) | -16:30:03.830 (0.003) | 0.008, 0.007 | 24.25, 26.00 | 25.00 | 4.33 (0.01) |
| | $H_2O$ | 2019.6 | CM1-W1 | 18:19:15.220 (0.018) | -16:30:03.97 (0.01) | 0.025, 0.011 | 21.50, 31.00 | 29.25 | 2.29 (0.02) |
| | | | NC-W1 | 18:19:15.542 (0.031) | -16:29:45.78 (0.02) | 0.027, 0.020 | 15.75, 21.00 | 16.50 | 1.20 (0.02) |
| G18.89–0.47 | $CH_3OH$ | 2018.4 | NC-M1 | 18:27:06.313 (0.004) | -12:41:46.52 (0.01) | 0.005, 0.006 | 73.25, 74.75 | 73.50 | 0.194 (0.006) |
| | | | NC-M2 | 18:27:07.823 (0.006) | -12:41:36.90 (0.01) | 0.012, 0.026 | 72.00, 73.25 | 72.50 | 0.271 (0.006) |
| | | | NC-M3 | 18:27:07.837 (0.001) | -12:41:35.870 (0.004) | 0.003, 0.014 | 52.50, 65.50 | 55.50 | 2.59 (0.01) |
| | $H_2O$ | 2018.1 | CM4-W1 | 18:27:07.910 (0.009) | -12:41:37.98 (0.01) | 0.024, 0.029 | 54.75, 57.50 | 56.50 | 0.024 (0.007) |
| | | | NC-W1 | 18:27:06.292 (0.008) | -12:41:46.79 (0.01) | 0.023, 0.021 | 42.25, 66.50 | 47.50 | 1.29 (0.01) |
| | | | NC-W2 | 18:27:07.789 (0.019) | -12:41:36.12 (0.03) | 0.024, 0.032 | 64.75, 66.50 | 65.25 | 0.052 (0.006) |
| | | | NC-W3 | 18:27:07.824 (0.008) | -12:41:36.88 (0.01) | 0.027, 0.026 | 51.25, 57.75 | 52.25 | 0.711 (0.007) |
| | | | NC-W4 | 18:27:07.973 (0.014) | -12:41:37.73 (0.01) | 0.031, 0.017 | 55.50, 57.00 | 56.50 | 0.292 (0.008) |
| | $NH_3$ | 2018.1 | NC-A1 | 18:27:07.705 (0.010) | -12:41:37.01 (0.01) | 0.015, 0.021 | 65.50, 66.50 | 66.00 | 0.224 (0.008) |
| | | | NC-A2 | 18:27:07.978 (0.014) | -12:41:35.63 (0.03) | 0.024, 0.037 | 63.50, 66.25 | 65.00 | 0.113 (0.009) |
| G19.36–0.03 | $CH_3OH$ | 2018.4 | CM3-M1 | 18:26:25.781 (0.042) | -12:03:53.27 (0.02) | 0.043, 0.014 | 24.00, 30.25 | 25.25 | 23.0 (0.1) |
| | $H_2O$ | 2018.1 | CM3-W1 | 18:26:25.788 (0.007) | -12:03:53.31 (0.01) | 0.023, 0.039 | 19.50, 21.25 | 20.25 | 0.71 (0.02) |
| | | | NC-W1 | 18:26:24.086 (0.003) | -12:03:09.58 (0.01) | 0.003, 0.009 | 12.25, 13.00 | 12.75 | 0.416 (0.018) |
| | | | NC-W2 | 18:26:25.131 (0.026) | -12:04:04.29 (0.03) | 0.032, 0.028 | 25.50, 26.00 | 25.75 | 0.077 (0.014) |
| | | | NC-W3 | 18:26:25.409 (0.011) | -12:03:47.25 (0.02) | 0.026, 0.035 | 24.50, 25.75 | 25.00 | 0.371 (0.017) |
| | | | NC-W4 | 18:26:27.150 (0.010) | -12:03:54.91 (0.01) | 0.015, 0.024 | -9.25, 6.00 | -5.75 | 0.580 (0.017) |
| | $H_2O$ | 2019.6 | CM3-W1 | 18:26:25.787 (0.027) | -12:03:53.32 (0.01) | 0.030, 0.009 | 19.75, 24.25 | 21.75 | 0.235 (0.023) |
| | | | NC-W1 | ··· | ··· | ··· | ··· | ··· | |
| | | | NC-W2 | ··· | ··· | ··· | ··· | ··· | |
| | | | NC-W3 | 18:26:25.421 (0.014) | -12:03:47.090 (0.004) | 0.014, 0.005 | 16.75, 26.00 | 25.25 | 0.27 (0.02) |
| | | | NC-W4 | 18:26:27.148 (0.010) | -12:03:54.920 (0.003) | 0.013, 0.003 | -5.00, -4.00 | -4.50 | 0.230 (0.018) |
| | $NH_3$ | 2018.1 | NC-A1 | 18:26:25.148 (0.006) | -12:03:52.88 (0.01) | 0.007, 0.008 | 25.50, 26.25 | 25.75 | 0.094 (0.011) |
| | | | NC-A2 | 18:26:25.609 (0.006) | -12:03:47.65 (0.02) | 0.010, 0.033 | 25.75, 26.50 | 26.00 | 0.198 (0.010) |

*Table 6 continued*

Table 6 *(continued)*

| EGO | Maser[a] Species | Obs[b] Epoch | Group[c] Name | Centroid Position[d] RA (h m s) ($''$) | Dec (° ′ ″) ($''$) | Angular Spread[e] (RA, Dec) ($''$) | $V_{Min}$, $V^f_{Max}$ (km/s) | $V^g_{Peak}$ (km/s) | $S^h_{Peak}$ (Jy) |
|---|---|---|---|---|---|---|---|---|---|
| | | | NC-A3 | 18:26:26.018 (0.069) | -12:03:57.27 (0.14) | 0.079, 0.162 | 26.50, 28.00 | 27.50 | 0.34 (0.01) |
| | $NH_3$ | 2019.6 | NC-A1 | 18:26:25.148 (0.033) | -12:03:52.89 (0.01) | 0.109, 0.007 | 25.50, 26.00 | 25.50 | 0.061 (0.007) |
| | | | NC-A2 | 18:26:25.607 (0.009) | -12:03:47.67 (0.01) | 0.013, 0.007 | 25.75, 26.50 | 26.00 | 0.178 (0.018) |
| | | | NC-A3 | 18:26:26.017 (0.047) | -12:03:57.23 (0.14) | 0.071, 0.210 | 26.75, 28.00 | 27.25 | 0.31 (0.01) |
| G22.04+0.22 | $CH_3OH$ | 2018.4 | CM1-M1 | 18:30:34.698 (0.045) | -09:34:47.00 (0.14) | 0.075, 0.200 | 44.50, 55.50 | 54.50 | 9.31 (0.01) |
| | $H_2O$ | 2018.1 | CM1-W1 | 18:30:34.691 (0.057) | -09:34:47.20 (0.04) | 0.091, 0.093 | 43.75, 54.75 | 52.75 | 10.1 (0.1) |
| | | | NC-W1 | 18:30:34.394 (0.006) | -09:34:42.46 (0.01) | 0.015, 0.019 | 54.25, 58.25 | 55.00 | 0.840 (0.010) |
| | | | NC-W2 | 18:30:34.561 (0.001) | -09:34:50.53 (0.01) | 0.001, 0.002 | 50.00, 52.25 | 51.50 | 2.07 (0.01) |
| | | | NC-W3 | 18:30:34.776 (0.029) | -09:34:45.00 (0.04) | 0.035, 0.048 | 47.00, 48.25 | 47.25 | 0.08 (0.01) |
| | $H_2O$ | 2019.6 | CM1-W1 | 18:30:34.691 (0.075) | -09:34:47.21 (0.06) | 0.135, 0.071 | 44.25, 60.00 | 46.25 | 7.40 (0.02) |
| | | | NC-W1 | ... | ... | ... | ... | ... | |
| | | | NC-W2 | ... | ... | ... | ... | ... | |
| | | | NC-W3 | ... | ... | ... | ... | ... | |
| G28.83–0.25 | $CH_3OH$ | 2018.4 | CM1-M1 | 18:44:51.084 (0.104) | -03:45:48.44 (0.16) | 0.109, 0.137 | 79.50, 94.00 | 83.50 | 61.4 (0.1) |
| | $H_2O$ | 2018.1 | CM1-W1 | 18:44:51.086 (0.081) | -03:45:48.45 (0.10) | 0.077, 0.094 | 72.75, 101.75 | 85.00 | 2.47 (0.01) |
| | | | NC-W1 | 18:44:50.678 (0.010) | -03:45:47.41 (0.01) | 0.011, 0.014 | 87.75, 88.00 | 87.75 | 0.034 (0.007) |
| | $NH_3$ | 2018.1 | NC-A1 | 18:44:52.794 (0.061) | -03:45:49.67 (0.09) | 0.080, 0.117 | 86.00, 86.75 | 86.50 | 0.192 (0.007) |
| | | | NC-A2 | 18:44:52.876 (0.040) | -03:45:47.43 (0.04) | 0.048, 0.045 | 85.00, 88.00 | 86.75 | 0.177 (0.007) |

[a] The molecular species of the maser emission. Emission denoted "$NH_3$" refers exclusively to the $NH_3$ (3,3) metastable state, as we did not detect any $NH_3$ (6,6) emission with these observations.

[b] The epoch in which the listed observations were taken. Exact dates for each observation can be found in Tables 2 & 3.

[c] Maser groups are named for the centimeter continuum source with which they are associated, and for their maser species. If a maser group is not associated with any centimeter continuum emission, it is named "NC" (for "no centimeter continuum") and the maser species. The "NC" numbering scheme is separate for each species and increases with increasing Right Ascension within a given EGO. Maser groups with " ··· " listed in the data columns were non-detections in that epoch.

[d] The intensity-weighted position of each maser group. Uncertainties are the standard deviation of the difference between the maser spots in each group and the centroid position, weighted by flux density. Uncertainties are in arcseconds for both RA and Dec.

[e] A measure of the physical extent of the maser group. Each RA, Dec value is the standard deviation of the unweighted differences between each individual maser spot and the centroid position for the maser group.

[f] The minimum and maximum velocity at which $>4\sigma$ maser emission of a given species is detected.

[g] The velocity at which the brightest maser spot in each group appears, as determined by the integrated flux density of its 2D Gaussian fit.

[h] The integrated flux density of the brightest maser spot in each group, as determined by its 2D Gaussian fit. Uncertainties, returned by the Gaussian fit, are in parentheses.

### 3.2.3. *Properties of the Maser Emission*

We detect at least two sites of 22 GHz $H_2O$ maser emission and at least one site of 6.7 GHz $CH_3OH$ maser emission in every EGO, for detection rates of 100% for both species. For the 4 sources which were observed twice at 1.3 cm, the detection statistics described in this section treat the two epochs separately for these 4 sources. (See § 4.4 for a discussion of how these sources' maser emission varies with time.) In most cases, 22 GHz $H_2O$ maser emission had previously been detected toward these EGOs in single-dish data (Cyganowski et al. 2013), but had not been observed interferometrically. The exceptions are G19.36–0.03, which was not detected by Cyganowski et al. (2013), and G14.63–0.58, for which NC-W1 was detected in the VLBI observations of Sanna et al. (2018).

Likewise, 6.7 GHz $CH_3OH$ masers have been previously detected toward most sources with either the VLA (Cyganowski et al. 2009) or Australia Telescope Compact Array (ATCA; Green et al. 2010). The exception is G14.33–0.64, for which the observations reported in this work are, to the best of our knowledge, the first reported interferometric observations of 6.7 GHz $CH_3OH$ masers in this source. We detect $NH_3$ (3,3) maser emission in five of our nine EGOs (56% detection rate), but do not detect any $NH_3$ (6,6) maser emission in any EGOs at the current sensitivity and angular resolution. There is no trend in $NH_3$ (3,3) maser presence/absence with EGO luminosity, mass, $L/M$, or 870 μm or mid-infrared flux density. Our surface bright-ness sensitivities (see table notes in Tables 2 and 3) preclude the possibility of detecting thermal line emission for any of these three molecules.

The number of $H_2O$ maser groups per EGO ranges from 2 to 7, with a median of 5; the number of $NH_3$ (3,3) maser groups per EGO is 2-3, with a median of 2. In most EGOs, we detect only one 6.7 GHz $CH_3OH$ maser group; the lone exception is G18.89–0.47, which has three 6.7 GHz maser groups. This is consistent with the results of Cyganowski et al. (2009), who detect only one site of 6.7 GHz $CH_3OH$ maser emission for all but one EGO (G11.92–0.61) in their sample of 20, at $\sim3''$ resolution. A comparison between our data and previous maser observations of each EGO can be found in § 4.4.

The majority of the $H_2O$ masers in our sample are not associated with detectable centimeter continuum emission, but strong centimeter continuum emission *is* strongly associated with $H_2O$ masers: either CM1 or CM2 is associated with $H_2O$ maser emission in 8 out of 9 EGOs. Likewise, the 6.7 GHz $CH_3OH$ masers are nearly always associated with either CM1 or CM2 for a given EGO (8/9 cases). The lone exception to both the $H_2O$ and $CH_3OH$ trends is G18.89–0.47, in which only CM4 has associated $H_2O$ masers, and none of the three $CH_3OH$ maser detections is associated with detectable centimeter continuum emission.

The spatial distribution of the masers in our sample varies by molecular species. $H_2O$ masers are typically found $\sim3''$-$8''$ (3400 to 38400 au) away from the center of the ATLASGAL clump hosting the EGO (defined as the peak of the 870 μm emission), while $NH_3$ (3,3) masers are more bimodally distributed, with about half the detections $\sim4''$ from the ATLASGAL peak and half $\sim20''$ away. The 6.7 GHz $CH_3OH$ masers are generally located at or near the ATLASGAL peak, and are usually also coincident with the brightest 37 μm source in the field. The exception to all of these trends is again G18.89–0.47, in which the $CH_3OH$ maser NC-M1 is located $\sim20''$ to the southwest of the nearest 37 μm source, and well away from the ATLASGAL peak. G10.29–0.13 was not detected at 37 μm in Towner et al. (2019), so it is excluded from the trends related to 37 μm emission.

We report detailed, per-channel fit results for all maser species and epochs in our online tables. Each maser species and epoch of observation is listed in a separate table for each EGO. Fitted sizes are not reported, since all masers were fit as point sources. Table 7 shows an abbreviated version of the per-channel fit results for G10.29–0.13 as an example of the information contained in the online tables.

**Table 7**. 2D Gaussian Fit Results[a]: G10.29–0.13, Epoch 2018.1

| N[b] | Group[c] Name | Velocity[d] (km/s) | RA[e] (h m s) ($''$) | Dec[e] (° ′ ″) ($''$) | Flux[f] (Jy) |
|---|---|---|---|---|---|
| 1 | NC-W2 | -12.00 | 18:08:49.2973 (0.017) | -20:06:03.532 (0.006) | 0.0639 (0.0127) |
| 2 | NC-W2 | -11.75 | 18:08:49.3008 (0.017) | -20:06:03.538 (0.005) | 0.0535 (0.0116) |
| 3 | NC-W2 | -11.50 | 18:08:49.3000 (0.019) | -20:06:03.545 (0.006) | 0.0279 (0.0088) |
| 4 | CM1-W1 | +5.50 | 18:08:49.3651 (0.011) | -20:05:58.992 (0.004) | 0.136 (0.015) |
| 5 | CM1-W1 | +5.75 | 18:08:49.3649 (0.004) | -20:05:58.990 (0.001) | 0.303 (0.014) |
| 6 | CM1-W1 | +6.00 | 18:08:49.3652 (0.005) | -20:05:58.993 (0.002) | 0.303 (0.015) |
| 7 | CM1-W1 | +6.25 | 18:08:49.3653 (0.003) | -20:05:58.983 (0.001) | 0.466 (0.014) |
| 8 | NC-W1 | +6.50 | 18:08:46.6678 (0.011) | -20:05:51.469 (0.004) | 0.113 (0.012) |
| 9 | CM1-W1 | +6.50 | 18:08:49.3653 (0.002) | -20:05:58.984 (0.001) | 0.640 (0.014) |
| 10 | NC-W4 | +6.50 | 18:08:52.4618 (0.021) | -20:06:03.087 (0.007) | 0.0714 (0.0123) |

[a] The print version of this table is truncated at 10 lines; full tables for all sources can be found in the online materials.

[b] Fits are listed on a per-channel basis. If a channel has three maser spots, all three spots for that channel will be listed before any spots for the next channel. See N = 8-10 in this table: there are three masers (NC-W1, CM1-W1, and NC-W4) at +6.50 km $s^{-1}$, so all three masers are listed before any maser data for +6.75 km $s^{-1}$.

[c] The name of the maser group with which each maser is associated. See text for naming convention.

[d] The velocity of the channel in which the fit was performed. If multiple maser spots were fit within the same channel, the fit for each spot is listed separately, in order of increasing RA.

[e] J2000 coordinates of each maser spot, as returned by `imfit`. Uncertainties are those returned by `imfit`, in arcseconds for both RA and Dec.

[f] The integrated flux density of the 2D Gaussian fit to the maser spot.

Figure 3 and the associated figure set show "spot maps" of the maser emission in each EGO. Each EGO has one overview panel, labeled with the EGO name only, which shows the 37 μm emission as a greyscale background with 37 μm contours in light yellow, and maser positions overlaid. We then show additional, "high-zoom" images for select maser groups; these images are labeled with both the EGO name and the maser group name. We show high-zoom images for a maser group if either of the following conditions is met: a) the maser group is associated with centimeter continuum emission in one or both bands, or b) the projected linear separation of any two individual maser spots is significant (defined as the separation being greater than the sum of the relative position uncertainties). For well-calibrated interferometric images, the relative position uncertainty of an unresolved (point) source relative to other unresolved sources is:

$$\Delta\theta \sim \frac{\theta_{synth.beam}}{2 \times (S/N)} \quad (2)$$

where $S/N$ is the signal-to-noise of the source in question (see Cyganowski et al. 2009, and references therein). For a $10\sigma$ detection with a synthesized beam of $\sim0.''30$, this gives a relative position uncertainty of 15 mas.

The high-zoom images in Figure 3 and the associated online Figure Set show a separate symbol for every maser spot in a maser group, color-coded by velocity, on top of the centimeter continuum (shown in greyscale). The $NH_3$ masers are not shown on these spotmaps in order to ensure that the $H_2O$ and $CH_3OH$ emission is always visible. The scaling of the greyscale images - both 37 μm and centimeter continuum - is arbitrary and meant only to aid the eye; source intensity and morphology are best seen by examining the contour levels, and specific source flux densities can be found in Table 4. Likewise, the size of the maser symbols is not reflective of the positional uncertainty for any individual maser spot. As a rule of thumb, the relative position uncertainty between two maser spots of S/N = $4\sigma$ (our detection cutoff) will be $\sim$33 mas, which is much smaller than the effective symbol size in the high-zoom images.

## 4. ANALYSIS

In this section we discuss the nature of the continuum and maser emission in our sample as a whole. In § 4.1, we discuss the distribution of spectral indices among our sample, how this compares to similar samples, and what this implies about the general nature of weak, compact/unresolved continuum emission in our EGOs. In § 4.2, we examine how our continuum detections compare to established observational relations between various source properties. § 4.3 and 4.4, we discuss the characteristics of the maser emission and its variation with time.

### 4.1. *Nature of the Continuum Emission*

Figure 4 shows the distribution of spectral indices ($\alpha$) in our sample in steps of $\Delta\alpha = 0.25$. Uncertainties for each bin are listed in the figure caption.

#### 4.1.1. *Spectral Index Statistics, and Properties of Sources with Constrained Sizes*

Of the sources detected at both 1.3 and 5 cm, the minimum and maximum $\alpha$ are −0.42 and 1.58, respectively, with a median of 0.59±0.45 and a mean of 0.50±0.59. The uncertainty on the median is the MAD, and the uncertainty on the mean is the standard deviation. The median and mean of the uncertainties for individual measurements are 0.17 and 0.15, respectively. Our derived lower limits range from 0.43 to 2.25 with a median of 0.77, and our derived upper limits range from -0.35 to 0.89, with a median of 0.08. It is important to note that, since we used a $4\sigma$ upper limit value for each nondetection, our upper and lower limits are not solely a reflection of source properties but are also dependent on the noise in our data.

The range and statistical properties of our derived spectral indices are consistent with the results reported by other teams for similar MYSO samples. For their sample of 58 high-mass star-forming regions spanning a range of evolutionary states, Rosero et al. (2016) find a total range of $-1.2 \leq \alpha_{6cm-1.3cm} \leq 1.8$, with the majority in the range $-0.1 < \alpha < 1.1$ and a median of $\alpha = 0.5$. Purser et al. (2016) find a median of $\alpha \sim 0.6$ for their (sub)sample of 26 jets observed at 5.5, 9, 17, and 22.8 cm, as expected for those sources, and a median of $\alpha \sim 0.08$ for their sample as a whole (see the supplemental material for that paper). The minimum and maximum $\alpha$ they report are -2.48 and 1.71, respectively. Sanna et al. (2018) find a total range of $-0.1 \leq \alpha \leq 2.5$ for their POETS sample of 36 massive star-forming regions (targeted for their weak radio-continuum and rich 22 GHz $H_2O$ maser emission), with a median of $\alpha = 0.8$.

Table 8 shows the angular and physical sizes for those sources in Table 4 whose sizes are either fully constrained or have upper limits. Eight sources (22%) have `imfit`-derived sizes in at least one band, and a further nine sources (25%) have upper limits on their sizes. Our remaining sources are either unresolved or had some parameters fixed during the fitting procedure. At the distances of our sample, the high number of unresolved sources suggests that most of these emitting regions are quite small ($\lesssim$500 au). Our ability to derive deconvolved, fitted sizes depends on both the intrinsic physical size of each source and its signal-to-noise ratio; sources whose `imfit`-derived sizes are listed as either 'unres' or 'fixed' are preferentially the weaker sources in our sample. Consequently, the statistics on source size discussed in this section should be assumed to be biased toward stronger and/or larger sources.

Sources with fully-constrained sizes are always less than 1000 au in any dimension; the minimum and maximum fitted sizes in the sample are 150 au × 75 au (G19.36–0.03 CM1 at 1.3 cm) and 756 au × 497 au (G14.33–0.64 CM7 at 1.3 cm). These sizes meet both the Hoare et al. (2007) and Kurtz (2002) size criteria for HCH II regions (<0.05 pc and $\leq$0.01 pc, respectively), but are too large in one or both dimensions to be consistent with either gravitationally-trapped HCH II regions (a few tens to $\sim$100 au Keto 2003; Lund et al. 2019) or outflow-confined H II regions (286 au at 22 GHz and 115 au at 6 GHz on breakout, following the equation in §3.2.2, paragraph 2 of Tanaka et al. 2016). The source G19.36–0.03 CM1 could potentially fit the size criteria of an outflow-confined H II region if the 1.3 cm major axis is an overestimate, but the source is underluminous by approximately two orders of magnitude at both 6 and 22 GHz compared to the predictions shown in Tanaka et al. (2016, Figure 11) and is therefore unlikely to be a good candidate for this object type.

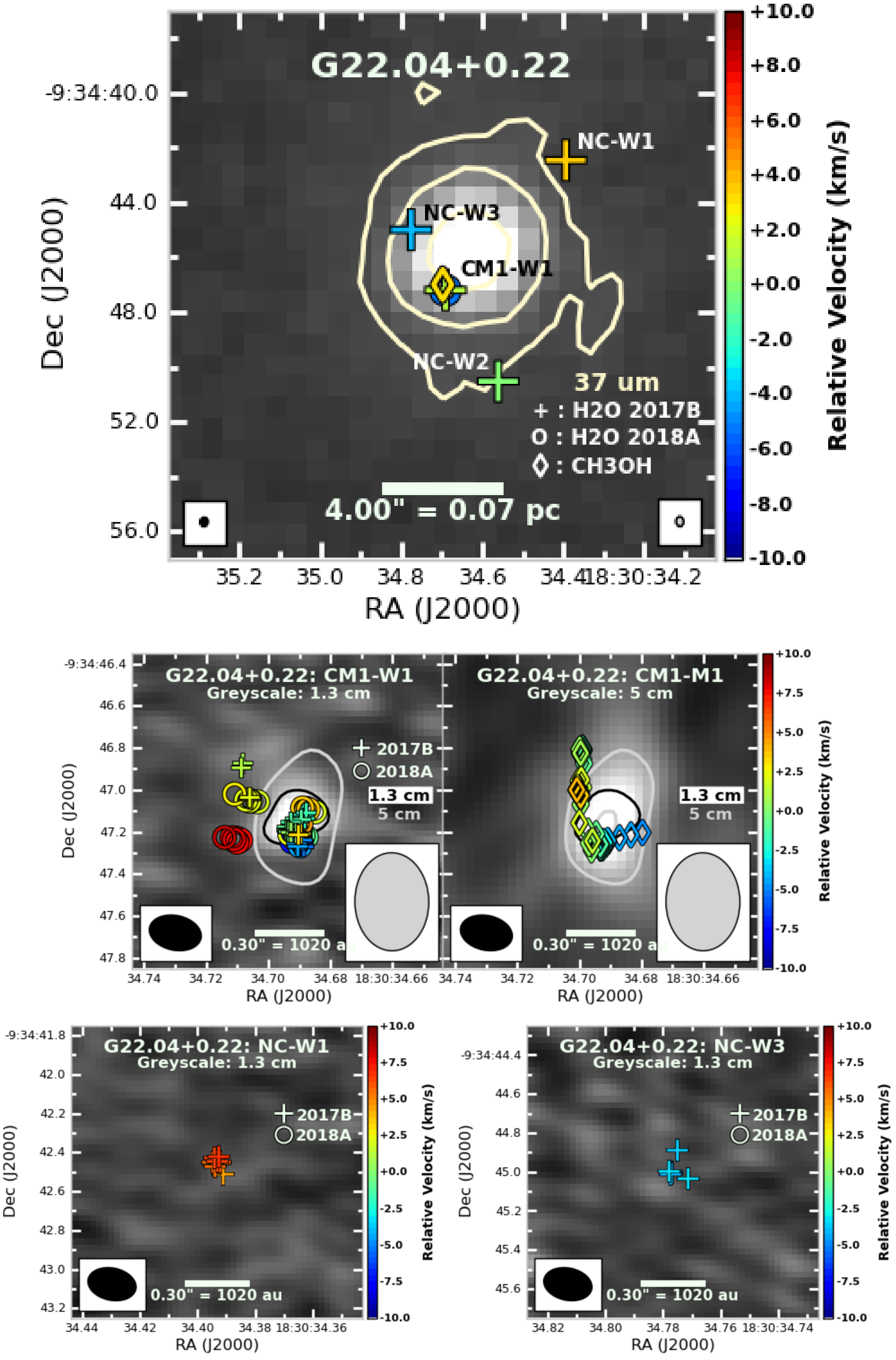

**Figure 3**. Maser spot maps for G22.04+0.22. *Full FOV image (20″)*: background greyscale is 37 μm continuum from Towner et al. (2019), and 37 μm contour levels (yellow) are [5,15,45]×$\sigma$, where $\sigma$ = 0.21 Jy beam$^{-1}$ is the 37 μm scaled MAD. We show one symbol for each maser group, color-coded by the velocity of the peak emission in that group. *CM1-W1 & -M1:* FOV is 1.″5, and contour levels are [4]×$\sigma$ at both 1.3 cm (black) and 5 cm (silver), where $\sigma_{1.3cm}$ = 12.3 μJy beam$^{-1}$ and $\sigma_{5cm}$ = 7.5 μJy beam$^{-1}$. The background greyscale is 1.3 cm in the left-hand panel, and 5 cm in the right-hand panel. All $H_2O$ maser spots from both observing epoch 2018.1 (+ symbols) and epoch 2019.6 are shown (○ symbols); $CH_3OH$ maser spots (epoch 2018.4) are shown with ◇. Each maser spot is color-coded according to its velocity relative to source $V_{LSR}$ (51.0 km s$^{-1}$; see Table 1). *NC-W# Images:* Same FOV and maser-spot symbol and color system as CM1-W1. Background greyscale is 1.3 cm in all cases; no contours are shown because there is no continuum emission ≥4 $\sigma$ in these FOV. The complete figure set (46 images) is available in the online journal.

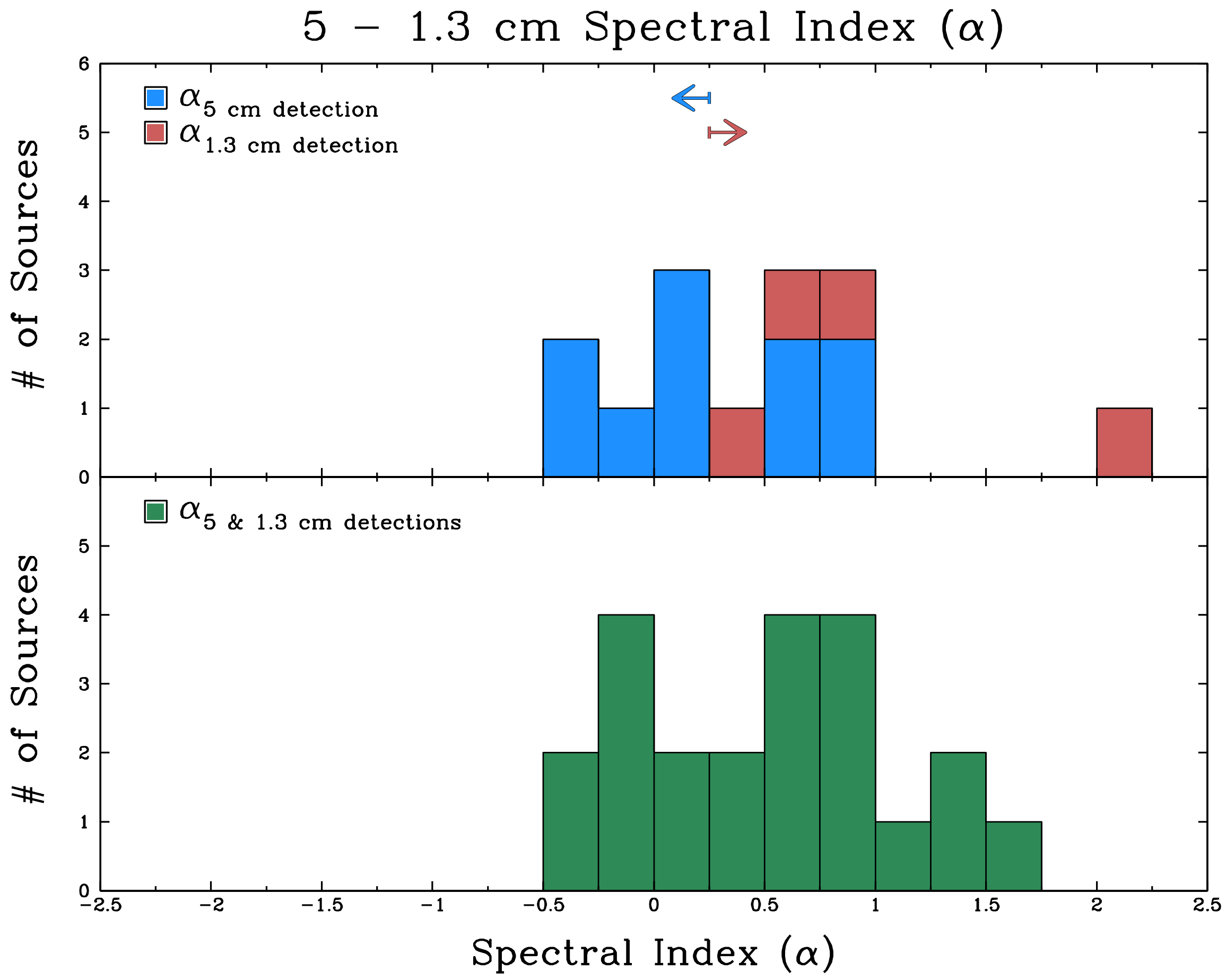

**Figure 4**. Histograms of spectral indices ($\alpha$) for all centimeter continuum detections in the EGO-9 sample. The bottom panel shows the distribution of $\alpha$ for sources detected at both 1.3 and 5 cm (green). The top panel shows a color-coded distribution of spectral indices which are upper limits (blue) or lower limits (red), indicated by colored arrows. These bars are "stacked": for example, there are three spectral indices within $0.75 < \alpha < 1.0$ - two are upper limits (blue) and one is a lower limit (red). For sources detected at only one wavelength, we adopt the $4\sigma$ limit as the flux density "measurement" for the non-detected band to derive the upper/lower limit for $\alpha$. Histogram bin counts and their uncertainties for the lower panel, for $-0.75 \leq \alpha \leq 2.0$, are: $0^{+1}_{-0}$, $2^{+2}_{-2}$, $4^{+3}_{-3}$, $2^{+4}_{-2}$, $2^{+3}_{-2}$, $4^{+2}_{-3}$, $4^{+3}_{-3}$, $1^{+4}_{-1}$, $2^{+2}_{-2}$, $1^{+2}_{-1}$, $0^{+0}_{-0}$.

Our sources with derived sizes do favor steeper spectral indices, with $\alpha > 1.1$ in four out of eight cases. In fact, these sources account for all $\alpha > 1.1$ in our sample, both constrained and lower limits. Such spectral indices are indicative of some appreciable optical depth; in the case of a jet, these spectral index values require acceleration or recombination within the flow (see Reynolds 1986, and additional discussion below). The remaining four sources have spectral indices ranging from −0.33 to 1.03, i.e. there is no bias toward or away from the spectral indices canonically associated with emission from an ionized jet or stellar wind($\alpha \sim$ 0.6 Reynolds 1986; Panagia & Felli 1975).

**Table 8**. Properties of Sources with Constrained Physical Sizes

| Source[a] | 1.3 cm Size[b] | | 5 cm Size[b] | |
|---|---|---|---|---|
| Name | (mas)×(mas) | au × au | (mas)×(mas) | au × au |
| G10.34 CM1 | 357 (111) × 146 (68) | 571 (197) × 230 (110) | fixed | ··· |
| CM2 | 379 (152) × 225 (181) | 610 (260) × 360 (295) | fixed | ··· |
| G12.91 CM1 | <360 | <1620 | unres | ··· |
| G14.33 CM1 | <1300 | <1500 | <2400 | <1810 |
| CM2 | <5500 | <6200 | <5600 | <6300 |
| CM4 | 301 (57) × 225 (45) | 340 (74) × 250 (58) | ··· | ··· |
| CM5 | 395 (194) × 175 (93) | 446 (225) × 200 (110) | <1800 | <2000 |
| CM7 | 669 (334) × 440 (316) | 756 (386) × 497 (361) | fixed | ··· |
| G14.63 CM1 | 154 (48) × 36 (31) | 282 (89) × 66 (57) | fixed | ··· |
| CM2 | <280 | <510 | <870 | <1600 |
| G19.36 CM1 | 69 (11) × 34 (16) | 150 (33) × 75 (37) | 73 (28) × 47 (42) | 160 (66) × 100 (94) |
| CM2 | <1800 | <4000 | <1950 | <4300 |
| CM3 | 192 (66) × 145 (111) | 420 (160) × 319 (249) | unres | ··· |
| CM4 | <680 | <1500 | <880 | <1900 |
| G22.04 CM1 | <600 | <2040 | <1100 | <3700 |
| G28.83 CM1 | <760 | <3600 | unres | ··· |
| CM2 | <890 | <4300 | <1150 | <5500 |

[a] Positions, flux densities, and spectral indices are listed in Table 4; see § 3.1.2 for source naming convention.

[b] Uncertainties on angular size (where given) are those returned by `imfit`; uncertainties on physical size reflect both angular-size uncertainty and distance uncertainty. Distances and their uncertainties are listed in Table 1. Sources which were non-detections at a particular wavelength have their angular size listed as ‘ ··· .’

The derived upper limits on source size range from 510 au (G14.63–0.58 CM2 at 1.3 cm) to 6300 au (G14.33–0.64 CM2 at 5 cm). Our sources with upper limits on size include two known or candidate H II regions. G14.33–0.68 CM2 is coincident with the known H II region IRAS 18159-1648 (Jaffe et al. 1982), and its size limits (6200 au at 1.3 cm and 6300 au at 5 cm) are consistent with the upper limits derived by those authors. G19.36–0.03 CM2 was observed by Cyganowski et al. (2011) at 3.6 cm and Towner et al. (2017) at 1.3 cm, and Cyganowski et al. (2011) identify it as an expanding H II region candidate. Its size limits (4000 au at 1.3 cm and 4300 au at 5 cm) are consistent with either a HCH II region or a small UCH II region, depending on the classification system used. Both sources have negative spectral indices; the spectral index for G14.33–0.64 CM2, $\alpha$ = −0.16 ± 0.06, is consistent within errors with that of optically thin free-free emission, but the spectral index of G19.36–0.03 CM2, −0.22 ± 0.09, is not. While H II regions with non-thermal spectral indices have been observed (Sgr B2(DS), attributed to Fermi acceleration in shocks; see Meng et al. 2019), it is also possible that we are missing some flux at 1.3 cm due to spatial filtering. Further investigation of this source, including combining these data with data taken in more compact VLA configurations, is needed.

#### 4.1.2. *Thermal Emission*

Given the prevalence of compact sources (and thus the absence of clear morphological markers of source type), we additionally examine our distribution of $\alpha$ to see if we can distinguish dominant emission mechanisms in our sample.The distribution of $\alpha$ in the bottom panel of Figure 4 ranges from –0.42 to 1.58, with the majority of emission between –0.25 $< \alpha <$ 1.0 and a tail from 1.0 $< \alpha <$ 1.25.

In general, free-free (thermal) emission in star-forming regions can produce $\alpha$ anywhere between –0.1 (completely optically thin emission) and 2 (optically thick emission). The canonical value for a partially-ionized spherical, isothermal, constant-velocity stellar wind is $\alpha$ = 0.6 (Panagia & Felli 1975), but this value can also be produced by, e.g., a conical, partially-ionized jet (Reynolds 1986). That work showed that $\alpha$ for such a jet can easily range from $-0.1 \leq \alpha \leq 1.1$ depending on the properties of the flow, such as gradients in temperature, ionization fraction, and velocity. Spectral indices >1.1 are possible, but require acceleration or recombination within the flow. Of the sources we detected at both wavelengths, the majority (14/22, 64%) fall in the range $-0.1 < \alpha < 1.1$. That is, they are consistent with the standard values for thermal free-free emission from an ionized or partially-ionized source (Reynolds 1986; Rosero et al. 2019b, and references therein). The sources with $\alpha > 1.1$ are the sources with resolved sizes, discussed above. All four sources detected at only 1.3 cm have derived lower limits of $\alpha \geq$0.43, i.e. they are consistent with thermal emission either from free-free interactions or warm dust.

#### 4.1.3. *Non-thermal Emission*

Seven sources (19% of the sample) have $\alpha < -0.1$, of which two are upper limits and five are detected at both wavelengths. As $\alpha$ = –0.1 is the minimum spectral index which can be produced by free-free emission (for the special case of completely optically-thin free-free emission), this suggests a small population of sources whose spectral indices are consistent with non-thermal emission. It is unlikely that all seven sources can be attributed to contamination from background galaxies, and in fact, two sources are extended (G14.33–0.64 CM2 and G19.36–0.03 CM2, both discussed above). One additional source is coincident with 6.7 GHz $CH_3OH$ masers; the other two fall well inside the extended 4.5 μm emission in their respective fields, making it less likely that they are field sources. The two non-thermal sources whose spectral indices are upper limits do lie well outside both the 4.5 μm extended emission and the ATLASGAL clumps; these detections may indeed be background sources. If this is the case, then this reduces our number of EGO-associated non-thermal detections to five (14% of the sample), which is still consistent with non-thermal detection rates reported by other teams.

Rosero et al. (2019b) conclude that at least 10% of the continuum detections in their sample have spectral indices consistent with non-thermal emission (which they define as $\alpha <$ -0.25). Purser et al. (2016) suggest that synchrotron emission may be commonplace within their jet sample specifically; they identify non-thermal emission sources associated with 10 out of 26 jets or jet candidates (38%), with an average spectral index of –0.55. From this, they conclude that synchrotron radiation is relatively commonplace within their jet sample overall. Sanna et al. (2018) notably do not detect any sources with $\alpha < -0.1$ within their sample. This is possibly due to their slightly more expansive selection criteria, as they are targeting very young protostellar outflows based in part on plentiful $H_2O$ maser emission, which the other samples do not consider.

#### 4.1.4. *Comparison with Prior Observations*

Of the 9 EGOs presented in this paper, only G28.83–0.25 had bright enough emission that Cyganowski et al. (2011) could examine its spectral index. For this field, Cyganowski et al. (2011) determine a 3.6-1.3 cm spectral index of $\alpha_{3.6-1.3} < 1.9$ for our source CM1, and $\alpha_{3.6-1.3} <1.1$ for our source CM2. Our derived spectral indices are consistent with these results in both cases ($\alpha_{5-1.3cm}$ = 0.93 ± 0.16 for CM1 and $\alpha_{5-1.3cm}$ = -0.42 ± 0.10 for CM2).

#### 4.1.5. *Limitations*

Both Reynolds (1986) and Purser et al. (2016) note the strong effect that spatial resolution can have on derived spectral indices. Reynolds (1986) suggests a model for a bipolar conical jet powered by a spherical region of hot, dense ionized material (see their Figure 2). In this model, different physical components of the jet have different spectral indices and will dominate the emission at different frequencies: the optically-thick central spherical region will dominate at higher frequencies and has a spectral index of $\alpha$ = 2, the (confined) inner jet will become prominent at slightly lower frequencies with a shallower spectral index ($\alpha \sim$0.2 depending on jet properties), and the (unconfined) outer jet will dominate at the lowest frequencies with a slightly steeper spectral index ($\alpha \sim$ 0.6 in this model; see Purser et al. (2016) for a thorough discussion). Fully spatially-resolved observations will recover the spectral index of each component, but observations which blend emission from all three components will derive a single spectral index, which obscures the internal jet structure.

Since most of our sources are unresolved, we must consider whether and how our angular resolution may be affecting our data. However, we have no sources which are so well-resolved that we can derive different spectral indices for different locations within the source, as discussed by Reynolds (1986). Instead, we turn to an analysis by Purser et al. (2016), who examine how the spectral indices of two of their well-resolved jet+lobe sources would change if those sources be-came unresolved. They found that degrading the angular resolution would, in both cases, flatten the measured spectral index of the target, but that the emission would still be fit well with a simple power law. They suggest that significant numbers of unresolved sources could lead to a flattening of the spectral indices in a jet population overall. Extrapolating this to our own sample, we conclude that blended thermal and non-thermal emission in our unresolved sources would likely have the same effect.

Contamination by warm dust, on the other hand, would have the opposite effect of resolution limitations. Dust contamination will decrease with increasing wavelength (assuming greybody emission), so shorter wavelengths will be more strongly affected, and the measured spectral indices will become steeper than they would be if dust were not contributing to the observed emission. Brogan et al. (2016) found, in their examination of the massive star-forming region NGC6334I, that the spectral energy distributions for two of their sources (MM1B and MM1D) could only be accurately fit with a combination of thermal dust and free-free emission. For these objects, the $\tau \sim 1$ warm dust emission components (438 and 309 K, respectively) produced flux densities of ~200 µJy at 22 GHz (1.3 cm). If we assume that similar source types are likely in our EGOs, then at the noise levels (14 µJy beam$^{-1}$) and typical flux densities (tens to a few hundred µJy) of our sources, even moderate dust contamination could significantly steepen the derived spectral indices. Our high percentage of $\alpha > 1.1$ in our sources with fully-constrained sizes - some of the strongest in the sample - raises the possibility that these sources in particular may exhibit both free-free emission from ionized gas and thermal emission from warm dust.

The best solution to both the blending/resolution and potential dust-contamination problems is data at additional, shorter wavelengths which, when combined with the centimeter data, can be fit with SED models which account for multiple emission mechanisms (free-free, dust, synchrotron). We discuss our planned future work in this area in greater detail in § 5.

#### 4.1.6. *Summary of Continuum Analysis*

In summary, the $\alpha$ values that we derive for our centimeter continuum detections are broadly consistent with the range of spectral indices predicted for thermal free-free emission and with the results of other teams (Purser et al. 2016; Rosero et al. 2016; Sanna et al. 2018). However, for the majority of our sample (19/36 sources, 53%) whose sizes are not well-constrained by our observations and due to the possibility of dust contamination at 1.3 cm, we are unable to distinguish between jets/stellar winds (e.g. Reynolds 1986) and radiatively-ionized regions such as gravitationally-trapped or bipolar H II regions (Keto 2003; Tanaka et al. 2016). Our spatially-resolved sources have a tendency toward elongated emission (aspect ratio >2 in 80% of cases) which is consistent with both shock-ionized jets and bipolar H II regions. Their spectral index values (typically $\alpha > 1.1$) are may indicate a small population of very young H II regions in these EGOs, but are also consistent with partially-ionized jets in which either acceleration or recombination are occurring within the flow or with emission from warm dust contributing additional flux at 1.3 cm. We also find a small population of detections with $\alpha < -0.1$, which is consistent with non-thermal emission and with results presented in the literature for other samples of high-mass star-forming regions (Rosero et al. 2016; Purser et al. 2016).

Towner et al. (2019) conducted a multiwavelength photometric analysis of a sample of 12 EGOs, including all 9 EGOs examined in this work. They found that the EGO population they examined occupied a specific, narrow range of luminosity-to-mass ratio values ($L/M$). $L/M$ is often taken as a proxy for evolutionary state; Urquhart et al. (2018) give a general range of 1-100 for massive star-forming regions, with compact H II regions becoming common at $L/M > 40$. The EGO-12 sample examined in Towner et al. (2019) spans the range 5-60 $L_\odot/M_\odot$, and the authors note that the $L/M$ space occupied by the EGO-12 sample as opposed to other similar samples suggests that EGOs may occupy the evolutionary stage in which the central massive protostar is about to reach or has just reached $M = 8M_\odot$ (see Towner et al. 2019, § 4.5 for the full, detailed discussion of this possibility). If so, this would place EGOs in a very specific, relatively short-lived evolutionary state. These centimeter continuum results, combined with the $L/M$ findings of Towner et al. (2019), suggest that there are a range of centimeter continuum-emitting processes present in EGOs *simultaneously* even in a single evolutionary stage. In some cases, such as G14.33–0.64 and G19.36–0.03, both of which contain both H II regions or candidates and continuum emission more consistent with ionized protostellar jets, this could be indicative of multiple stages of high-mass star formation occurring simultaneously within a single massive protocluster.

### 4.2. *Radio versus Bolometric and $H_2O$-Maser Luminosities*

In addition to spectral index, correlations between radio distance-luminosity ($L_{radio}$), clump bolometric luminosity ($L_{bol}$), and total water-maser luminosity ($L_{H_2O}$) for each EGO can help elucidate the nature of the continuum sources. By combining the observations presented in this work with the $L_{bol}$ values derived in Towner et al. (2019), we are able to analyze centimeter continuum emission along with $L_{bol}$ for the first time for a population of EGOs.

Anglada (1995), using observations of low-mass star-forming regions, derive a power-law relationship between $L_{radio}$ and $L_{bol}$ for low-luminosity molecular outflows and determined that shock-ionization associated with the molecular outflows and jets was sufficient to produce the observed levels of continuum emission in their sample. Later works (Anglada et al. 2018; Sanna et al. 2018; Purser et al. 2016;

Rosero et al. 2016) have extended this examination into the high-mass (high-bolometric luminosity) regime. Purser et al. (2016) and Rosero et al. (2016) found the $L_{radio}$ versus $L_{bol}$ relationship derived for low-mass jets to be valid for a majority of their high-mass sources as well, indicating that the centimeter continuum emission observed in their sample predominantly arises from protostellar jets, and not HC or UCH II regions. Sanna et al. (2018) conduct a similar analysis - with similar findings - for the POETS sample, but they additionally examine the relationship between $L_{H_2O}$ and $L_{radio}$. Their goal was to test the shock-ionization hypothesis specifically, arguing that if 22 GHz $H_2O$ masers - assumed to be collisionally excited - and the radio-continuum emission both arise due to shocks in protostellar jets then the radio-continuum and $H_2O$-maser luminosities ought to be correlated as well. Sanna et al. (2018) find a power-law correlation between $L_{H_2O}$ and $L_{radio}$ for their sample, with a similar slope to the $L_{radio}$ versus $L_{bol}$ relationship but with greater scatter.

In the following subsections, we describe the relationships between $L_{radio}$, $L_{bol}$, and $L_{H_2O}$ for our sample, and compare these to the observationally-derived correlations published in Purser et al. (2016), Rosero et al. (2016), and Sanna et al. (2018). These results are presented in the subsections 4.2.2, 4.2.3, and 4.2.4, below. We use the standard definition of radio distance-luminosity as $L_{radio} = S_\nu \times D^2$, where $D$ is the heliocentric distance in kpc, $S_\nu$ is the flux density of each continuum source in mJy, and $\nu$ is the observation frequency. $L_{radio}$ has units of mJy kpc$^2$. $L_{H_2O}$ is determined using the standard formula for isotropic luminosity for 22 GHz $H_2O$ masers:

$$\left[\frac{L_{H_2O}}{L_\odot}\right] = 2.30 \times 10^{-8} \left[\frac{\int S_\nu dV}{\mathrm{Jy\ km\,s^{-1}}}\right] \left[\frac{D}{\mathrm{kpc}}\right]^2 \qquad (3)$$

where $\int S_\nu dV$ is the total maser flux density in Jy km s$^{-1}$ for a maser group or set of maser groups integrated over all channels, and $D$ is in kpc (see, e.g. Anglada et al. 1996; Cyganowski et al. 2013). $L_{bol}$ for this analysis comes from Towner et al. (2019), who use multiwavelength photometry and Spectral Energy Distribution (SED) modeling to derive four $L_{bol}$ values for each source using three radiative-transfer models and a one-component greybody fit; the $L_{bol}$ we use here is the median of these four values for each source.

In order to better compare our results to the literature (e.g., Purser et al. 2016; Sanna et al. 2018), we interpolate 8 GHz flux densities for each source using its spectral index and 6 GHz flux density. For sources detected only at one wavelength, we use the 4$\sigma$ upper limit (with an uncertainty of $\pm$ 1$\sigma$) as the "flux" for the interpolation, and treat the interpolated 8 GHz flux density as an upper limit. In the plots in § 4.2.3 and 4.2.4 below, we specifically compare our 8 GHz results to those of the Protostellar Outflows at the EarliesT Stages (POETS) survey (Moscadelli et al. 2020; Sanna et al. 2018). Typical $L_{bol}$ for the POETS sample is $10^3 - 10^4$ $L_\odot$, which is comparable to those in our EGO sample; see § 1 for additional details about the POETS sample selection criteria.

#### 4.2.1. *A Note on Relevant Physical Scales for $L_{bol}$*

Comparing $L_{bol}$ to either $L_{radio}$ or $L_{H_2O}$ implicitly assumes that the flux densities contributing to $L_{bol}$ arise from either similar physical scales or the same individual sources (e.g. protostars) as the other two luminosities. However, previous work has shown the assumption that $L_{bol}$ for each EGO is dominated by a single protostellar source is likely erroneous for our sample (Towner et al. 2019), in part because the $L_{bol}$ were derived using far-infrared data which had projected spatial resolutions $\geq$ 6,500 – 27,000 au (5.″8 angular resolution of the HiGAL PACS 70 μm images at the distances listed in Table 1). Millimeter observations of massive protoclusters are increasingly detecting multiple protostars within $\leq$10,000 au of each other (2.″1 to 8.″8 at the distances to our sample), including in EGOs (see Beuther et al. 2018; Cyganowski et al. 2017; Hunter et al. 2017, for examples of massive star-forming regions with these small protostellar separations). As discussed in § 3, nearly all EGOs in this sample show multiple centimeter-continuum sources within the bounds of the extended 4.5 μm emission, which supports the possibility of significant contributions to EGO $L_{bol}$ from multiple separate protostars. Therefore, a direct comparison of $L_{bol}$ with emission from very small physical areas (e.g. $L_{radio}$ for an individual centimeter continuum source) may not be valid in all cases.

In order to both account for and explore the possibility of multiple individual sources contributing significantly to $L_{bol}$, we have chosen to derive *three* values for $L_{radio}$ and $L_{H_2O}$ at each wavelength. These values differ based on the apertures over which we sum our flux densities to derive $L_{radio}$ and $L_{H_2O}$. First, we determine $L_{radio}$ ($L_{H_2O}$) by summing the flux density of all continuum sources (maser groups) within the ATLASGAL 870 μm 5$\sigma$ contour used by Towner et al. (2019). This is the largest region, and represents the physical area corresponding to the complete star-forming clump. Second, we determine $L_{radio}$ ($L_{H_2O}$) by summing the continuum (maser group) flux densities within the SOFIA 37 μm 5$\sigma$ contour[8], as the shape of each SED (and thus the total $L_{bol}$) in Towner et al. (2019) was most strongly dependent on the flux density at 37.1 μm for nearly all EGOs. This is an intermediate physical scale, corresponding to emission from hot dust and ionized gas in outflows in each EGO. Third, we determine $L_{radio}$ and $L_{H_2O}$ on the scale of individual protostars. This is the smallest region. In the case of $L_{radio}$ versus $L_{bol}$, we use $L_{radio}$ for each centimeter continuum source which is associated with a 6.7 GHz maser; this approach implicitly assumes that $L_{bol}$ for each EGO is dominated by the

[8] For this analysis, we only consider the 37 μm 'a' sources - the brightest 37 μm sources within each EGO - as those are the 37 μm flux densities on which the $L_{bol}$ are based.

6.7 GHz maser-hosting source, which is assumed to be massive (Minier et al. 2003). In the case of $L_{radio}$ versus $L_{H_2O}$, we examine $L_{radio}$ and $L_{H_2O}$ for all centimeter continuum sources which have $H_2O$ masers within 1000 au, based on the findings of maser-host separations in Moscadelli et al. (2020) (see § 4.2.3 for a more detailed discussion).

Using these three spatially-selected luminosity values at each frequency, we explore both the luminosity correlations themselves and whether and how these correlations change with frequency and spatial scale.

#### 4.2.2. $L_{radio}$ vs $L_{bol}$

Figure 5 shows $L_{radio}$ plotted against $L_{bol}$ for our EGO sample. The left, middle, and right columns show plots for the ATLASGAL-, SOFIA-, and $CH_3OH$ maser-based selection criteria, respectively, and the top, middle, and bottom rows correspond to $L_{radio}$ derived from 6 GHz, interpolated 8 GHz, and 22 GHz flux densities, respectively. For each plot, we use the python package `scipy.odr` to perform Orthogonal Distance Regression to fit our data using the equation $\log(L_{radio}) = a\log(L_{bol}) + b$ while accounting for uncertainties in both the independent ($L_{bol}$) and dependent ($L_{radio}$) variables. (This equation is equivalent to $L_{radio} = BL_{bol}{}^{a}$, where $B = 10^{b}$; both forms are used in the literature.) We also use the python package `stats.kendalltau` to derive two- and and three-variable Kendall's Tau ($\tau$) values and statistical-significance ($p$) values for each plot, where the two-variable $\tau$ compares $L_{radio}$ and $L_{bol}$ directly and the three-variable (i.e. partial) $\tau$ compares $L_{bol}$ and radio continuum flux density ($S_\nu$) while also accounting for distance $D$. The ODR and $\tau$ results for each fit are shown in the corresponding panels of Figure 5. The 6 GHz and 8 GHz plots are typically very similar to each other, which is expected as the 8 GHz $L_{radio}$ are derived in part from 6 GHz data and are close in frequency space.

In all cases, we derive a positive correlation between $L_{radio}$ and $L_{bol}$. This relationship statistically significant at the $2\sigma$ level ($p < 0.05$) in three cases: the SOFIA-selected 22 GHz $L_{radio}$, and the $CH_3OH$ maser-selected 6 and 8 GHz $L_{radio}$. Our derived values of $a$ agree within errors with those previously published by at least one other team ($a \sim 0.63$ plus or minus $\lesssim$10% in each of Purser et al. 2016; Rosero et al. 2019b; Sanna et al. 2018) in all three significant cases. This suggests that the dominant source of emission in these cases is the same as assumed by these other teams, i.e., thermal emission from ionized protostellar jets.

Our remaining six plots all have $p \leq 0.18$. While these do not meet the criteria for statistical significance, they are low enough to allow us to investigate trends in $a$ with frequency and aperture size. Examining our variation in $a$ with frequency, we find that the 6 GHz and 8 GHz $L_{radio}$ show no preference for either shallower or steeper correlations than those found by other teams, while our 22 GHz-derived $L_{radio}$ consistently produce lower $a$ than either the 6 or 8 GHz data or the literature values. This is due to larger $L_{radio}$ at lower $L_{bol}$ as compared to the other two frequencies. Additionally, the 22 GHz correlations show little variation with aperture size ($a = 0.58$ in the ATLASGAL aperture versus $a = 0.55$ in the maser-selected plot), unlike the 6 and 8 GHz $L_{radio}$. Examination of Table 4 shows that the 6.7 GHz maser-hosting continuum sources are responsible for the majority of 22 GHz flux density in 5 out of 8 cases (median contribution: 59.25% of total flux density), whereas they are responsible for the majority of 6 GHz flux density in only 2 out of 8 cases (median contribution: 32.85% of total flux density). This suggests that the 22 GHz $L_{radio}$ are both a) largely insensitive to aperture size, i.e. dominated by centrally-concentrated emission in a way that the lower frequencies are not, and b) display an excess of continuum emission at lower $L_{bol}$ as compared to the 6 and 8 GHz $L_{radio}$. Both trends are consistent with the possibility of non-trivial flux contributions at 22 GHz from processes other than thermal free-free emission, such as thermal emission from warm dust. With our current sample size, we cannot definitively confirm this scenario, but do suggest that it bears further investigation.

As for variation of $L_{radio}$ with aperture size, we find that $\tau$ increases and $p$ decreases (i.e. relations become tighter and more significant) for the 6 and 8 GHz $L_{radio}$ as we move from larger to smaller apertures, though $a$ shows no obvious trend with aperture size. The derived $a$ for our SOFIA-selected data ($0.52 < a < 0.62$) are the closest to those derived in the literature by other teams; the 22 GHz correlation is statistically significant with $p = 0.03$, and the 6 and 8 GHz correlations both have $p = 0.07$. The $CH_3OH$ maser-selected correlations are by far the tightest, but the two statistically-significant correlations have derived slopes $a$ which are a factor of $\sim$1.3 larger (steeper) than those derived in the literature for other samples. This is due primarily to a decrease in $L_{radio}$ for the lower bolometric-luminosity sources as we move from the SOFIA-selected to maser-selected apertures, i.e., the 6 GHz flux densities are less centrally-concentrated than the 22 GHz, as previously discussed. Essentially, when we consider only the 6.7 GHz maser-hosting sources, we find less 6 GHz flux density for a given $L_{bol}$ (or, alternately, higher $L_{bol}$ for a given $L_{radio}$) than we would expect for $a = 0.62$ to 0.64.

Based on the results of this work alone, the null hypothesis for a correlation between $L_{radio}$ and $L_{bol}$ can be rejected at the $2\sigma$ level (95% confidence); in combination with other works, there is evidence of a strong correlation between $L_{radio}$ and $L_{bol}$ in massive star-forming regions. Given the derived $a$ for our statistically-significant results, we conclude that the primary driver of $L_{radio}$ in our EGO sample is similar to that of other teams, i.e. thermal free-free emission in ionized protostellar jets. However, the overall trends of our data suggest that a) there may be some contribution to the 22 GHz $L_{radio}$ from non-jet sources, most likely warm dust, and b) there is a possibility that assuming that a single dominant continuum source is responsible for the majority of the clump $L_{bol}$ may miss smaller, but non-trivial, contributions from additional protostellar sources. Given our sample size, we cannot make a definitive statement on either of these points at this time, but we plan to address both possibilities with additional multiwavelength data in the near future.

#### 4.2.3. $L_{radio}$ vs $L_{H_2O}$

In addition to $L_{radio}$ versus $L_{bol}$, we also examine the relationship between $L_{radio}$ and $L_{H_2O}$ following the method of Sanna et al. (2018). Sanna et al. (2018) found generally good agreement between the $L_{radio}$ versus $L_{H_2O}$ relation for their sample of $H_2O$ maser-hosting MYSO continuum sources compared to published relations for low-mass sources (Anglada et al. 2018). Based on this result, Sanna et al. (2018) conclude that the mechanism of ionization in jets from high-mass YSOs is the same as that in low-mass YSOs, i.e., shock-ionization rather than photoionization, though they do note that it is possible $H_2O$ masers are simply a preferred signpost of ionized jets.

Figure 6 shows $L_{radio}$ versus $L_{H_2O}$ for our sample. The row and column properties follow the same system as in Figure 5, except for the right-hand column, in which we show $L_{radio}$ versus $L_{H_2O}$ for all centimeter continuum sources with $H_2O$ maser emission within 1000 au. The choice of 1000 au as the distance cutoff comes from Moscadelli et al. (2020), who find that ~84% of $H_2O$ masers in the POETS sample which can be positively kinematically associated with a specific centimeter source lie within 1000 au of that source[9]. We do not require that a continuum source be detected at both 1.3 cm and 5 cm in order to include it in this analysis; for sources detected at only one wavelength, $L_{radio}$ for the non-detected wavelength is plotted in red as an upper limit based on the $4\sigma$ flux density limit in that band.

We do not find any statistically-significant correlation between $L_{radio}$ and $L_{H_2O}$ for the ATLASGAL apertures ($p \sim 0.9$ at all three frequencies). $L_{radio}$ versus $L_{bol}$ within the SOFIA-based apertures do just meet our criteria for significance with $p = 0.05$ at all three frequencies, but the $\tau$ and $\tau_p$ values for these data are not consistent with a significant correlation: $\tau \sim 0.5$ for all frequencies but $\tau_p \lesssim -0.4$ in all cases. We also find no correlation between $L_{radio}$ and $L_{H_2O}$ when we select only $H_2O$-maser emission within 1000 au of a continuum source; $p = 0.11$ at 22 GHz and $p = 0.41$ at 6 and 8 GHz, again with a sign change between $\tau$ and $\tau_p$.

In order to test whether our lack of robust, statistically-significant correlations is due to our small sample size (15 data points at most, when including both epochs of observation for the $H_2O$ masers), we additionally compare our data to the $L_{radio}$-$L_{H_2O}$ relation derived by Sanna et al. (2018). The relationship they derive is $L_{radio} = 10^{(3.8\pm0.4)} L_{H_2O}{}^{(0.74\pm0.07)}$. For our values of $L_{H_2O}$, we calculate the predicted $L_{radio}$ according to this equation, and determine the standard deviation of our measured $L_{radio}$ versus $L_{radio\,predicted}$. Sanna et al. (2018) have a $1\sigma$ dispersion about the best-fit line of 0.47 (A. Sanna, private communication). In all cases, the standard deviation of our data from the Sanna et al. (2018) best-fit line ranges from 0.71 to 0.92, i.e. ~1.5 to 2 × the standard deviation of the Sanna et al. (2018) sample. This significant difference in dispersion suggests that the lack of correlation between $L_{radio}$ and $L_{H_2O}$ in our data is not merely due to small sample size, but to a more fundamental difference between the two populations.

It is possible that, as Sanna et al. (2018) suggest, $H_2O$ masers are simply a preferred signpost of massive YSO jets, and the correlation they find in their data is not due to a common excitation mechanism for continuum and maser emission. Alternately, it is possible that the selection criteria for their sample and ours were sufficiently different that the two data sets are not comparable with respect to the maser emission; our sources were selected on the basis of extended 4.5 μm emission with no particular preference for $H_2O$ maser activity, whereas the POETS sample was selected partially on the basis of prolific $H_2O$ maser emission. Given our small sample size, we cannot definitively state which of these two possibilities is correct, though we consider the latter to be more likely.

#### 4.2.4. $L_{H_2O}$ vs $L_{bol}$

In Figure 7 we show $L_{H_2O}$ versus $L_{bol}$ for our sample, where $L_{H_2O}$ is the total $H_2O$-maser luminosity within the ATLASGAL aperture for each EGO. We compare our results to those of Urquhart et al. (2011), who examine a sample of ~600 massive YSOs ($L_{bol} > 10^3 L_\odot$) drawn from the Red MSX Source (RMS) catalog (Urquhart et al. 2011, see their Figure 16). For our data, we derive a relation between $L_{H_2O}$ and $L_{bol}$ of the form $L_{H_2O} = b(L_{bol})^a$, where $b = 1.3 \pm 10.6 \times 10^{-14}$ and $a = 2.46 \pm 1.00$. Our $a$ value agrees within errors with Urquhart et al. (2011), who find $a = 1.47 \pm 0.76$ for the RMS data.

Our $b$ value differs from that derived by Urquhart et al. (2011) by approximately an order of magnitude, likely as a result of the difference in $a$ values.

We note that the majority of our sources lie above the best-fit line for $L_{H_2O}$ versus $L_{bol}$ for the RMS sample. It is possible that this trend is simply a consequence of small-number statistics with regards to our sample. However, given the initial selection criterion for EGOs and its current interpretation (i.e. extended emission at 4.5 μm due to active accretion and ejection processes; see § 1.1), it may also be the case that the typical protostars in our EGOs are in a more active state - and thus producing more or stronger $H_2O$ maser emission - than the typical RMS sources in Urquhart et al. (2011). Overall these results suggest that, as with the Urquhart et al. (2011) sample, there is a strong link between the driving source of the $H_2O$ maser emission in the clump and the dominant source of $L_{bol}$. Whether the driving source is a single massive protostar or the combination of $H_2O$ maser and bolometric flux from multiple protostellar sources cannot be determined at the scales considered for this relationship ($\gtrsim$0.5 pc).

[9] Moscadelli et al. (2020) examined $H_2O$ maser sources out to a distance of ~18,000 au from their continuum targets and conducted a multi-epoch VLBI proper-motion analysis.

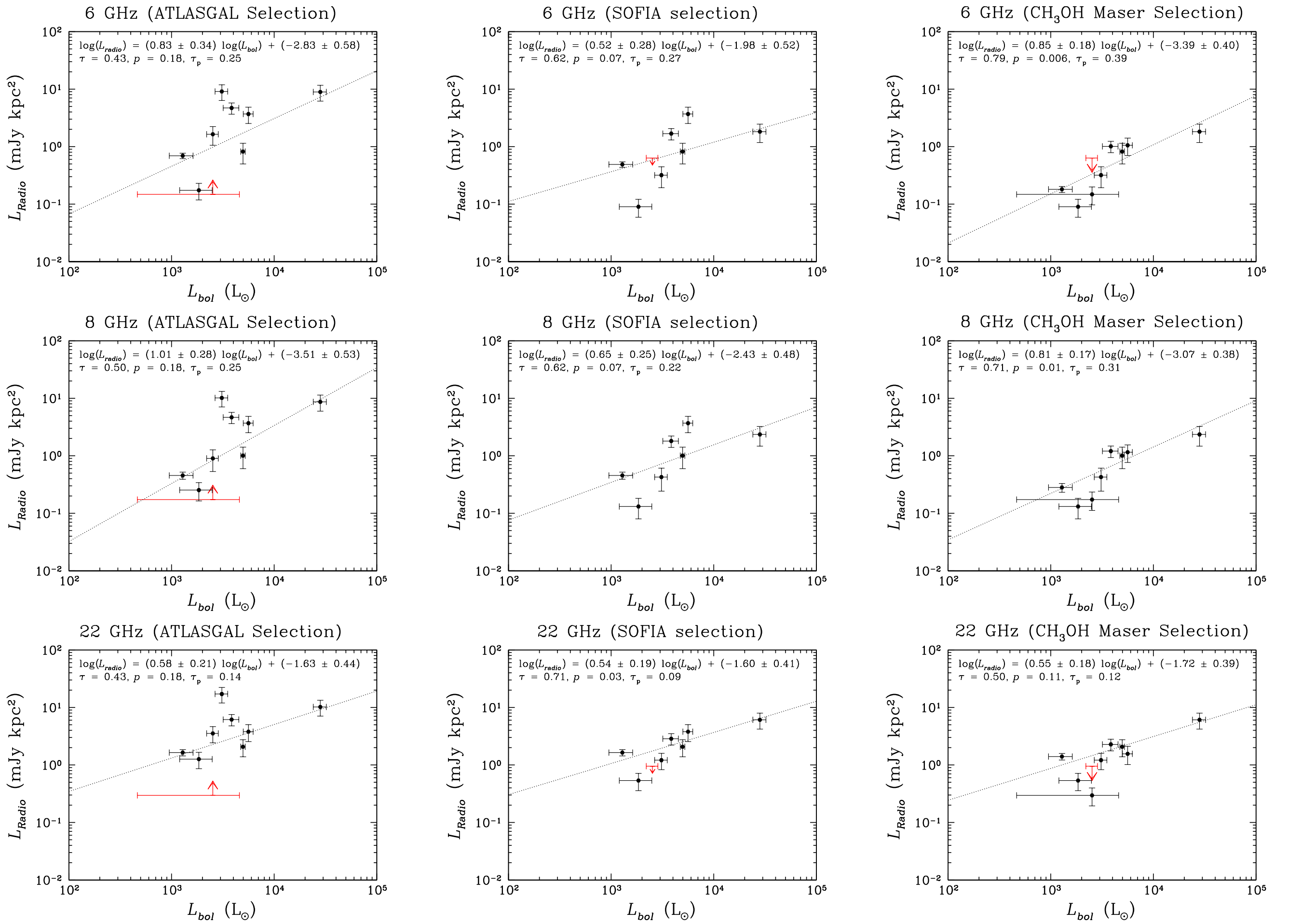

**Figure 5**. $L_{radio}$ versus $L_{bol}$ for the 9 EGO targets, where $L_{radio}$ is the radio distance-luminosity ($S_\nu D^2$), and $L_{bol}$ is the median of the four luminosities for each EGO derived by Towner et al. (2019). $L_{radio}$ is derived from the sum of all continuum emission within each aperture at each wavelength; a source need not be detected at 1.3 cm to be included in $L_{5\ cm}$, etc. Non-detections/upper limits in Table 4 are not included in the flux density totals. The left-hand column shows $L_{radio}$ vs $L_{bol}$ where $L_{radio}$ is derived from all radio continuum emission within the ATLASGAL aperture; the middle column shows the same where $L_{radio}$ is determined from all continuum flux within the SOFIA aperture, and in the right-hand column, $L_{radio}$ is for those continuum sources which have associated 6.7 GHz masers. In the left-hand column, G10.29–0.13 is shown as a lower limit in the left-hand column as the ATLASGAL aperture contains some partially resolved-out continuum emission which may be reflected in $L_{bol}$ but will not be reflected in $L_{radio}$. G18.89–0.47 is shown as an upper limit in the middle and right-hand columns as it neither has continuum emission within the SOFIA aperture nor associated with 6.7 GHz $CH_3OH$ maser emission.

### 4.3. *Characteristics of the Maser Emission*

In the following subsections we briefly analyze the characteristics of the maser emission in our sample. As deep continuum imaging was the primary aim of these observations, our maser data lack the sensitivity and velocity resolution necessary to perform detailed, small physical-scale kinematic analyses, such as would be possible with VLBI. Instead, we here limit ourselves to more general descriptions of the maser data, and make note of which sources may be good candidates for high-resolution follow up observations of the maser line emission.

#### 4.3.1. *Peak Velocity Offsets & Velocity Extents*

Here we present basic statistics about the maser emission in our sample. Statistics are calculated for individual maser groups and presented for the groups as a whole (i.e. the analysis is not per-channel). "Velocity extent" refers to the total velocity range over which maser emission is detected in a given maser group ($V_{max}$ - $V_{min}$). Note that this does not indicate that there is *continuous* maser emission within that range; there may instead be multiple distinct velocity components within a given maser group. "Peak velocity offset" refers to the difference between the peak velocity of each maser group and the source $V_{LSR}$ (see Table 1).

The $H_2O$ maser groups in our sample have a median and mean velocity extent of 7.38 and 12.51 km s$^{-1}$, respectively,

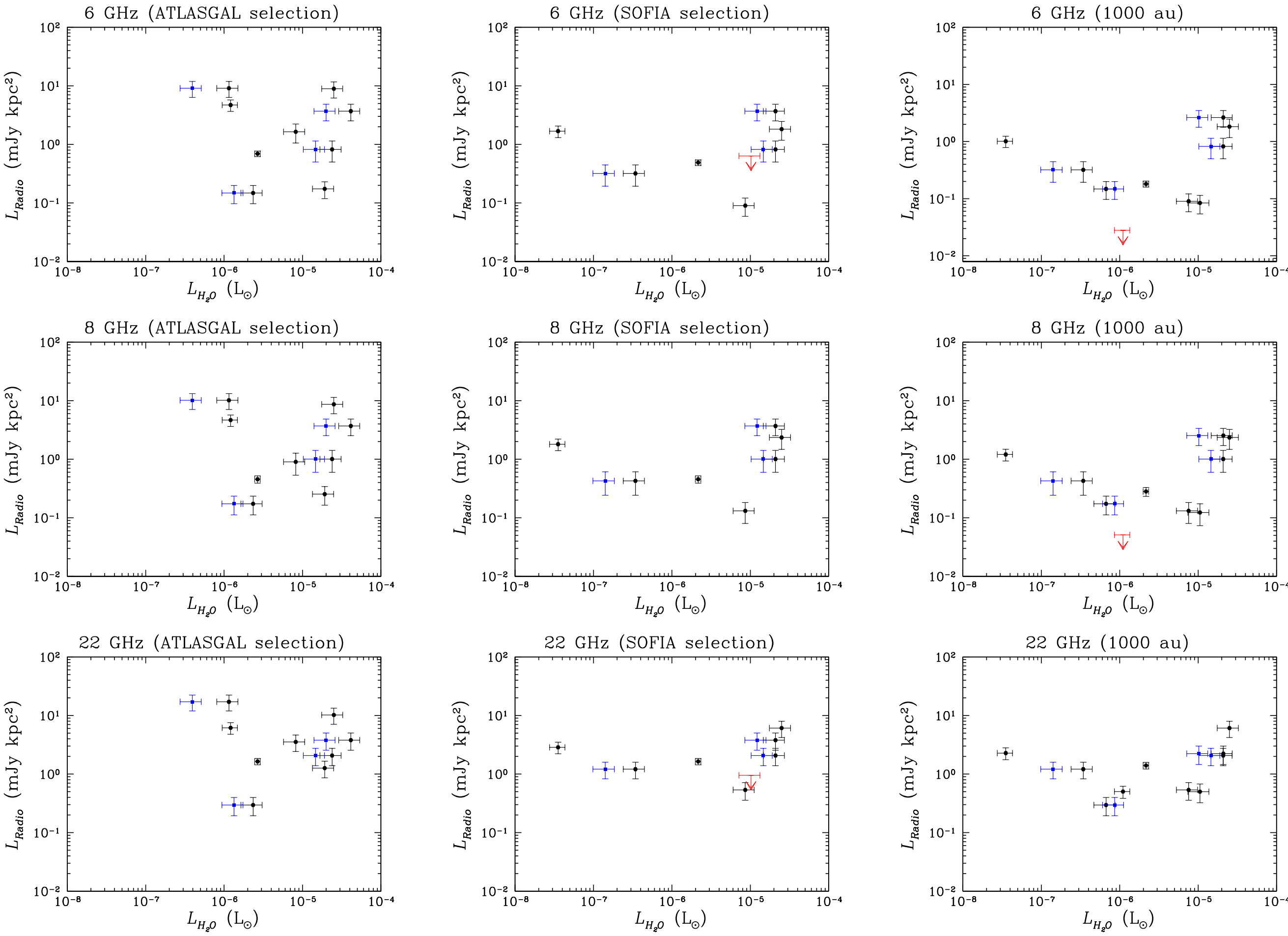

**Figure 6**. $L_{radio}$ versus $L_{H_2O}$ for the EGO-9 targets, where $L_{radio}$ is the radio distance-luminosity ($S_\nu D^2$) and $L_{H_2O}$ is the isotropic $H_2O$ maser luminosity in $L_\odot$ (see Eq. 3). The row and column scheme is the same is in Figure 5. $L_{H_2O}$ is determined by summing the flux density in all maser spots within a given region. For those EGOs observed twice at K-band, the first epoch of observations is shown in black and the second in blue. G14.33–0.64 CM3, which was detected only at 1.3 cm, is shown as an upper limit (in red) in the 6 and 8 GHz panels in the right-hand column.

with a minimum of 0.25 km $s^{-1}$ (a 2-channel detection) and a maximum of 101.25 km $s^{-1}$. The median and mean peak velocity offsets are −1.75 and −4.51 km $s^{-1}$, respectively, with minimum and maximum values of −43.50 km $s^{-1}$ and 29.25 km $s^{-1}$. In other words, the $H_2O$ maser emission is typically centered on the $V_{LSR}$ for each source but spans a very broad range in velocity, with a slight preference for blueshifted emission. The $NH_3$ maser emission, in contrast, is also well-centered on the $V_{LSR}$ of each source but spans a very narrow range in velocity compared to the $H_2O$ emission. The $NH_3$ maser groups have median and mean velocity extents of 0.88 and 1.21 km $s^{-1}$, respectively, with minimum and maximum values of 0.5 and 3.0 km $s^{-1}$. They have mean and median peak velocity offsets of only −0.38 and −0.34 km $s^{-1}$, respectively, with minimum and maximum values of only −2.00 and 2.50 km $s^{-1}$. The 6.7 GHz $CH_3OH$ masers span a median and mean velocity range of 11.0 and 9.1 km $s^{-1}$, respectively - comparable to a typical $H_2O$ maser group - with minimum and maximum values of 1.25 and 19.5 km $s^{-1}$, respectively. However, the $CH_3OH$ peak velocity offsets are more similar to the $NH_3$ emission, with median and mean offsets of −0.5 and 0.27 km $s^{-1}$, respectively; minimum and maximum values are −10.5 and 7.5 km $s^{-1}$. Overall, the $NH_3$ and $CH_3OH$ masers are consistent with gas which has a relatively low peculiar velocity compared to the EGO $V_{LSR}$, while $H_2O$ masers are consistent with both co-moving gas and gas which is moving quite rapidly compared to the systemic velocity. This latter point is consistent with $H_2O$ maser emission which can arise from a wide range of sources, but definitely includes sources of emission which involve high relative velocities, such as outflows and jets.

#### 4.3.2. *Morphology of the 6.7 GHz $CH_3OH$ Maser Emission as a Probe of Disks or Other Structures*

While 6.7 GHz $CH_3OH$ masers are known to be exclusively associated with MYSOs (Minier et al. 2003), there is some debate as to which specific physical structures host these masers: accretion disks, infalling envelopes, base of a jet/outflow, expanding shocks, etc. (see De Buizer &

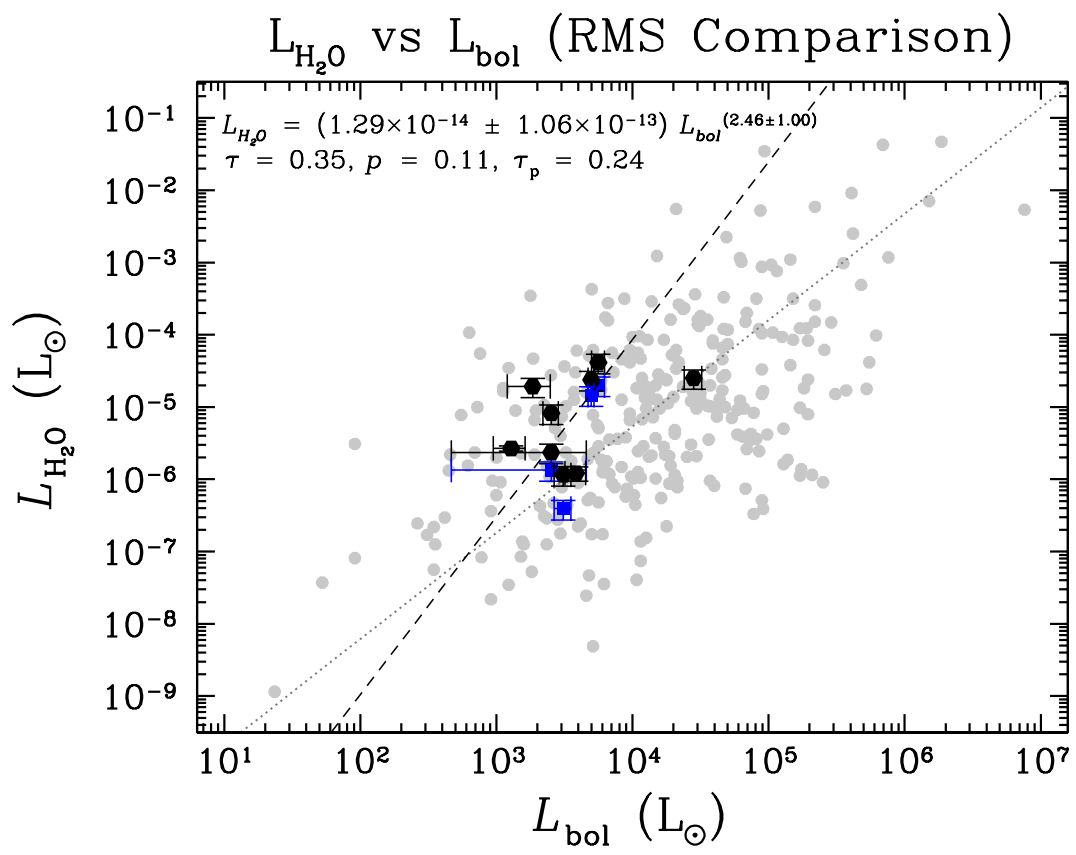

**Figure 7**. $L_{H_2O}$ versus $L_{bol}$ for our sample (black and blue points) compared to the sources in the Red MSX Source (RMS) survey (gray points; see Urquhart et al. 2011, Figure 16). The dashed line shows the best-fit relation for our data: $L_{H_2O} = (1.29 \times 10^{-14} \pm 1.06 \times 10^{-13}) L_{bol}{}^{(2.46\pm1.00)}$. The dotted line shows the best-fit relation for $L_{H_2O}$ versus $L_{bol}$ for the RMS-selected sources: $L_{H_2O} = 7.1\times10^{-12} \times L_{Bol}^{1.47}$. For our sources, $L_{H_2O}$ is the luminosity of all $H_2O$ maser emission within the ATLASGAL 870 µm contour used in Towner et al. (2019) (typically $5\sigma_{870\mu m}$).

Minier 2005; Sanna et al. 2010a,b; Moscadelli et al. 2011; Goddi et al. 2011). The morphology of 6.7 GHz $CH_3OH$ maser spots in observations to date is quite diverse, boasting linear, curved, ring-like, complex, and paired structures. Bartkiewicz et al. (2009) and Bartkiewicz et al. (2016) present a systematic examination of VLBI-observed 6.7 GHz maser morphologies for a sample of 63 MYSO targets, and a detailed discussion of the possible physical origins of the observed 6.7 GHz maser structures. Below, we briefly examine the morphology of the 6.7 GHz maser spots in our data, as well as projected maximum size scales.

Overall, we find that our 6.7 GHz maser data display a similar diversity of morphology and kinematic structure as those described by Bartkiewicz et al. (2009, 2016) and elsewhere in the literature. Some maser groups display relatively simple kinematic structure (e.g. a single velocity peak) and little to no position shift between channels (G14.63–0.58, all three 6.7 GHz sources in G18.89–0.47), while others display linear or possibly arched or curved structures (G10.29–0.13, G10.34–0.14, G12.91–0.03, G14.33–0.64, G19.36–0.03), and still others display much more complex kinematic behavior and morphologies (G22.04–0.22, G28.83–0.25).

In order to examine the significance of these morphologies - especially the more linear morphologies which some teams have concluded trace disk rotation in other sources (Sanna et al. 2010a) - we calculate the maximum (projected) physical size of the maser-hosting region for each 6.7 GHz maser group. First, we derive $\Delta\theta$ for the fitted position of each maser spot within a maser group using Eq. 2. We then determine the (angular) distance between a single maser spot and all other spots in that maser group, and evaluate whether these positions are identical within $\Delta\theta$. If the maximum position-difference within a maser group is greater than the sum of the position uncertainties of the two individual maser spots in question, we calculate the (projected) physical separation of the two maser spots in au using the angular position-difference and the distance to the EGO. We also calculate the difference in velocity between the two maser spots. These projected maximum separations, which we take as a proxy for the size of the maser-hosting structure, range from 64 ± 58 to 2932 ± 23 au, with a median value of 370 au. The $\Delta V$ over which these projected distances were measured range from 0.5 to 9.25 km s$^{-1}$, with median $\Delta V$ = 4 km s$^{-1}$.

The median (projected) size of our maser groups (370 au) is small enough to be consistent with the measured sizes of MYSO accretion disks; Ilee et al. (2018) examine an accretion disk around the massive protostar G11.92–0.61 MM1 and determine a size of $R \sim 800$ au. However, it is also possible that this could merely be a projection effect. Additionally, the maximum projected size (2932 ± 23 au in G28.83–0.25) is far more consistent with the size scales of protostellar envelopes or outflows than disks. Finally, some of the smallest projected sizes (e.g. G14.63–0.58 CM1-M1) do not display strong velocity gradients with position, and have both kinematic structures and apparent morphologies more consistent with the "simple" classification used by Bartkiewicz et al. (2009) than with Keplerian disks or other ordered rotation.

Bartkiewicz et al. (2016) note that, even with their high-spatial and -velocity resolution VLBI data, it is difficult to relate maser morphology and kinematic structure to a specific physical structure/phenomenon in an individual source, and the diversity of structure indicates that there is not one single physical phenomenon responsible for 6.7 GHz $CH_3OH$ maser emission around MYSOs (Bartkiewicz et al. 2009). Given the generally good $S/N$ of our 6.7 GHz detections, and the typical (projected) physical sizes derived in this analysis, we identify these masers as excellent candidates for future high angular- and velocity-resolution follow-up observations for those teams wishing to address the question of which physical structures host 6.7 GHz $CH_3OH$ maser emission in MYSOs.

### 4.4. *Maser Variation with Time*

In this subsection, we examine the variation in our maser emission with time. All nine of our EGO targets were previously observed by Cyganowski et al. (2013), who used the 45-m Nobeyama telescope to observe 22 GHz $H_2O$ and $NH_3$ (1,1), (2,2), and (3,3) emission in all 94 EGOs visible from the northern hemisphere. Additionally, six of our nine targets were included in the $CH_3OH$ maser survey of Cyganowski et al. (2009), who used the VLA to observe 6.7 GHz Class II and 44 GHz Class I $CH_3OH$ masers toward 20 EGOs with $\sim3''$ resolution. We compare our epoch 2018.1 and 2019.6 data to both of these literature data sets and examine what variations, if any, exist between our spectra and those previously published by Cyganowski et al. We also examine the variation in 22 GHz $H_2O$ and $NH_3$ (3,3) maser emission between epochs 2018.1 and 2019.6 in our data for the four sources with two epochs of observations (G10.29–0.13, G12.91–0.03, G19.36–0.03, G22.04+0.22).

#### 4.4.1. *6.7 GHz $CH_3OH$ Masers - 10.5-year Epoch*

This work has six EGO targets in common with Cyganowski et al. (2009): G10.29–0.13, G10.34–0.14, G18.89–0.47, G19.36–0.03, G22.04+0.22, and G28.83–0.25. In order to account for the difference in both angular and velocity resolution between the two data sets, we have regridded the Cyganowski et al. (2009) data to match the coarser velocity resolution of our current data (0.25 km $s^{-1}$), and smoothed our current data to match the parameters of the lower-resolution Cyganowski et al. (2009) cubes (see Table 1 in Cyganowski et al. (2009) for the relevant synthesized beam parameters for each source, and their § 2.1 for further observational details). We then compared the spectra for each "maser group" reported in Cyganowski et al. (2009) Table 9, limiting our analysis to the velocities originally examined by Cyganowski et al. (2009), which can be found in their Table 1. These results are shown in Figure 8. In most cases the maser groups directly comparable (similar or identical locations and velocities) across the two data sets. We do note that we detect three 6.7 GHz $CH_3OH$ maser groups in the EGO G18.89–0.47 while Cyganowski et al. (2009) report only one. This is likely due in part to angular resolution differences ($\sim 3''$ in their data versus $0\farcs3$ in ours), which would blend emission from our maser groups NC-M2 and NC-M3. However, we also note that Cyganowski et al. (2009) only search from maser emission in the velocity range 44.4 to 71.8 km $s^{-1}$ for this source, and two of our three $CH_3OH$ maser groups have no emission in this range ($V_{min}$ = 73.25 and 72.00 for G18.89–0.47 NC-M1 and NC-M2, respectively). Therefore, while it is possible that the number of 6.7 GHz $CH_3OH$ maser groups increased between 2008 and 2018, we proceed with the analysis below on the assumption that the difference in number of maser groups is due to the differing velocity ranges between their analysis and ours.

In our sample, the overall velocity structure of each spectrum - the velocities at which maser peaks are present or absent - is largely unchanged between the two epochs in all cases. The only exception is G18.89–0.47, in which the Cyganowski et al. (2009) line profile is somewhat confused, and it is unclear whether a new velocity component has appeared in the 2018 data or an existing velocity component has increased in brightness while the others dimmed.

However, while the velocity structures are relatively unchanged, all six sources do show amplitude variations to one degree or another. In at least three sources (G10.29–0.13, G19.36–0.03, G22.04+0.22) an existing velocity component has increased in intensity by $\geq 2\times$, and in at least three more sources (G10.29–0.13, G10.34–0.14, G18.89–0.47), one or more velocity components have decreased in intensity in the 2018 observations. Some sources (notably G10.34–0.14) exhibit both characteristics simultaneously (dimming of some components and brightening of others). In no case do we observe a global increase or decrease in intensity for all velocity components in a given maser group, such as would be consistent with the more spectacular accretion-flare events recently reported in the literature (e.g. S255IR-NIRS3, NGC6334I, G358.93-0.03; see Caratti o Garatti et al. 2017; Hunter et al. 2017; Brogan et al. 2019, respectively).

Overall, the behaviors observed in this limited subsample are consistent with what has been previously reported in long-term monitoring studies of 6.7 GHz $CH_3OH$ masers. Goedhart et al. (2004) monitored 54 6.7 GHz $CH_3OH$ masers from January 1999 to March 2003 using the Hartesbeethoek Radio Astronomy Observatory (HartRAO) 26-meter telescope at biweekly intervals or shorter, and find that their sample displays nearly the full range of possible variability behaviors: from non-varying masers to monotonic increases/decreases in brightness, aperiodic variations, and periodic variations. Szymczak et al. (2018) find a similar incidence rate of periodic/aperiodic/non-varying behaviors in their sample of 166 6.7 GHz $CH_3OH$ masers in the Torun Methanol Source Catalog (Szymczak et al. 2012), except that they observe a higher rate of monotonically decreasing sources than Goedhart et al. (2004). Both teams find that numerous sources display a range of behaviors over time or a combination of behaviors concurrently (e.g. aperiodic short-term variations as well as overall monotonic increase/decrease), with little to no velocity drift ($\Delta V \lesssim 0.25$ km $s^{-1}$) of individual maser peaks within the roughly 4-year periods of their monitoring programs.

In the case of comparatively low-level variations, especially aperiodic ones, both Goedhart et al. (2004) and Szymczak et al. (2018) suggest that small changes in either the masing column or the line-of-sight between maser and observer are responsible for the amplitude variations, rather than significant changes in the pumping mechanism itself (i.e. the infrared radiation produced by the protostar). Both teams cite localized turbulence as the most likely culprit for such changes in these cases. With only two time points for comparison, it is not possible to assess whether the amplitude variations in our spectra are periodic or aperiodic, or whether different velocity components are flaring in sequence. However, both the diversity of amplitude variations in our sample and the apparent lack of velocity drift in our spectral components are well in line with the results of both Goedhart et al. (2004) and Szymczak et al. (2018). Given these similarities and the lack of evidence of significant flare events in these sources, we suggest that small-scale localized turbulence, resulting in relatively small-scale changes to the properties of the masing column itself which are not correlated across the different velocity features, is the most likely explanation for the amplitude variations in our data. However, observations from at least one additional epoch would be needed in order to evaluate the periodicity/aperiodicity of these variations in order to confirm or reject this hypothesis.

#### 4.4.2. *22 GHz $H_2O$ Masers - 10.5-year epoch*

22 GHz $H_2O$ masers have long been known to be highly variable (Breen et al. 2010; Felli et al. 2007; Trinidad et al. 2003; Valdettaro et al. 2002; Furuya et al. 2001). Long-term monitoring projects have shown that variations occur on timescales ranging from weeks to years, and include fluctuations in flux density up to two orders of magnitude (Valdettaro et al. 2002) or, in some cases, the complete appearance/disappearance of individual velocity components altogether (Trinidad et al. 2003; Hunter et al. 1994). In this subsection and the following subsection, we compare the multiple epochs of 22 GHz $H_2O$-maser data for our sample, and place our results in the broader context of the $H_2O$ maser-variation literature.

All 9 EGOs examined in this paper were observed by Cyganowski et al. (2013) in 2008-2010 using the 45-m Nobeyama telescope near Minamimaki, Nagano, Japan. The observations have $\sim$73$''$ angular resolution, 0.5 km s$^{-1}$ velocity resolution, and $\sim$500 km s$^{-1}$ bandwidth. Cyganowski et al. (2013) define a water maser detection as $>4\sigma$ emission in at least two adjacent channels. With these criteria, Cyganowski et al. (2013) detect $H_2O$ maser emission in 7 of 9 sources. The two sources with nondetections were G19.36–0.03 and G28.83–0.25, with 4$\sigma$ upper limits of 0.64 and 0.68 Jy, respectively. We detect 22 GHz $H_2O$ masers in G19.36–0.03 with a peak integrated flux density of 0.71 $\pm$ 0.02 Jy in epoch 2018.1 and 0.27 $\pm$ 0.02 Jy in epoch 2019.6, and in G28.83–0.25 with a peak integrated flux density of 2.47 $\pm$ 0.01 Jy. Although our $S_{Peak}$ values are greater than the Cyganowski et al. (2013) 4$\sigma$ values in two of these three cases, our results are not necessarily inconsistent with theirs given the well-documented variability of 22 GHz $H_2O$ masers in general (Hunter et al. 1994; Trinidad et al. 2003; Valdettaro et al. 2002) and the long time scale between observations.

#### 4.4.3. *22 GHz $H_2O$ Maser Emission - 1.5-year epoch*

In this subsection we examine the $H_2O$ maser emission in the four EGOs for which we have two epochs of observations in our VLA data (2018.1 and 2019.6). Figures 9 through 12 show the 2018.1 and 2019.6 spectra for the maser groups common to both observation epochs in these four sources. Epoch 2018.1 spectra are shown in black, and epoch 2019.6 spectra are shown in red. All plots for a given EGO are shown on the same velocity axis.

Each of these sources exhibits at least some variation in its $H_2O$ maser emission. In G10.29–0.13, no individual maser groups completely appear or disappear between the two observations, but every maser group exhibits a change in the peak velocity or peak intensity of at least one velocity component. One maser group (G10.29–0.13 NC-W3) has dimmed by more than an order of magnitude between the two epochs, and velocity components in several other maser groups have strengthened or weakened by a factor of 2 or more. Still other velocity components show only small ($<0.5\times$) variations in peak intensity. In G12.91–0.13, three new $H_2O$ maser groups appear in the epoch 2019.6 data which were not present in epoch 2018.1, and the velocity structures of the two maser groups common to both epochs exhibit a range of changes on par with those observed in G10.29–0.13: some components disappear completely, some exhibit order-of-magnitude dimming, and some display only minor changes in peak intensity and peak velocity. In G19.36–0.03, two maser groups from the epoch 2018.1 data have disappeared in the 2019.6 data (with comparable sensitivity limits), and in G22.04+0.22, three maser groups are no longer visible in 2019.6 which were present in 2018.1 (but with $\sim$2$\times$ worse sensitivity). Again, for both sources, the maser groups which are visible in both epochs show variation in their overall kinematic structure, and individual velocity components present in both epochs exhibit a range of variations in peak intensity. We do not detect any proper motions (i.e. $\geq$beamsize) for these $H_2O$ maser groups between the first and second epochs.

As we have only two epochs of observation, we cannot test for periodicity in the maser intensities or peak velocities. However, we do find that the variations we observe in these data overall are well in line with those previously observed and published by other teams. Trinidad et al. (2003) report four types of variation for their sample of 11 sources monitored for 8 months with the Haystack 37-m telescope: the strongest component appears or disappears, the weakest component appears or disappears, one or more components changes its peak flux density by more than a factor of two, or the maser components show small variations. Hunter et al. (1994) likewise find that some velocity components in their sample of $H_2O$ masers in the UC H II region W75N appear or disappear completely over the course of the monitoring period. All four of these behaviors are present in our sample; in fact, in most cases, they are all present in a single EGO. The high angular resolution of our data suggest that this variability arises on extremely small scales ($\lesssim$500 au); Trinidad et al. (2003) suggest that the $H_2O$ maser variability they observe can be explained by small Gaussian fluctuations in the line opacity.

#### 4.4.4. *$NH_3$ (3,3) Maser Emission - 1.5-year Epoch*

G19.36–0.03 is the only source with two epochs of observation which has any detectable $NH_3$ (3,3) maser emission[10], and we find that this emission is effectively unchanged be-tween the two epochs. Each maser group has decreased in flux density between the first and second epochs, but these decreases are slight: 33 mJy, 20 mJy, and 30 mJy (35%, 10%, and 9%) for maser groups NC-A1, NC-A2, and NC-A3, respectively. At the same time, our image noise has increased 63% between the two observations, from 5.4 to 8.8 mJy beam$^{-1}$. A change of 30 mJy is consistent with only a $\sim$3.4$\sigma$ change in the epoch 2018.1 data.

[10] The other three EGOs observed at both epochs - G10.29–0.13, G12.91–0.03, G22.04+0.22 - have no detectable $NH_3$ (3,3) emission at either epoch.

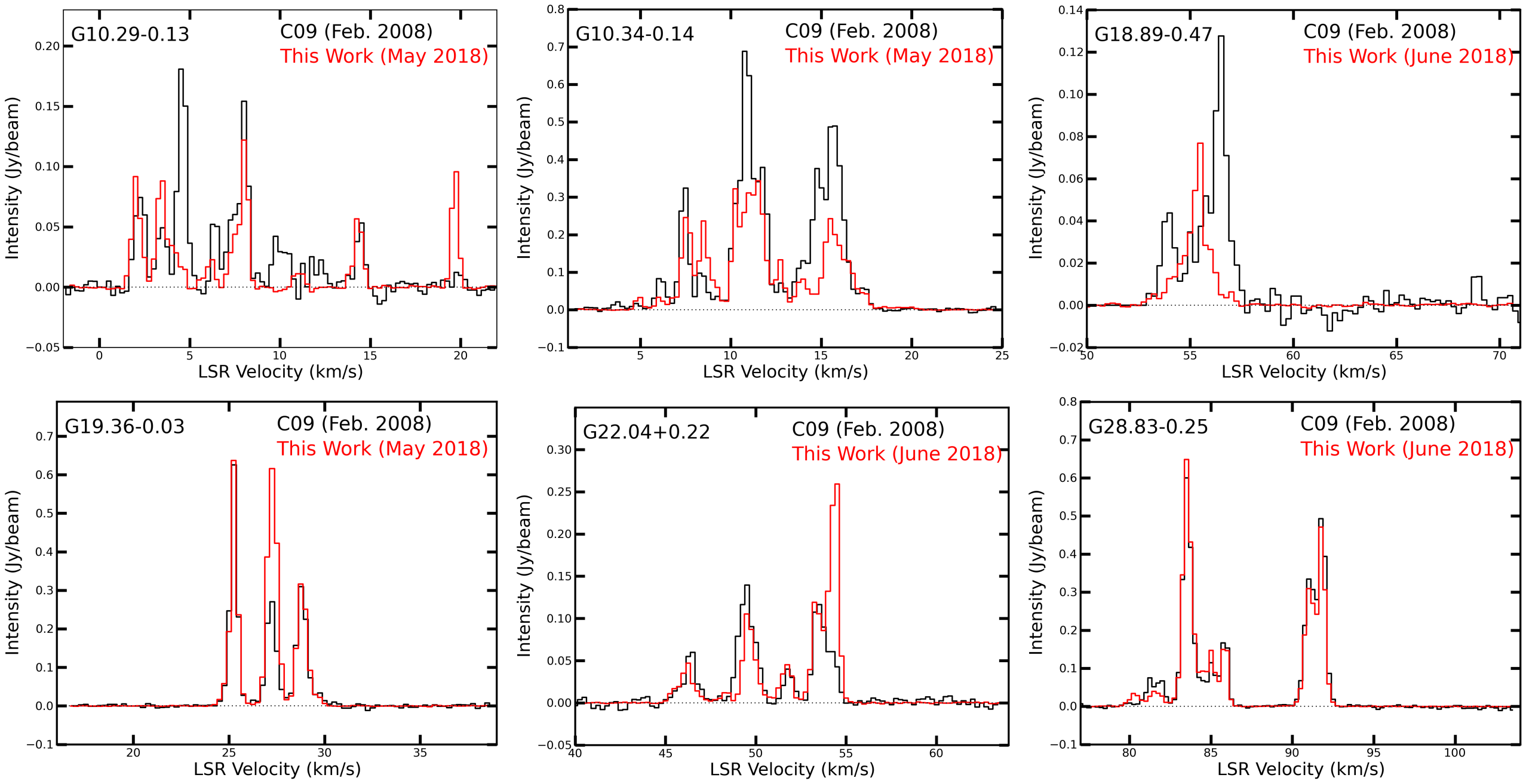

**Figure 8**. 6.7 GHz $CH_3OH$ maser line profiles for 2008 (Cyganowski et al. 2009, black, abbreviated as 'C09' in the plots) and 2018 (this work, red). Line profiles were extracted over identical regions for the two epochs of each source. Images were regridded and convolved to have identical angular and velocity resolutions before the profiles were extracted. $\Delta V = 0.25$ km s$^{-1}$ for all plots. The angular resolution of each image is $\sim 3''$; see Cyganowski et al. (2009) for exact values. These profiles suggest that some non-trivial variation in the 6.7 GHz $CH_3OH$ maser kinematics over the 10-year period is not uncommon, but the lack of any extreme ($>10\times$) changes in flux density suggests that these variations are more likely due to variations in the masing gas (i.e. a change in the projected or actual length of the masing column) rather than any significant variations in the pumping mechanism itself (i.e. the luminosity of the central MYSO).

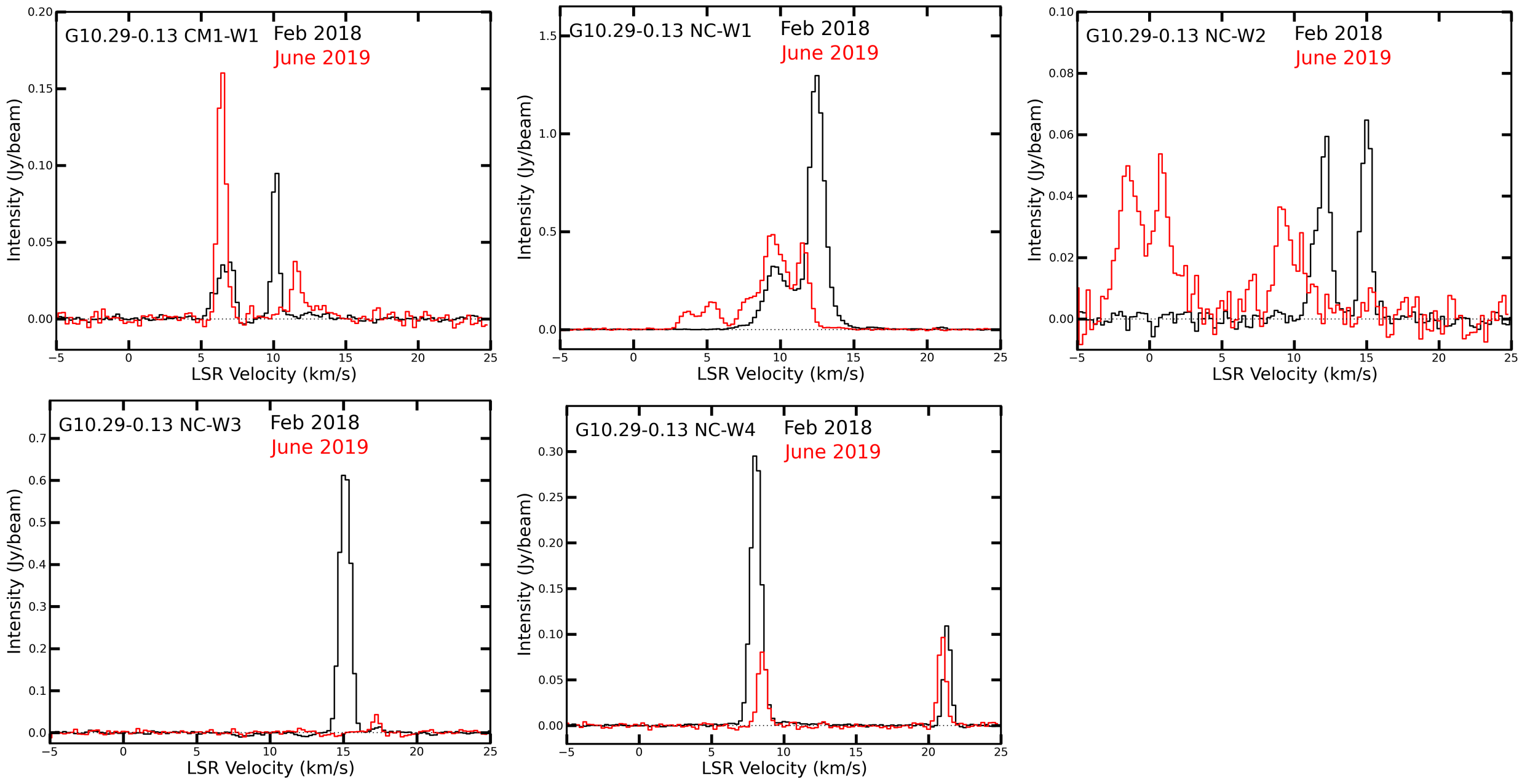

**Figure 9**. Epoch 2018.1 (black) and 2019.6 (red) $H_2O$ spectra for G10.29–0.13.

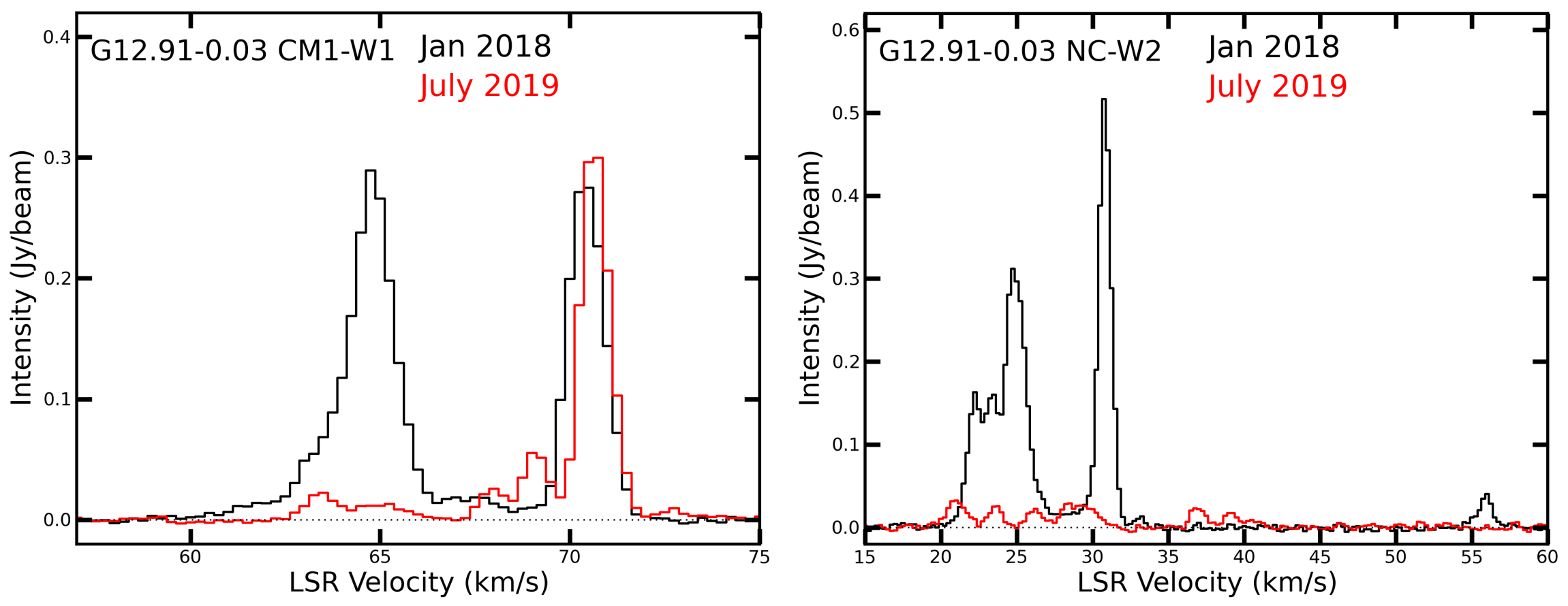

**Figure 10**. Epoch 2018.1 (black) and 2019.6 (red) $H_2O$ spectra for G12.91–0.03.

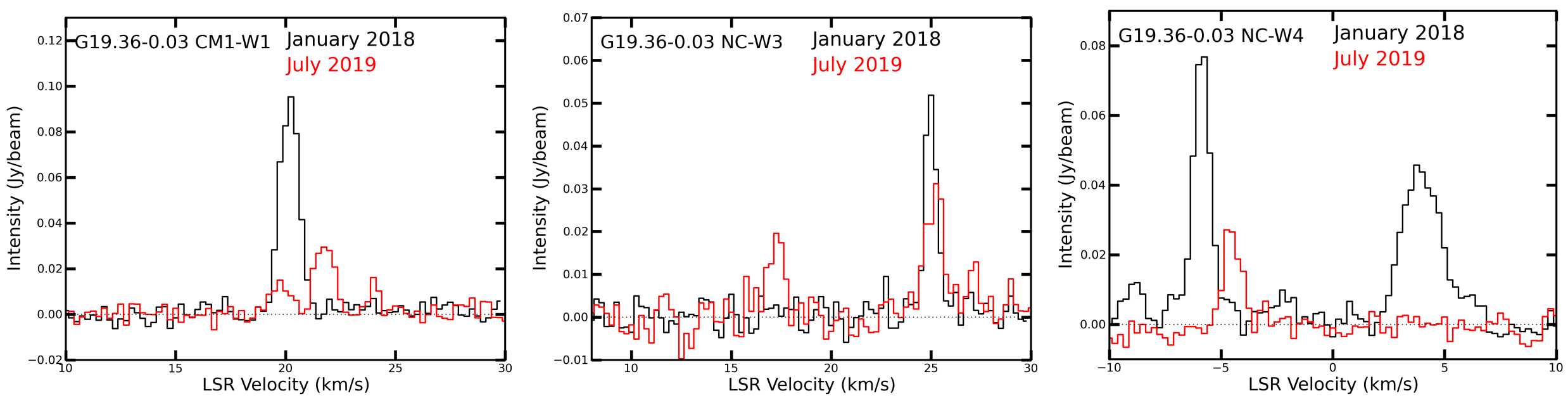

**Figure 11**. Epoch 2018.1 (black) and 2019.6 (red) $H_2O$ spectra for G19.36–0.03.

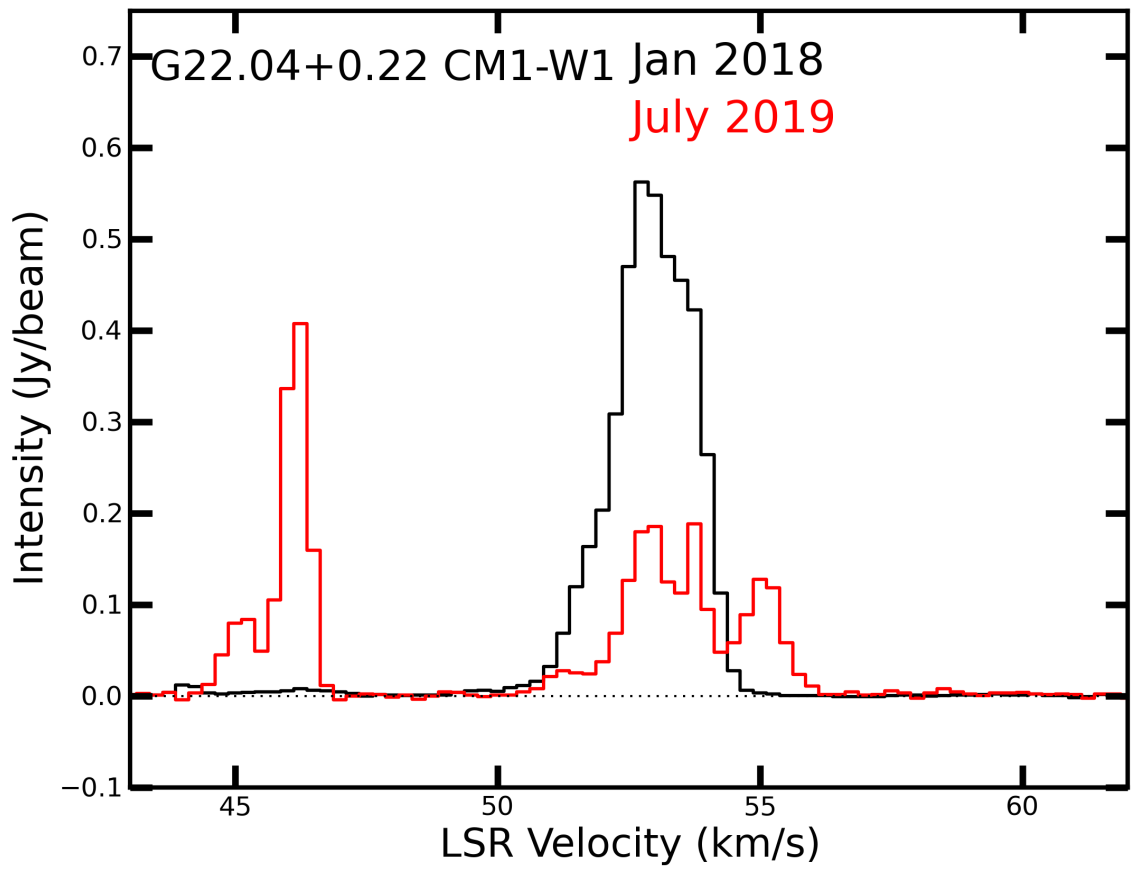

**Figure 12**. Epoch 2018.1 (black) and 2019.6 (black) $H_2O$ spectra for G22.04+0.22.

The velocity structure of each maser group is also largely unchanged between the two epochs: peak velocities and total velocity extents shift no more than 0.25 km s$^{-1}$ (1 channel) between the two epochs. The relative stability of these masers between the two observations is consistent with the observed behavior of Class I (collisionally-excited) $CH_3OH$ masers (Leurini et al. 2016). Both Class I $CH_3OH$ masers and the (few) known examples of $NH_3$ (3,3) masers are commonly associated with outflows (see, e.g., Mangum & Wootten 1994; Hunter et al. 2008, and references therein for examples of known $NH_3$ (3,3) masers).

## 5. CONCLUSIONS

We have examined a sample of 9 Extended Green Objects (EGOs) at 1.3 and 5 cm with sub-arcsecond angular resolution and 7-14 μJy continuum sensitivities, and found ubiquitous weak (<500 μJy), compact continuum emission ($\lesssim$2000 au). We detect a total of 41 continuum sources within the 1.3 cm HBPW in our 9 EGOs, of which five are partially resolved out and 36 have well-constrained flux densities or upper limits. The majority of our sources (53%) are unresolved at the current angular resolution ($\sim$0.″3-0.″5), while eight (22%) have fully-constrained sizes derived by `imfit` and nine (25%) have upper limits on their size, determined via aperture photometry. We identify two UC/HCH II regions previously published in the literature, one of which is a candidate for an expanding HCH II region and is potentially indicative of multiple stages of MYSO formation present within a single massive protocluster (Cyganowski et al. 2011). The remainder of our sources with derived or upper-limit sizes have sizes consistent with UC or HCH II regions, but are also consistent with ionized jets. They are typically underluminous compared to predicted H II region flux densities, but many have spectral indices indicative of non-trivial opacity so the possibility cannot be ruled out at this time. In all cases, sources with fully-constrained sizes are too large in at least one dimension to be consistent with gravitationally-trapped or outflow-confined H II regions; for sources which remain unresolved, we cannot rule out this possibility, although their typically-lower flux densities make this possibility unlikely in most cases.

Most EGOs have $\geq$3 continuum sources, of which one (usually) is associated with 6.7 GHz $CH_3OH$ maser emission. This is an increase in overall centimeter-continuum source multiplicity from previous 1.3 and 3.6 cm studies of EGO populations (Towner et al. 2017; Cyganowski et al. 2011). In most cases, the total continuum flux density in each EGO is clearly dominated by only one continuum source, and this is usually the continuum source which has associated 6.7 GHz $CH_3OH$ maser emission. As 6.7 GHz $CH_3OH$ masers are exclusively associated with massive protostars, this is consistent with the findings of Towner et al. (2019), who find massive-source multiplicities of 1-2 per EGO based on mid-infrared (19.7 and 37.1 μm) photometry.

We examine the relationship between radio distance-luminosity ($L_{radio}$) and bolometric luminosity ($L_{bol}$) for our sample at 6, 8, and 22 GHz using interpolated 8 GHz flux densities, and using three different spatial-selection criteria, in order to explore if and how this relationship changes with frequency and physical scale. We find a positive correlation between $L_{radio}$ and $L_{bol}$ in all cases, but the relationship is only statistically-significant in three cases: at 22 GHz using SOFIA-based apertures to determine $L_{radio}$, and at 8 GHz (interpolated) and 6 GHz using only the flux from the continuum sources which host 6.7 GHz $CH_3OH$ masers. Our derived relationships are also consistent with the relationships derived for radio emission dominated by thermal emission from jets (Anglada et al. 1996; Sanna et al. 2018; Rosero et al. 2019b) in all but one case. We find that these correlations are always dominated by the 6.7 GHz maser-hosting source at 22 GHz, but less so at 6 and 8 GHz. The 22 GHz correlation is also always shallower than the 6 and 8 GHz correlations. Considering the possibility of emission from warm dust at 22 GHz and its effect on derived spectral indices, we conclude that these correlations should be viewed with caution: the may describe emission from the dominant continuum sources well, but should not be taken as an indication that all continuum emission in EGOs arises from protostellar jets or from massive sources.

We find no statistically-robust relationship between $L_{radio}$ and $L_{H_2O}$ for our sample, regardless of the frequency observed or the spatial selection criteria used. It is possible that this is a consequence of small-number statistics, but compar-ison between our data and the fitted relationship of Sanna et al. (2018) suggest that our sample shows more intrinsic scatter than their sample does. Our lack of agreement with Sanna et al. (2018) is likely attributable to the different selection criteria for our two samples, as Sanna et al. (2018) specifically sought sources with plentiful 22 GHz $H_2O$ maser emission while our sample was selected first on the basis of extended 4.5 µm emission.

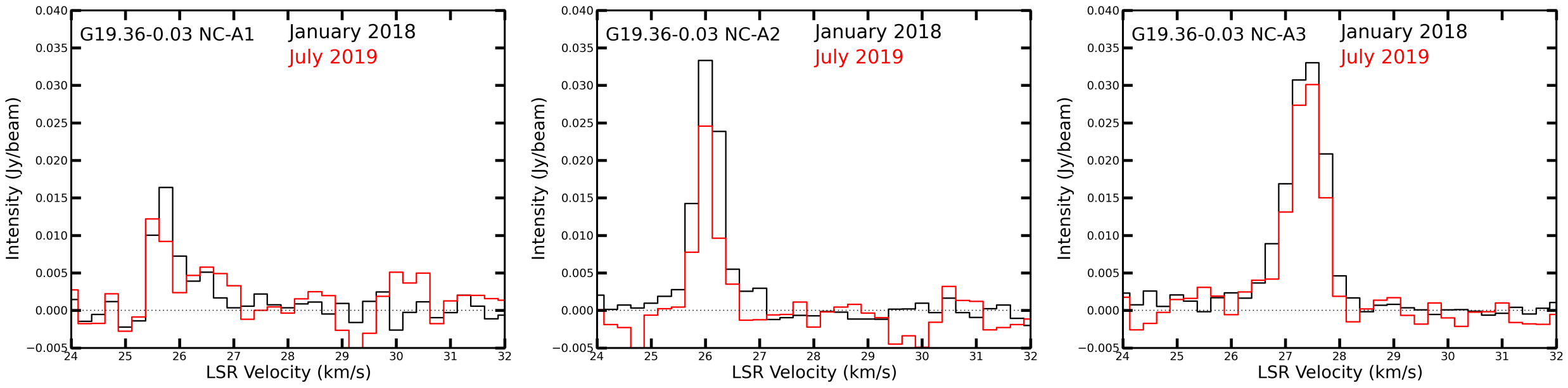

**Figure 13.** $NH_3$ (3,3) spectra for G19.36–0.03, for observation epochs January 2018 (black) and July 2019 (red). Our data show little variation at a statistically-significant level ($>5\sigma$) between the two epochs of observation. Typical rms values are 5.4 mJy beam$^{-1}$ (2018.1) and 8.8 mJy beam$^{-1}$ (2019.6), respectively.

Our $L_{bol}$-$L_{H_2O}$ values agree within errors with the relationship derived by Urquhart et al. (2011), as expected for massive clumps.

We find that the spatial distributions of 6.7 GHz $CH_3OH$ masers in our sample are consistent with the diverse morphologies observed in the literature (e.g. Bartkiewicz et al. 2016, 2009), as are variations in their spectra over a 10.5-year epoch (Cyganowski et al. 2009; Szymczak et al. 2018; Goedhart et al. 2004). We do not find $CH_3OH$ maser-flux variations of an order of magnitude or more, as has recently been observed in MYSO episodic-accretion candidates (Brogan et al. 2019; Caratti o Garatti et al. 2017; Hunter et al. 2017). We find that the 22 GHz $H_2O$ maser emission variation is much more volatile than the $CH_3OH$ maser emission, which is also consistent with the literature (Trinidad et al. 2003; Hunter et al. 1994). We do not find any significant variation in the kinematic structure of the $NH_3$ (3,3) maser data for which we have two epochs of observation, and all variations in peak intensity are limited to $\leq 4\sigma$. The overall stability of our $NH_3$ (3,3) masers is consistent with the stability of Class I (collisionally-excited) $CH_3OH$ masers which, like $NH_3$ (3,3) masers, are commonly associated with outflows.

The radio-continuum spectral indices we derive are largely consistent with free-free emission in protostellar jets, with a small population potentially representing very young hypercompact (HC) H II regions and a small population consistent with non-thermal emission (14% to 19%). We find that the spectral indices of our continuum sources span a broad range: $-0.42 < \alpha < 1.58$ for sources detected at both 1.3 and 5 cm. This broad range of spectral indices suggests that multiple emission mechanisms are present *simultaneously* in this EGO population. As EGOs are thought to occupy a specific, narrow evolutionary state (Towner et al. 2019; Cyganowski et al. 2009), this suggests that centimeter continuum emission in massive protoclusters may arise from multiple processes even at a single evolutionary stage– which may be indicative of multiple stages of high-mass star formation present simultaneously in these regions.

The sum total of our findings in this paper is consistent with a view of EGOs as young massive star-forming regions which are forming one to a few massive stars and are in an active stage of outflow and accretion prior to the emergence of H II regions. The presence of one known and one candidate HC/UC H II region in the sample suggest that EGOs may span a slightly broader range of states than previously thought, but the majority of our centimeter-continuum detections are consistent with less-luminous emission mechanisms such as jets.

Distinguishing between the ionized-jet and HC H II region origins for radio-continuum emission requires additional observations in the millimeter regime in order to constrain the source SEDs at (sub)millimeter wavelengths, and to account for the possibility of multiple emission mechanisms being present simultaneously in a single source, such as both free-free emission from an ionized jet and thermal emission from warm dust. We have recently obtained sub-arcsecond resolution observations of these EGOs at 1.3 and 3 mm using the Atacama Large Millimeter/submillimeter Array (ALMA). We will combine these data with the centimeter continuum data presented in this work in order to model the SED of each continuum detection and distinguish between dust, H II region, ionized jet, and non-thermal emission. These results will be published in our next paper.

## 6. ACKNOWLEDGEMENTS

We thank the referee for their thorough and helpful comments, which have helped improve the paper. Support for this work was provided by the NSF through the Grote Reber Fellowship Program administered by Associated Universities, Inc./National Radio Astronomy Observatory. Support for this work was provided by the NSF through award SOSP18A-007 from the NRAO. The National Radio Astronomy Observatory is a facility of the National Science

Foundation operated under agreement by the Associated Universities, Inc. This research made use of NASA's Astrophysics Data System Bibliographic Services and the SIMBAD database operated at CDS, Strasbourg, France. This research made use of APLpy, an open-source plotting package for Python (Robitaille & Bressert 2012).

*Facilities:* VLA, SOFIA